\documentclass[12pt,titlepage,letterpaper]{article}

\usepackage{amsmath}
\usepackage{amsfonts}
\usepackage{amssymb}
\usepackage{amsthm}
\usepackage{mathtools}
\usepackage{graphicx}
\usepackage{color}
\usepackage[height=8.8in,width=6.45in]{geometry}
\usepackage{xparse}
\usepackage{natbib}
\bibpunct{\textbf\bgroup[}{]\egroup}{,}{n}{}{} 
\usepackage{hyperref}
\usepackage{longtable}

\usepackage{tikz}
\def\includegraphics[width=#1]#2{\vcenter{\vspace*{2pt}\hbox{\resizebox{#1}{!}{\let\empty\varnothing\input{#2.tikz.tex}}}\vspace*{2pt}}}
\tikzset{>=latex}

\definecolor{aqua}{rgb}{0, 1.0, 1.0}
\definecolor{fuschia}{rgb}{1.0, 0, 1.0}
\definecolor{gray}{rgb}{0.502, 0.502, 0.502}
\definecolor{lime}{rgb}{0, 1.0, 0}
\definecolor{maroon}{rgb}{0.502, 0, 0}
\definecolor{navy}{rgb}{0, 0, 0.502}
\definecolor{olive}{rgb}{0.502, 0.502, 0}
\definecolor{purple}{rgb}{0.502, 0, 0.502}
\definecolor{silver}{rgb}{0.753, 0.753, 0.753}
\definecolor{teal}{rgb}{0, 0.502, 0.502}

\makeatletter
\newdimen\itex@wd%
\newdimen\itex@dp%
\newdimen\itex@thd%
\def\itexspace#1#2#3{\itex@wd=#3em%
\itex@wd=0.1\itex@wd%
\itex@dp=#2ex%
\itex@dp=0.1\itex@dp%
\itex@thd=#1ex%
\itex@thd=0.1\itex@thd%
\advance\itex@thd\the\itex@dp%
\makebox[\the\itex@wd]{\rule[-\the\itex@dp]{0cm}{\the\itex@thd}}}
\makeatother

\makeatletter
\newif\if@sup
\newtoks\@sups
\def\append@sup#1{\edef\act{\noexpand\@sups={\the\@sups #1}}\act}%
\def\reset@sup{\@supfalse\@sups={}}%
\def\mk@scripts#1#2{\if #2/ \if@sup ^{\the\@sups}\fi \else%
  \ifx #1_ \if@sup ^{\the\@sups}\reset@sup \fi {}_{#2}%
  \else \append@sup#2 \@suptrue \fi%
  \expandafter\mk@scripts\fi}
\def\tensor#1#2{\reset@sup#1\mk@scripts#2_/}
\def\multiscripts#1#2#3{\reset@sup{}\mk@scripts#1_/#2%
  \reset@sup\mk@scripts#3_/}
\makeatother

\makeatletter
\newbox\slashbox \setbox\slashbox=\hbox{$/$}
\def\itex@pslash#1{\setbox\@tempboxa=\hbox{$#1$}
  \@tempdima=0.5\wd\slashbox \advance\@tempdima 0.5\wd\@tempboxa
  \copy\slashbox \kern-\@tempdima \box\@tempboxa}
\def\slash{\protect\itex@pslash}
\makeatother

\def\clap#1{\hbox to 0pt{\hss#1\hss}}

\let\oldroot\root
\def\root#1#2{\oldroot #1 \of{#2}}
\renewcommand{\sqrt}[2][]{\oldroot #1 \of{#2}}

\DeclareSymbolFont{symbolsC}{U}{txsyc}{m}{n}
\SetSymbolFont{symbolsC}{bold}{U}{txsyc}{bx}{n}
\DeclareFontSubstitution{U}{txsyc}{m}{n}

\DeclareSymbolFont{stmry}{U}{stmry}{m}{n}
\SetSymbolFont{stmry}{bold}{U}{stmry}{b}{n}

\makeatletter
\def\re@DeclareMathSymbol#1#2#3#4{%
    \let#1=\undefined
    \DeclareMathSymbol{#1}{#2}{#3}{#4}}
\re@DeclareMathSymbol{\neArrow}{\mathrel}{symbolsC}{116}
\re@DeclareMathSymbol{\neArr}{\mathrel}{symbolsC}{116}
\re@DeclareMathSymbol{\seArrow}{\mathrel}{symbolsC}{117}
\re@DeclareMathSymbol{\seArr}{\mathrel}{symbolsC}{117}
\re@DeclareMathSymbol{\nwArrow}{\mathrel}{symbolsC}{118}
\re@DeclareMathSymbol{\nwArr}{\mathrel}{symbolsC}{118}
\re@DeclareMathSymbol{\swArrow}{\mathrel}{symbolsC}{119}
\re@DeclareMathSymbol{\swArr}{\mathrel}{symbolsC}{119}
\re@DeclareMathSymbol{\nequiv}{\mathrel}{symbolsC}{46}
\re@DeclareMathSymbol{\Perp}{\mathrel}{symbolsC}{121}
\re@DeclareMathSymbol{\Vbar}{\mathrel}{symbolsC}{121}
\re@DeclareMathSymbol{\sslash}{\mathrel}{stmry}{12}
\re@DeclareMathSymbol{\bigsqcap}{\mathop}{stmry}{"64}
\re@DeclareMathSymbol{\biginterleave}{\mathop}{stmry}{"6}
\re@DeclareMathSymbol{\invamp}{\mathrel}{symbolsC}{77}
\re@DeclareMathSymbol{\parr}{\mathrel}{symbolsC}{77}
\makeatother

\makeatletter
\DeclareRobustCommand\widecheck[1]{{\mathpalette\@widecheck{#1}}}
\def\@widecheck#1#2{%
    \setbox\z@\hbox{\m@th$#1#2$}%
    \setbox\tw@\hbox{\m@th$#1%
       \widehat{%
          \vrule\@width\z@\@height\ht\z@
          \vrule\@height\z@\@width\wd\z@}$}%
    \dp\tw@-\ht\z@
    \@tempdima\ht\z@ \advance\@tempdima2\ht\tw@ \divide\@tempdima\thr@@
    \setbox\tw@\hbox{%
       \raise\@tempdima\hbox{\scalebox{1}[-1]{\lower\@tempdima\box
\tw@}}}%
    {\ooalign{\box\tw@ \cr \box\z@}}}
\makeatother

\makeatletter
\NewDocumentCommand\mathraisebox{moom}{%
\IfNoValueTF{#2}{\def\@temp##1##2{\raisebox{#1}{$\m@th##1##2$}}}{%
\IfNoValueTF{#3}{\def\@temp##1##2{\raisebox{#1}[#2]{$\m@th##1##2$}}%
}{\def\@temp##1##2{\raisebox{#1}[#2][#3]{$\m@th##1##2$}}}}%
\mathpalette\@temp{#4}}
\makeatletter

\makeatletter
\def\udots{\mathinner{\mkern2mu\raise\p@\hbox{.}
\mkern2mu\raise4\p@\hbox{.}\mkern1mu
\raise7\p@\vbox{\kern7\p@\hbox{.}}\mkern1mu}}
\makeatother

\newcommand{\gt}{>}
\newcommand{\lt}{<}

\theoremstyle{plain}

\theoremstyle{definition}

\theoremstyle{remark}

\numberwithin{equation}{section}
\def\cN{\mathcal{N}}

\def\SU{\mathrm{SU}}
\def\SO{\mathrm{SO}}
\def\UU{\mathrm{U}}
\def\OO{\mathrm{O}}
\def\Spin{\mathrm{Spin}}
\def\Sp{\mathrm{Sp}}
\def\BP#1{{\color{red}{#1}}}
\begin{document}


\begin{titlepage}

\begin{flushright}
ICTP-SAIFR/2012-010\\
IPMU-12-0226\\
UT-12-43\\
UTTG-24-12\\
TCC-023-12
\end{flushright}
\vskip 1cm

\begin{center}
{\Large \bfseries
Gaiotto Duality for the Twisted $A_{2N-1}$ Series
}

\vskip 1.2cm

Oscar Chacaltana$^\natural$,
 Jacques Distler$^\sharp$, 
and Yuji Tachikawa$^\flat$

\bigskip
\bigskip

\begin{tabular}{ll}
$^\natural$ & ICTP South American Institute for Fundamental Research,\\
& Instituto de F\'isica Te\'orica, Universidade Estadual Paulista,\\
& 01140-070 S\~{a}o Paulo, SP, Brazil
\\[.5em]
$^\sharp$ & Theory Group and Texas Cosmology Center,\\
&
Department of Physics, University of Texas at Austin, Austin, TX 78712, USA \\[.5em]
$^\flat$  & 
Department of Physics, Faculty of Science, \\
& University of Tokyo,  Bunkyo-ku, Tokyo 133-0022, Japan, and \\
&Institute for the Physics and Mathematics of the Universe, \\
& University of Tokyo,  Kashiwa, Chiba 277-8583, Japan
\end{tabular}

\vskip 1.5cm

\textbf{Abstract}
\end{center}

\medskip
\noindent
We study 4D $\mathcal{N}=2$ superconformal theories that arise from the compactification of 6D $\cN=(2,0)$ theories of type $A_{2N-1}$ on a Riemann surface $C$, in the presence of punctures twisted by a $\mathbb{Z}_2$ outer automorphism. We describe how to do a complete classification of these SCFTs in terms of three-punctured spheres and cylinders, which we do explicitly for $A_3$, and provide tables of properties of twisted defects up through $A_9$. 
We find atypical degenerations of Riemann surfaces that do not lead to weakly-coupled gauge groups, but to a gauge coupling pinned at a point in the interior of moduli space.

As applications, we study: i) 6D representations of 4D superconformal quivers in the shape of an affine/non-affine $D_n$ Dynkin diagram, ii) S-duality of $\SU(4)$ and $\Sp(2)$ gauge theories with various combinations of fundamental and antisymmetric matter, and iii) realizations of all rank-one SCFTs  predicted by Argyres and Wittig.

\bigskip
\vfill
\end{titlepage}

\setcounter{tocdepth}{2}
\tableofcontents
\newpage

\hypertarget{introduction_1}{}
\section{{Introduction and Summary}}\label{introduction_1}

Considerable progress has been made recently in the program of understanding the 4D theories that arise from the compactification of 6D $\cN=(2,0)$ theories on a Riemann surface, $C$, possibly in the presence of codimension-two defects of the (2,0) theories, which correspond to punctures on $C$ \cite{Gaiotto:2009we,Tachikawa:2009rb,Nanopoulos:2009xe,Chacaltana:2010ks,Chacaltana:2011ze}\footnote{To avoid confusion, we note that, in this paper, $\textbf{[1]}$ denotes a reference, $\BP{[5]}$ stands for a B-partition (an embedding $\mathfrak{sl}(2)\to \mathfrak{so}(2N+1)$), and $[3]$ is a generic partition.}. Much of the richness of the construction stems from the variety of available defects. 
When an $\cN=(2,0)$ theory of type $J=A$, $D$, or $E$ has a nontrivial outer-automorphism group, 
there exists, in addition to untwisted defects, a sector of twisted defects equipped with an action of a element of the outer-automorphism group of $J$ as one goes around the defect. The general local properties of twisted and untwisted defects were studied in \cite{Chacaltana:2012zy}. 
In particular, the $A_{2N-1}$ series has a sector of defects twisted by the $\mathbb{Z}_2$ outer automorphism of $A_{2N-1}$. In this paper, we study the \emph{global} properties of theories of type $A_{2N-1}$ in the presence of such $\mathbb{Z}_2$-twisted defects.

Just as \emph{untwisted} defects of the $A_{2N-1}$ series are classified by embeddings of $\mathfrak{sl}(2)$ in $\mathfrak{sl}(2N)$, \emph{twisted} defects in this series are classified by embeddings of $\mathfrak{sl}(2)$ in $\mathfrak{so}(2N+1)$. Equivalently, untwisted defects are classified by partitions of $2N$, while twisted defects are classified by certain partitions of $2N+1$, called B-partitions. So, for instance, the twisted sector contains a ``maximal'' twisted puncture, denoted by the B-partition $\BP{[1^{2N+1}]}$ and with flavour group $\SO(2N+1)$, and a ``minimal'' twisted puncture, denoted $\BP{[2N+1]}$ and with trivial flavour group. The local properties of these and other twisted punctures can be computed following \cite{Chacaltana:2012zy}.  In this paper we will provide some new, explicit algorithms to make these calculations easier.

One especially interesting twisted defect is the one with B-partition $\BP{[2N-1,1^2]}$, which arises from the collision of a minimal untwisted and a minimal twisted defect. Such defect is unique in that it can be continuously deformed into a pair of defects $\BP{[2N+1]}$ and $[2N-1,1]$.
In the global picture, this property leads to a number of elements that were absent in the untwisted story:\begin{itemize}%
\item three-punctured spheres that correspond to gauge theories fixed at a point in the interior (not a cusp) of their moduli space.
\item cylinders whose pinching leave a gauge coupling at a point in the interior of moduli space.
\item cylinders whose pinching decouples a semisimple gauge group.
\end{itemize}
These phenomena part ways with the usual understanding that a degeneration of a Riemann surface corresponds to the weakly-coupled limit of a simple gauge group in the 4D theory and vice versa.

\bigskip

We then study three applications of our constructions:\begin{itemize}
\item It is well known that Lagrangian 4D $\mathcal{N}=2$ superconformal quivers whose gauge group is a product of special-unitary groups can be constructed only in the shapes of (affine and non-affine) Dynkin diagrams of type A-D-E \cite{Katz:1997eq}. The $A_n$-shaped quivers were used originally by Gaiotto to deduce local properties of untwisted defects in the $A_{N-1}$-series, and are realized as compactifications on untwisted spheres. In this paper we find a realization of the affine and non-affine $D_n$-shaped quivers as compactifications of the $A_{2N-1}$ series on twisted spheres. An equivalent expression for the Seiberg-Witten curve for the \emph{affine} $D_n$-shaped quivers was found long ago by Kapustin \cite{Kapustin:1998fa} from a IIA brane construction. (Quite recently, a uniform way to derive the Seiberg-Witten solutions for these ADE quiver theories were found by \cite{Fucito:2012xc,Nekrasov:2012xe} using instanton calculus.)
\item We present a full tinkertoy representation of the twisted $A_3$ theory, and, as an application, study the S-dual frames of $\SU(4)$ and $\Sp(2)$ gauge theories with matter in the fundamental and antisymmetric representations. 
\item We show how to construct the rank-one 4D SCFTs studied by Argyres and Wittig in \cite{Argyres:2007tq,Argyres:2010py}.\footnote{These were called \emph{non-maximal} mass-deformations of the $E_{6,7,8}$ theories in \cite{Argyres:2007tq,Argyres:2010py}, but we prefer to call them just \emph{new} rank-one theories, since physically they are not deformations of the $E_{6,7,8}$ theories, although their Seiberg-Witten curves are.} 
These are SCFTs whose only Coulomb branch operator has scaling dimension $\Delta=3,4,6$ respectively, and which are \emph{not} the more familiar Minahan-Nemeschansky theories with $E_{6,7,8}$ flavour symmetry. 
The $\Delta=3$ theory will be found in the context of the twisted $A_2$ theory, although we leave a systematic analysis of the twisted $A_{2N}$ series for later, due to the subtle issues  pointed out in \cite{Tachikawa:2011ch}.  We will be able to pin down numerical invariants of this $\Delta=3$  theory left undetermined in \cite{Argyres:2007tq,Argyres:2010py}
\end{itemize}

The rest of the paper is organized into two parts. 
The first part consists of Sec.~2,~3~and~4. In Sec.~\ref{local_properties_of_twisted_punctures_2}, after recalling the general method for obtaining a 4D theory from a 6D $\cN=(2,0)$ theory on a Riemann surface, we describe the algorithms to compute the local properties of twisted punctures of type $A_{2N-1}$, elaborating on \cite{Chacaltana:2012zy}. In Sec.~\ref{irregular_punctures_6}, we develop the method to identify the behaviour of the theory when two defects are brought together. 
In Sec.~\ref{atypical_degenerations_11}, we study atypical degenerations in detail, where the degeneration of the Riemann surface does not correspond to the emergence of a weakly-coupled gauge group. 

The second part of the paper deals with applications. In Sec.~\ref{dshaped_quivers_15} we show how a\linebreak $D_n$-shaped quiver gauge theory can be realized in terms of the 6D $\cN=(2,0)$ theory of type $A_{2N-1}$ on a sphere with twisted punctures. In Sec.~\ref{_and__gauge_theories_21} we study the S-duality properties of all superconformal $\SU(4)$ and $\Sp(2)$ gauge theories. In Sec.~\ref{rank1_scfts_33} we discuss rank-one SCFTs and their realizations in terms of 6D  $\cN=(2,0)$ theory.  In  Appendix~\ref{list_of_twisted_punctures_35} we list all twisted fixtures and cylinders for $A_3$, and tabulate the properties of twisted punctures for $A_{5,7,9}$.

\hypertarget{local_properties_of_twisted_punctures_2}{}

\section{4D Theories and Punctures}\label{local_properties_of_twisted_punctures_2}
In Sec.~\ref{intro_to_section2} we recall the construction of 4D theories from the compactification of the 6D $\cN=(2,0)$ theory of type $A_{2N-1}$ on Riemann surfaces with punctures.  Sec.~\ref{localproperties} through \ref{constraints_4} detail algorithms to compute local properties of punctures. We show extensive tables of local properties in Appendix~\ref{list_of_twisted_punctures_35}. After going over Sec.~\ref{intro_to_section2}, a busy reader can skip the rest of this section, and continue directly to Sec.~\ref{irregular_punctures_6}.

\subsection{Punctures, the fields $\phi_k(z)$ and the Hitchin field $\Phi(z)$}\label{intro_to_section2}
Consider the 6D (2,0) theory of type $A_{2N-1}$, compactified on a Riemann surface $C$ with a partial twist to preserve supersymmetry \cite{Gaiotto:2009we,Gaiotto:2009hg}. We allow for the possibility of having codimension-two defects of the (2,0) theory, localized at punctures on $C$. This construction  leads at low energies to a 4D $\mathcal{N}=2$ SCFT. We are interested in classifying and characterizing the 4D SCFTs that arise for various choices of $C$ and defects on it.

Usually, the moduli space of the 4D SCFT can be identified with the complex-structure moduli space of $C$, so that cusps in the latter correspond to weakly-coupled limits of the theory, where a certain gauge group almost decouples. We will see in Sec.~\ref{atypical_degenerations_11} that there exist counterexamples to this statement when twisted punctures are included.

The Seiberg-Witten curve $\Sigma$ of the theory can be realized as a ramified cover of $C$. To describe $\Sigma$ explicitly, we should consider the VEVs of certain protected operators in the 6D theory, which, upon compactification on $C$, give rise to meromorphic $k$-differentials $\phi_k$ on $C$, where the $k$ are the dimensions of the Casimirs of $A_{2N-1}$, i.e., $k=2,3,\dots, N$. The $\phi_k$ have poles at the positions of the punctures on $C$. We then have the following equation for $\Sigma$:
\begin{equation}
\Sigma: \lambda^{2N}+\sum_{k=2}^{2N}\lambda^{2N-k}\phi_k = 0.
\label{genericswcurve}\end{equation}
Here $\lambda$ is the Seiberg-Witten differential, which is a meromorphic 1-form on $\Sigma$.

In \cite{Chacaltana:2012zy}, we discussed a classification of codimension-two defects and how to compute their properties. Defects are grouped into sectors that are in 1-to-1 correspondence with the conjugacy classes of the outer-automorphism group of the simply-laced Lie algebra of the same type as the (2,0) theory that one chooses. In our case, this Lie algebra is $A_{2N-1}=\mathfrak{sl}(2N)$, which has a $\mathbb{Z}_2$ outer-automorphism group generated by an element $o$, whose action on the $k$-differentials is:
\begin{equation}
o: \phi_k \mapsto (-1)^k\phi_k.
\label{actionouterautmorphism}\end{equation}
Then, the sector of untwisted punctures is the one corresponding to the identity element, while the twisted sector corresponds to $o$. As one goes around a twisted puncture on $C$, \eqref{actionouterautmorphism} tells us that the $k$-differentials of odd degree $k$ must change sign.

Now, untwisted punctures are classified by $\mathfrak{sl}(2)$ embeddings in $\mathfrak{sl}(2N)$, whereas twisted punctures are classified by $\mathfrak{sl}(2)$ embeddings in $\mathfrak{so}(2N+1)$.
More practically, recall that $\mathfrak{sl}(2)$-embeddings in $\mathfrak{sl}(2N)$ are in bijection with partitions of $2N$. Similarly, $\mathfrak{sl}(2)$-embeddings in $\mathfrak{so}(2N+1)$ are in bijection with \emph{B-partitions} of $2N+1$, which are defined as partitions of $2N+1$ where every even part has even multiplicity. For example, $\BP{[4^2,3^3,2^6]}$ is a B-partition, but $[6,5,4^2]$ is not.

If $z$ is a local coordinate on $C$ such that the puncture is at $z=0$, the $k$-differentials near $z=0$ have the behaviour:
\begin{equation}
\phi_k(z)=\frac{c^{(k)}_{p_k}}{z^{p_k}}+\cdots
\label{generickdifferentialexpansion}\end{equation}
We call the set $\{p_k\}$, for $k=2,\dots, 2N$, the \emph{pole structure} of the puncture. For an untwisted puncture, all the $p_k$ should be integer, while for a twisted one, the $p_k$ for odd (even) $k$ must be half-integer (integer) because of \eqref{actionouterautmorphism}.

Let us now relate the discussion to the Hitchin system. Following \cite{Gaiotto:2009hg}, the classical integrable system associated to our 4D $\mathcal{N}=2$ theories is a Hitchin system on $C$ with gauge group $\mathfrak{sl}(2N)$. Let $\Phi$ be the Higgs field for the Hitchin system, i.e., $\Phi$ is an $\mathfrak{sl}(2N)$-valued meromorphic 1-form on $C$ in the adjoint representation of $\mathfrak{sl}(2N)$. Then, the Seiberg-Witten curve $\Sigma$ of this system, \eqref{genericswcurve}, is given by the spectral curve for the Hitchin system,\begin{equation}
\Sigma: \text{det}(\Phi-\lambda I) = 0.
\label{spectralcurve}\end{equation}
Thus, comparing with \eqref{genericswcurve}, we see that the $\phi_k$ are polynomials in the trace invariants $\text{Tr}(\Phi^k)$ of the Higgs field.

In terms of the Hitchin system, an untwisted defect on $C$ corresponds to a local boundary condition for the Higgs field. Specifically, in local coordinates $z$ on $C$, let the untwisted puncture be at $z=0$. Then,  we have
\begin{equation}
\Phi(z)=\frac{X}{z}+\cdots,
\label{untwistedbdycondition}\end{equation}
where $X$ is an element in $\mathfrak{sl}(2N)$ specifying the puncture, 
and the ellipsis denotes a generic element of $\mathfrak{sl}(2N)$. 
Since $\Phi$ is not gauge invariant, the defect is actually characterized by the conjugacy class of $X$, known as a (co)adjoint orbit in $\mathfrak{sl}(2N)$.  When the mass parameters of the puncture are set to zero, $X$ is nilpotent, and the orbit is called a nilpotent orbit.

Nilpotent orbits in $\mathfrak{sl}(2N)$ are classified by $\mathfrak{sl}(2)$ embeddings in $\mathfrak{sl}(2N)$, or, equivalently, by partitions of $2N$. If a puncture is labeled by a partition $\rho$, the nilpotent orbit that defines its boundary condition \eqref{untwistedbdycondition} is the one corresponding to the transpose partition $\rho^T$ of $2N$.

The analogous boundary condition for a twisted puncture was given in \cite{Chacaltana:2012zy}. First, decompose the $\mathfrak{sl}(2N)$ Lie algebra as a direct sum of eigenspaces of the $\mathbb{Z}_2$ outer automorphism, $\mathfrak{j}=\mathfrak{j}_{1}+\mathfrak{j}_{-1}$, where $\mathfrak{j}_{1}\simeq\mathfrak{sp}(N)$ is the invariant subalgebra. Then, if the twisted defect is at $z=0$, the local boundary condition for the Higgs field is
\begin{equation}
\Phi(z) = \frac{X}{z}+\frac{A}{z^{1/2}} + A'.
\label{twistedbdycondition}\end{equation}
Here, $X$ is an element of a nilpotent orbit in $\mathfrak{sp}(N)$, and $A$ and $A'$ are generic elements of $\mathfrak{j}_{-1}$ and $\mathfrak{sp}(N)$, respectively.

As before, nilpotent orbits in $\mathfrak{sp}(N)$ are classified by $\mathfrak{sl}(2)$ embeddings in $\mathfrak{sp}(N)$, or, equivalently, by \emph{C-partitions} of $2N$, which are defined as partitions of $2N$ where every odd part has even multiplicity. (For example, $[6^2,3^4,2]$ is a C-partition, but $[5^2,3,1]$ is not.)
Then, a twisted puncture in the $A_{2N-1}$ series is labeled by a B-partition $\rho$ of $2N+1$, but its Higgs-field boundary condition is given by a C-partition $\rho'$ of $2N$.
There is a map, called the Sommers-Achar map \cite{Sommers,AcharSommers,Achar}, which is a generalization of the Spaltenstein map on nilpotent orbits, which gives us the Hitchin-system data associated to $\rho$:
\begin{equation}
d_{S}:\rho \mapsto (\rho',C(\rho)).
\end{equation}
Here, $C(\rho)$ is a discrete group. Then, $X$ is the nilpotent element $\rho'(\sigma^+)$, seeing $\rho'$ as an $\mathfrak{sl}(2)$ embedding in $Sp(N)$. In \cite{Chacaltana:2012zy}, $\rho'$ was called the \emph{Hitchin pole} of the puncture labeled by the \emph{Nahm pole} $\rho$. $\rho'$ is given by the C-collapse of the reduction of the transpose of $\rho$ (a $2N+1$ partition) to a $2N$ partition; see Sec.~2.2 and Sec.~3.4.4 of \cite{Chacaltana:2012zy}.

\subsection{Local properties of punctures}\label{localproperties}

The local properties of a twisted puncture that we can compute are:\begin{enumerate}%
\item the pole structure $\{p_k\}$,
\item the contributions $\{n_k\}$ to the number of Coulomb branch operators with scaling dimension $k$,
\item the constraints on the leading coefficients $c^{(k)}_{p_k}$,
\item the local flavour symmetry group,
\item the contribution to the conformal-anomaly central charges, $(a,c)$, or, equivalently, to the effective number of hypermultiplets and vector multiplets, $(n_h,n_v)$.
\end{enumerate}
Let us briefly explain the terminology. A \emph{constraint} refers to a relation among leading coefficients $c^{(k)}_{p_k}$, or to the fact that a leading coefficient can be expressed in terms of more basic gauge-invariants, as we will see in Sec. \ref{constraints_4}. (In principle, \emph{subleading} coefficients may have been constrained too, but it turns out that this does not occur.)

Once the local form of the Higgs field $\Phi$ for a specific puncture is known, as in \eqref{untwistedbdycondition} and \eqref{twistedbdycondition}, one can find the local form of the $k$-differentials from \eqref{genericswcurve} and \eqref{spectralcurve}, read off the the pole structure $\{p_k\}$, find the constraints, and compute the $\{n_k\}$.

However, carrying out this ``honest'' procedure is quite tedious in practice.  In what follows, we describe algorithms to compute these properties directly from the B-partition, which we found after looking at a large number of examples.   First, in Sec.~\ref{graded_coulomb_branch_dimensions_3}, we explain how to calculate the $\{n_k\}$, and then, in Sec.~\ref{constraints_4}, how to compute the constraints. Once these are known, the pole structure $\{p_k\}$ can be easily reconstructed.
We will see that the only twisted defect that gives rise to a Coulomb branch operator of dimension \emph{two} is the one with B-partition $\BP{[2N-1,1^2]}$. This occurs through a constraint of the form $c^{(4)}_{3}=(a^{(2)}_{3/2})^2$. This particular puncture will play an important role in Sec.~\ref{atypical_degenerations_11}.

For untwisted punctures in the $A$ series, it is well known that there are no constraints at all, and so the pole orders $\{p_k\}$ are exactly the same as the $\{n_k\}$, for each $k$. (By contrast, untwisted punctures in the $D$ series generically do exhibit constraints \cite{Chacaltana:2011ze}.)

The Lie algebra of the global symmetry group $G_{\text{flavour}}$ of a twisted puncture labelled by the embedding $\rho:\mathfrak{su}(2)\to \mathfrak{so}(2N+1)$ is the commutant of $\rho(\mathfrak{su}(2))$ in $\mathfrak{so}(2N+1)$. It is easier to give a formula for $G_{\text{flavour}}$ in terms of the B-partition $p$ corresponding to $\rho$:
\begin{equation}
G_{\text{flavour}}=S\left[\prod_{l \text{ odd}} \OO(n_l)\times \prod_{l \text{ even}} \Sp(n_l/2)\right]
\end{equation}
where $l$ runs over the parts of the partition $p$, and $n_l$ is the multiplicity of $l$ in $p$. For even $l$, $n_l$ must be even because $p$ is a B-partition, so the formula above makes sense. (In our notation, $\Sp(1)\simeq \SU(2)$.)

As for the contributions to $n_h$ and $n_v$ (and thus $a$ and $c$), these can be easily computed from the embedding $\rho:\mathfrak{su}(2)\to\mathfrak{so}(2N+1)$. The formulas were given in \cite{Chacaltana:2012zy},
\begin{equation}
\begin{aligned}
n_h(\rho)&=8\left(\rho_A\cdot\rho_A - \rho_B\cdot\frac{h}{2}\right)+\frac{1}{2}\dim\mathfrak{g}_{1/2},\\
n_v(\rho)&=8\left(\rho_A\cdot\rho_A - \rho_B\cdot\frac{h}{2}\right)+\frac{1}{2}(\text{rank}(A_{2N-1})-\dim\mathfrak{g}_{0}),
\end{aligned}
\end{equation}
Here, $\rho_A$ and $\rho_B$ are the Weyl vectors of $A_{2N-1}$ and $B_N$, respectively; $h=\rho(\sigma_3)$ is the Cartan of the embedded $\mathfrak{su}(2)$, and we have decomposed $\mathfrak{g}=\mathfrak{so}(2N+1)=\sum_{r\in \mathbb{Z}+1/2} \mathfrak{g}_{r}$, where $\mathfrak{g}_r$ is the eigenspace of $h$ with eigenvalue $r$. 
The contributions to $n_h$ and $n_v$ for the twisted sectors of the $A_{3,5,7,9}$ theories are given in Appendix~\ref{list_of_twisted_punctures_35}.

\hypertarget{graded_coulomb_branch_dimensions_3}{}\subsection{{Graded Coulomb branch dimensions}}\label{graded_coulomb_branch_dimensions_3}

Consider a twisted puncture in the $A_{2N-1}$ theory, specified by a B-partition $p$ of $2N+1$. We want to compute the contributions $\{n_k\}$ to the dimensions of the graded Coulomb branch. The formula for the $\{n_k\}$ is most easily expressed in terms of a number of auxiliary sequences, which we now define.

Let $q=p^t$ be the transpose partition to $p$. First, let us define a sequence $\nu$ by
\begin{equation}
\nu=(\underbrace{1,\ldots,1}_{q_1},\underbrace{2,\ldots,2}_{q_2},\ldots).
\end{equation}
Next, let $s$ be the sequence of partial sums of $q$,
\begin{equation}
s_i=\sum_{j=1}^i q_j.
\label{si}\end{equation}
Finally, define a sequence $r$ of ``{}corrections''{} by
\begin{equation}
r_k=\begin{cases}
1 & \text{if}\, k\leq N\,\,\text{and}\,\, 2k\notin s,\\
0 & \text{otherwise}.
\end{cases}
\end{equation}
Then, the contribution $n_k$ for the twisted puncture with B-partition $p$ is
\begin{equation}
n_k(p)=\left\lceil  \frac{k+1-\nu_{k+1}}{2} \right\rceil - r_k +\frac{k}{2}-\left\lfloor\frac{k-1}{N}\right\rfloor\, ,
\end{equation}
where $\left\lceil ... \right\rceil$ and $\left\lfloor ... \right\rfloor$ denote the ceiling and floor functions, respectively.

For example, the contributions $\{n_k\}$ for the minimal twisted puncture, $p=\BP{[2N+1]}$, can be shown to be
\begin{equation}
n_k(\BP{[2N+1]}) = \frac{k}{2}-\left\lfloor\frac{k-1}{N}\right\rfloor\, .
\end{equation}
Similarly, the $\{n_k\}$ for the full (or maximal) twisted puncture, $p=\BP{[1^{2N+1}]}$, are
\begin{equation}
n_k(\BP{[1^{2N+1}]}) = \frac{3k}{2} -\left\lfloor\frac{k}{2}\right\rfloor-1.
\end{equation}

\hypertarget{constraints_4}{}\subsection{{Constraints}}\label{constraints_4}

\subsubsection{General structure of constraints}

The structure of constraints for twisted punctures in the A series is relatively simple. These constraints satisfy some properties:
\begin{itemize}%
\item They are polynomials in the \emph{leading} coefficients $c^{(k)}_{l}$ (that is, $l=p_k$), and, possibly, \emph{new} parameters, $a^{(k')}_{l'}$.

\item All terms in a constraint should have the same total scaling dimension and pole order. (The $c^{(k)}_l$ and $a^{(k)}_l$ have scaling dimension $k$ and pole order $l$.) Hence, we can talk about the scaling dimension and pole order of a constraint.

\item For each $k=2,...,2N$, there is \emph{at most one} constraint of scaling dimension $k$.

\item A constraint of scaling dimension $k$, if it exists, should be \emph{linear} in $c^{(k)}_l$, i.e., it will be of the form $c^{(k)}_l=f^{(k)}_{l}(c,a)$, where $f^{(k)}_{l}(c,a)$ stands for a polynomial (in other coefficients $c^{(k')}_{l'}$ and parameters $a^{(k'')}_{l''}$) of scaling dimension $k$ and pole order $l=p_k$.

\item The polynomials $f^{(k)}_l$ always have the form of squares or cross-terms (as in the expansion of the square of a sum), or sums of these.

\end{itemize}
Let us look at some representative examples of constraints:
\[
\begin{array}{rr@{}l@{\quad\quad}rr@{}l}
\text{i)}&  c_{7}^{(12)}&=\frac{1}{4}\left(c_{7/2}^{(6)}\right)^{2},&
\text{ii)}& c_{7/2}^{(7)}&=\frac{1}{2}c_{2}^{(4)}c_{3/2}^{(3)}, \\
\text{iii)}& c_{6}^{(10)}&=\left(a_{3}^{(5)}\right)^{2},&
\text{iv)}& c_{5}^{(9)}&=a_{3}^{(5)}c_{2}^{(4)}.
\end{array}
\]
The first and third examples are ``{}squares''{}, while the second and fourth are ``{}cross-terms''{}. Also, the first and second examples involve only the $c^{(k)}_l$, while the third and fourth involve also new parameters $a^{(k)}_l$.

We call constraints that do not introduce any new parameters, such as the first two examples above, \emph{c-constraints}. A c-constraint of scaling dimension $k$ (which necessarily has pole order $l=p_k$) tells us that the leading coefficient $c^{(k)}_l$ is dependent on others, and so the local contribution to $n_k$ should be reduced by one.

By contrast, the third example, of the form $c^{(2k)}_{2l}=(a^{(k)}_{l})^{2}$, tells us that $c^{(2k)}_{2l}$ is the square of another, more basic gauge-invariant parameter, $a^{(k)}_{l}$. Thus, it effectively trades a parameter of scaling dimension $2k$ by a parameter of scaling dimension $k$; in other words, the contribution to $n_{2k}$ is reduced by one, while the contribution to $n_k$ is raised by one. We call this type of constraint, which introduces a new parameter, an \emph{a-constraint}.

Finally, the fourth example is a cross-term involving the parameter $a^{(k)}_{l}$. However, the $a^{(k)}_{l}$ will already have been introduced by an a-constraint as in the previous paragraph. Hence, this cross-term constraint should be taken to be a c-constraint, not an a-constraint. Generically, for every $a^{(k)}_{l}$, there will be exactly \emph{one} a-constraint (a square in $a^{(k)}_{l}$) that introduces it, and (possibly) many cross-term c-constraints (linear in $a^{(k)}_{l}$).

\subsubsection{Number of constraints}\label{numberofconstraints}
Now, for a given twisted puncture, let us explain the rule to find at which scaling dimensions $k$ there exists a constraint.
Denote by $p$ the B-partition of $2N+1$ that labels our twisted puncture. Consider, as before, the transpose partition, $q=p^{t}$, and let $s$ be the sequence of partial sums of $q$, as in \eqref{si} above. We will see that $s$ contains all the information about constraints.

Let us first note that a B-partition always has an odd number of parts, so suppose our B-partition $p$ has $2l+1$ parts, and let $p_{2l+1}$ be the last part of $p$. Then, an a-constraint of scaling dimension $k$ exists if and only if:
\begin{enumerate}%
\item $k$ belongs to $s$.
\item $k$ is even.
\item $k$ is not a multiple of $2l+1$.

\end{enumerate}
If $k=2m$ satisfies these conditions, the local contribution to $n_{2m}$ should be reduced by one, and the contribution to $n_{m}$ should be raised by one.

Similarly, a c-constraint of scaling dimension $k$ exists if and only if:
\begin{enumerate}%
\item $k$ belongs to $s$.
\item If $k$ is odd, it must satisfy a ``{}cross-term''{} condition. Let $j$ be the unique index such that $k=s_j$. Then: 1) $s_j$ must \emph{not} be the last element of $s$; 2) both $s_{j-1}$ and $s_{j+1}$ must be even, $s_{j-1}=2u$ and $s_{j+1}=2v$; and 3) $s_{j}=u+v$.
\item If $k$ is even, it must be a multiple of $2l+1$.
\item If $k$ is a multiple of $2l+1$, that is, $k=r(2l+1)$ with $r$ integer, then $\left\lfloor\frac{p_{2l+1}}{2}\right\rfloor+1\leq r\leq 2\left\lfloor\frac{p_{2l+1}}{2}\right\rfloor$.

\end{enumerate}
If $k$ satisfies these conditions, the local contribution to $n_k$ should be reduced by one.

\paragraph{Example:}
For the B-partition $p=\BP{[15,13^{2},9^{4}]}$, we have $2l+1=7$, $p_{2l+1}=9$, $q=[7^9,3^4,1^2]$, $s=[7,14,21,28,35,42,49,56,63,66,69,72,75,76,77]$. For c-constraints whose dimensions are of the form $k=r(2l+1)=7r$, we should allow only $5\leq r\leq 8$. Thus, we have a-constraints at the dimensions $k=66,72,76$, and c-constraints at the dimensions $k=35,42,49,56,69$. We can also compute the pole structure $\{p_k\}$. For instance, we first find $n_{35}=63/2$, and, since we just found that at $k=35$ we have a c-constraint, we must have $p_{35}=65/2$. Also, we can compute $n_{66}=60$ and $n_{33}=63/2$, and we know that at $k=66$ we have an a-constraint; hence, $p_{66}=61$ and $p_{33}=61/2$. Notice that, although the $k=66$ a-constraint introduces a new parameter, $a^{(33)}_{61/2}$, with the same dimension and pole order as the leading coefficient $c^{(33)}_{61/2}$, these are independent.

\subsubsection{Explicit form of constraints}
Now, the rules described above (to compute the dimensions at which a- and c-constraints occur) are sufficient for most purposes, but if we want to know what the constraints look like more specifically, which we need to compute explicit Seiberg-Witten curves, we should study the constraint structure of twisted punctures a little more systematically. We will do so below.

Recall our B-partition $p$ has $2l+1$ parts, $p=\{p_{1},\dots,p_{2l+1}\}$. Then, $q$ must be of the form
\begin{equation}
q=[(2l+1)^{p_{2l+1}},\dots].
\end{equation}
Hence, the first $p_{2l+1}$ parts of the set of partial sums, $s$, are multiples of $(2l+1)$ in arithmetic progression:
\begin{equation}
s=[(2l+1),2(2l+1),3(2l+1),\dots,p_{2l+1}(2l+1),\dots]
\end{equation}
(By construction, the entry to the right of $p_{2l+1}(2l+1)$ cannot be a multiple of $2l+1$.) This block of multiples of $2l+1$ in $s$ will be important, since it gives rise to a particular set of c-constraints for $p$. So, let us look at it in detail.

Consider the first $2\,\left\lfloor\frac{p_{2l+1}}{2}\right\rfloor$ multiples of $2l+1$ in $s$, split into two groups:
\begin{equation}
\begin{aligned}
s'&=\left\{r(2l+1),\,\, 1\leq r\leq \left\lfloor\frac{p_{2l+1}}{2}\right\rfloor\,\right\}\\
s''&=\left\{r(2l+1),\,\, \left\lfloor\frac{p_{2l+1}}{2}\right\rfloor+1\leq r\leq 2\,\left\lfloor\frac{p_{2l+1}}{2}\right\rfloor\,\right\}
\end{aligned}
\end{equation}
For completeness, let us call $s'''$ the set of entries of $s$ that are not in $s'$ or $s''$, so $s=s'\cup s''\cup s'''$ is a disjoint union. Notice that if $p_{2l+1}$ is odd, the term $p_{2l+1}(2l+1)$ is in $s'''$. This term never gives rise to a constraint.

\paragraph{Entries in $s'$:}
None of the entries in $s'$ correspond to constraints. Rather, they can be used to define certain quantities that make c-constraints more transparent; see the example of the minimal puncture, $p=\BP{[2N+1]}$, below.

\paragraph{Entries in $s''$:}
Each entry in $s''$ can be interpreted as a dimension for a c-constraint. Let us look at these constraints in more detail. We study first the even entries. Let $2k$ be in $s''$. The corresponding c-constraint is, schematically, a square:
\begin{equation}
c_{l}^{(2k)}=\left(f_{l/2}^{(k)}(c)\right)^{2}+...
\end{equation}
where $f_{l/2}^{(k)}(c)$ is a polynomial in leading coefficients, of total dimension $k$ and total pole order $l=p_{2k}$ (e.g., $f_{2}^{(4)}=c_{2}^{(4)}-\frac{1}{4}\left(c_{1}^{(2)}\right)^{2}$). The ellipsis above (and in the rest of this subsection) stands for possible cross-terms, which are of the form
\begin{equation}
2f_{l'}^{(k')}(c)f_{l''}^{(k'')}(c).
\label{cconstraint1}\end{equation}
Such a term would arise if and only if there exist c-constraints of dimensions $2k'$ and $2k''$, and if $k'+k''=2k$ and $l'+l''=2l$.
On the other hand, the odd entries, say $2k+1$, of $s''$ always yield c-constraints that are sums of cross-terms, as \eqref{cconstraint1}, but with $k'+k''=2k+1$.

\paragraph{Entries in $s'''$:}
Let us now study the constraints in $s'''$. Again, let us look at even and odd entries separately. Each even entry, $2k$, in $s'''$ is the dimension of an a-constraint,
\begin{equation}
c_{l}^{(2k)}=\left(a_{l/2}^{(k)}\right)^{2}+...
\end{equation}
Finally, let us look at the odd entries, $2k+1$, in $s'''$. If $2k+1$ satisfies the requirements of Sec.~\ref{numberofconstraints}, it yields a cross-term c-constraint involving parameters introduced by a-constraints,
\begin{equation}
c_{l}^{(2k+1)}=2a_{l'}^{(u)}a_{l''}^{(v)}+...
\end{equation}
where $u+v=2k+1$ and $l=l'+l''$. 
Also, if the first c-constraint dimension in $s'''$ is odd, the c-constraint will include a ``{}mixed''{} cross-term of the form
\begin{equation}
2f_{l'}^{(u)}(c)a_{l''}^{(v)}.
\end{equation}

\paragraph{Examples:} Consider the minimal twisted puncture, ${p=\BP{[2N+1]}}$. Then, $q=[1^{2N+1}]$ and $s=[1,2,\dots, 2N+1]$. So, we have only c-constraints, which, since $l=0$, are of scaling dimension $k=N+1, N+2,\dots,2N$. To write the c-constraints specifically, we define auxiliary quantities $r_k$, for $0\leq k\leq N$, by $r_0\equiv 1$, $r_1\equiv 0$, and the rest by ${r_k\equiv \tfrac{1}{2}(c^{(k)}_{p_k}-R^{(k)}_{p_k})}$, where $R^{(k)}_{p_k}$ is the sum of all terms of the form $(r_j)^2$ or $2r_jr_{j'}$, with $j,j'<k$, of total scaling dimension $k$ and total pole order $p_k$. Then, expressing the $c^{(k)}_{p_k}$ back in terms of the $r_j$, for $0\leq j\leq k\leq N$, reveals a nice pattern of squares of cross-terms that should be completed, in a unique way, by the sought c-constraints. For instance, for $N=5$, we define
\begin{equation}
\begin{aligned}
r_{0}&\equiv 1, &
r_{1}&\equiv 0,&
r_{2}&\equiv \tfrac{1}{2}c_{1}^{(2)},\\
r_{3}&\equiv \tfrac{1}{2}c_{3/2}^{(3)},&
r_{4}&\equiv \tfrac{1}{2}\left(c_{2}^{(4)}-\tfrac{1}{4}\left(c_{1}^{(2)}\right)^{2}\right),&
r_{5}&\equiv \tfrac{1}{2}\left(c_{5/2}^{(5)}-\tfrac{1}{2}c_{3/2}^{(3)}c_{1}^{(2)}\right).
\end{aligned}
\label{rdefs}\end{equation}
Then, we can write:
\[
\begin{aligned}
c_{0}^{(0)}&=(r_{0})^{2}, &
c_{0}^{(1)}&=2r_{0}r_{1}, &
c_{1}^{(2)}&=2r_{2}r_{0}+(r_{1})^{2},\\
c_{3/2}^{(3)}&=2r_{3}r_{0}+2r_{1}r_{2}, &
c_{2}^{(4)}&=2r_{4}r_{0}+(r_{2})^{2}+2r_{1}r_{3}, &
c_{5/2}^{(5)}&=2r_{5}r_{0}+2r_{3}r_{2}+2r_{1}r_{4},\\
c_{3}^{(6)}&=2r_{2}r_{4}+(r_{3})^{2}+2r_{1}r_{5}, &
c_{7/2}^{(7)}&=2r_{3}r_{4}+2r_{2}r_{5}, &
c_{4}^{(8)}&=(r_{4})^{2}+2r_{3}r_{5},\\
c_{9/2}^{(9)}&=2r_{4}r_{5}, &
c_{5}^{(10)}&=(r_{5})^{2}.
\end{aligned}
\]
Here, the expressions of scaling dimension $k$ for $0\leq k\leq 5$ are equivalent to \eqref{rdefs}, while those with $6\leq k\leq 10$ are the actual c-constraints. Thus, introducing the $r_k$ makes clear what the c-constraints should be.

Now, let us discuss the puncture $p=\BP{[2N-1,1^2]}$. We find that there are a-constraints (c-constraints) for every even (odd) scaling dimension $k$ in the range $4,5,\dots, 2N$. These constraints follow a pattern of squares and cross-terms in $a^{(j)}_{l}$ parameters; see Appendix~\ref{list_of_twisted_punctures_35} for examples. More importantly, our rules of Sec.~\ref{numberofconstraints} tell us that the $\BP{[2N-1,1^2]}$ puncture is the only one with an a-constraint of scaling dimension four, that is, the only one that gives rise to an independent parameter $a^{(2)}$ of scaling dimension two.

Finally, we study the puncture of type $p=\BP{[9^{3},5^{2}]}$. The scaling dimensions of the c-constraints are 15, 20, 31, while those of the a-constraints are 28, 34. We find:
\begin{equation}
\begin{aligned}
c^{(15)}&=\tfrac{1}{2}c^{(5)}(c^{(10)}-\tfrac{1}{2}(c^{(5)})^{2}), &
c^{(20)}&=\tfrac{1}{4}(c^{(10)}-\tfrac{1}{2}(c^{(5)})^{2})^{2},\\
c^{(28)}&=(a^{(14)})^{2},&
c^{(31)}&=2a^{(14)}a^{(17)},&
c^{(34)}&=(a^{(17)})^{2}.
\end{aligned}
\end{equation}

\hypertarget{irregular_punctures_6}{}
\section{{Collisions of Punctures}}\label{irregular_punctures_6}
In this section we study what happens when two or more punctures collide. We call this process the operator product expansion (OPE) of punctures. In Sec.~\ref{ope_of_punctures_on_a_plane_7} and Sec.~\ref{degeneration_of_a_curve_and_irregular_puncture_8}, we discuss the overall strategy for analyzing the OPE, by first considering the OPE on an infinite plane, and then on a compact curve. 
We then describe an explicit algorithm to compute the OPE in Sec.~\ref{higgs_ope_two_punctures}.

\hypertarget{ope_of_punctures_on_a_plane_7}{}\subsection{{OPE of punctures on a plane}}\label{ope_of_punctures_on_a_plane_7}

So far we have studied how to compute the properties of a single puncture. Let us now see what happens if two or more punctures come close together. First, we would like to study the simpler case of a non-compact Riemann surface, the complex plane. Consider a six-dimensional space of the form $\mathbb{R}^4\times \mathbb{C}$. We denote by $z$ the coordinate on $\mathbb{C}$, and consider $k$ punctures of types $p_1, p_2, \dots, p_k$ to be localized, respectively, at $z=z_1, z_2, \dots, z_k$. Now, at very large $|z|$, the system looks as if consisting of:
\begin{itemize}%
\item a puncture $q$ at $z=0$, with flavour symmetry $F$,
\item a 4D $\mathcal{N}=2$ superconformal theory $X$, which depends on the types and positions of the $k$ punctures, and such that a certain subgroup $H$ of its global symmetry group, $G_X$, can be identified with a subgroup of $F$, and
\item a dynamical gauge multiplet for $H$, with coupling constant $\tau$ depending on the types and positions of the $k$ punctures, which couples X to $q$.

\end{itemize}
We call this process the operator product expansion (OPE) of the $k$ punctures, and we call $X$ the coefficient of the OPE. We schematically represent the outcome of the OPE as \begin{equation}
X\xleftrightarrow{\quad H\quad}q.
\end{equation}

If $q$ is the full puncture and $H=G$, the theory $X$ is the same as the 4D theory obtained by compactifying the 6D theory on a sphere, with punctures of type $p_i$ at $z=z_i$, and a full puncture at $z=\infty$. Otherwise, we say that the theory $X$ is the 4D theory ``obtained by compactifying the 6D theory on a sphere, $S$, with punctures of type $p_i$ at $z=z_i$, and an \emph{irregular puncture} at $z=\infty$, determined by the choice of $p_i$,'' and say that ``the gauge group $H$ arises from the cylinder connecting the irregular puncture with the regular puncture of type $q$.'' \footnote{Conventionally, in the Hitchin system literature, a ``{}regular puncture''{} is a point on $C$ where the Higgs field, $\Phi$, is allowed to have at worst a simple pole, while an ``{}irregular puncture''{} is a point where $\Phi$ may have higher-order poles. Our use of the term ``{}irregular puncture''{} (as also used in \cite{Chacaltana:2010ks,Chacaltana:2011ze}) differs from such conventional use. In particular, in our nomenclature, the full puncture $p$ is a limiting case of an irregular puncture, $p=(p,G)$. No other regular puncture can be thought of as an irregular puncture in this manner.}

We denote such irregular puncture by the pair $(q,H)$, and, if there are inequivalent embeddings $H\hookrightarrow F$, we add a label to distinguish which embedding we mean; see Sec.~\ref{d4example} for an example. We call $q$ the ``{}regular puncture conjugate to''{} the irregular puncture $(q,H)$. While the detailed properties of the theory, $X$, depend on the punctures, $p_i$ (and the various cross-ratios of their positions), certain features are encoded purely in the pair $(q,H)$. For instance, $H$, seen as a subgroup of the global symmetry group $G_X$ of the theory $X$, has some level $k\geq 0$. ($k=0$ if and only if $X$ is the empty theory.) This level is strictly determined \cite{Chacaltana:2011ze} by demanding that the $H$ gauge theory on the cylinder $(q,H)\xleftrightarrow{\quad H\quad}q$ has vanishing $\beta$-function. Similarly, the local contribution of the irregular puncture to $n_h$, $n_v$ and to the graded Coulomb branch dimensions of $X$ are determined by the pair $(q,H)$ \cite{Chacaltana:2011ze}.

\hypertarget{degeneration_of_a_curve_and_irregular_puncture_8}{}
\subsection{{Degeneration of a curve via the OPE}}\label{degeneration_of_a_curve_and_irregular_puncture_8}

Let us now consider the OPE on a compact curve. Let $C$ be a sphere, with $k+k'$ regular punctures of types $p_1,\ldots,p_k; p'_1,\ldots,p'_{k'}$. We assume that the punctures are such that all the graded Coulomb branch dimensions are non-negative. (Otherwise, the theory is ``{}bad''{}, and taking the 4D limit is a more delicate issue.) Now consider the limit where $C$ degenerates into two spheres, $C_1\cup C_2$, with $p_1,\ldots,p_k$ on $C_1$ and $p'_1,\ldots,p'_{k'}$ on $C_2$. We would like to understand the behaviour of the 4D theory in this limit. We proceed as follows:
\begin{itemize}%
\item Replace the punctures $p_1,\ldots,p_k$ with their OPE, as in Sec.~\ref{ope_of_punctures_on_a_plane_7}, obtaining a regular puncture $q$, a gauged subgroup $H$ of the flavour symmetry of $q$, and the 4D theory $X$, which is the 4D limit of a sphere with $p_1,\ldots,p_k$ plus an (ir)regular puncture $(q,H)$.

\item Similarly, replace the punctures $p'_1,\ldots,p'_{k'}$ by their OPE, obtaining a regular puncture $q'$, a gauged subgroup $H'$ of its flavour symmetry, and the 4D theory $X'$, which is the 4D limit of a sphere with $p'_1,\ldots,p'_{k'}$ plus an (ir)regular puncture $(q',H')$.

\item Then we have the following system consisting of
\begin{equation}
X\xleftrightarrow{\quad H\quad}I(q,q')\xleftrightarrow{\quad H'\quad}X'
\label{degenerationwithtwosphere}\end{equation}
where $I(q,q')$ is a sphere with two regular punctures of type $q$ and $q'$, respectively.

\item As explained in \cite{Gaiotto:2011xs}, a sphere with two regular punctures is a supersymmetric hyperK\"ahler non-linear sigma model with global symmetry $F\times F'$, where, in our case, we gauge the subgroup $H\times H' \subset F\times F'$. Any point on the target space of the non-linear sigma model breaks $F\times F'$ to the stabilizer subgroup, $F''$ and hence the gauge symmetry $H\times H'$ is always Higgsed to $H''\subset F''$.

\item Sometimes, the D-term and the F-term constraints for $H''$ force the theory $X$ coupled to $q$ via $H$ to be Higgsed to a theory $Y$. Similarly, the theory $X'$ coupled to $q'$ may be Higgsed to a theory $Y'$.

\item In the end, we have a 4D system of the form:
\begin{equation}
Y\xleftrightarrow{\quad H''\quad}Y'
\end{equation}
\end{itemize}
Now, a sphere with $k$ regular punctures and an irregular puncture has a degeneration where we consecutively collide two punctures, so that the resulting 4D theory consists of several three-punctured spheres coupled to each other. These three-punctured spheres, which we call  \emph{fixtures} \cite{Chacaltana:2010ks,Chacaltana:2011ze}, contain either
\begin{itemize}%
\item three regular punctures, or
\item two regular punctures and one irregular puncture.
\end{itemize}
A table of all possible fixtures makes finding the 4D description of an arbitrary degeneration a simple task. Let us illustrate these ideas with a few examples, all with untwisted punctures for simplicity. 

\subsubsection{Example 1}\label{untwistedexample1}
Consider the $A_{2N-1}$ theory compactified on a 4-punctured sphere, with punctures $(p_1,p_2,p_3,p_4) = ([N^2], [N^2], [2,1^{2N-2}], [2N-1,1])$.
\begin{itemize}%
\item The OPE of $p_1$ with $p_2$ is a full puncture, $[1^{2N}]$, coupled via $H=\Sp(N)$ to the theory, $X$, which is $4N$ free hypermultiplets transforming as 2 copies of the fundamental $2N$-dimensional representation of $\Sp(N)$.
\item The OPE of $p_3$ with $p_4$ yields a full puncture, $[1^{2N}]$, coupled via $H'=\SU(2N-1)$ to the theory, $X'$, which is $(2N-1)(2N-2)$ free hypermultiplets transforming as $(2N-2)$ copies of fundamental representation of $\SU(2N-1)$.
\item The 2-punctured sphere, with two full punctures is $T^*\SU(2N,\mathbb{C})$, which Higgses $H\times H'= \Sp(N)\times \SU(2N-1)$ down to $H''= \Sp(N-1)$.

\end{itemize}
So, in the end, the 4D theory looks like
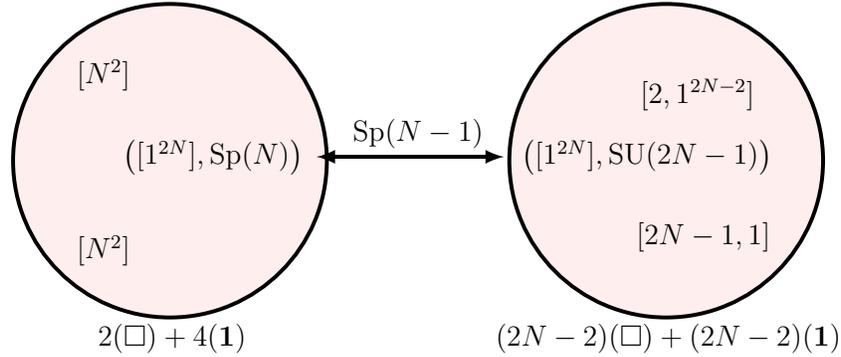
\begin{displaymath}
 \includegraphics[width=321pt]{irregExample1}
\end{displaymath}
i.e.|an $\Sp(N-1)$ gauge theory with $2N$ fundamentals plus $2(N+1)$ free hypermultiplets.  
Here the symbol $\square$ stands for the  $2(N-1)$-dimensional fundamental representation of $\Sp(N-1)$.
Note the cylinder $\left([1^{2N}],\Sp(N)\right)\xleftrightarrow{\quad \Sp(N-1)\quad }\left([1^{2N}],\SU(2N-1)\right)$ connecting two irregular punctures.

\subsubsection{Example 2}
As a second example. consider the $A_4$ theory, compactified on the 4-punctured sphere with punctures $(p_1,p_2,p_3,p_4) = ([4,1],[3,2], [2^2,1], [2^2,1])$.
\begin{itemize}%
\item The OPE of $p_1$ and $p_2$ yields the regular puncture $[2,1^3]$, coupled to the empty theory, $X$, via $H=\SU(2)$.
\item The OPE of $p_3$ and $p_4$ yields the full puncture, $q'=[1^5]$, coupled to the theory $X'=R_{2,5}$, via $H'=G=\SU(5)$. 
Here $R_{2,5}$ is a non-Lagrangian SCFT discussed in  \cite{Chacaltana:2010ks}.
Note that, since $q'$ is the full puncture and $H'$ is $G$, the 3-punctured sphere corresponding to $X'$ contains three regular punctures, $([2^2,1],[2^2,1], [1^4])$.
\item In between, we have the 2-punctured sphere with one full puncture and one $[2,1^3]$ puncture. This Higgses $H \times H'=\SU(2)\times \SU(5)$ down to $H''=\SU(2)$. However, in order to satisfy the F-term equations, the theory $X'$ is Higgsed down to the theory $Y'$, where the $[1^5]$ puncture is replaced by a $[2,1^3]$.

\end{itemize}
The end result is
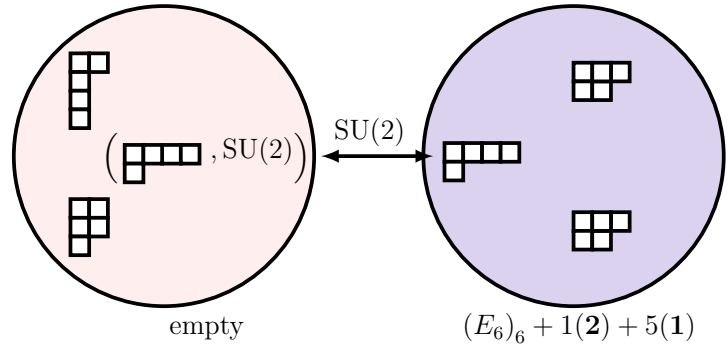
\begin{displaymath}
 \includegraphics[width=287.25pt]{irregExample2}
\end{displaymath}
an $\SU(2)$ gauging of the ${(E_6)}_6$ SCFT, with an additional doublet and 5 free hypermultiplets. Note that, in this case, the cylinder connects an irregular puncture with its conjugate regular puncture.

\subsubsection{Example 3}\label{d4example}
 Now, let us turn to an example from the $D_4$ theory. Consider the 4-punctured sphere
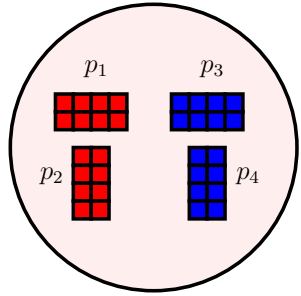
\begin{displaymath}
 \includegraphics[width=123pt]{D4example}
\end{displaymath}
Here, each ``{}very even partition''{} (e.g., $[2^4]$) corresponds to two nilpotent orbits in $\mathfrak{so}(8)$, and our sphere includes one of each type (indicated by the red/blue colour); see \cite{Chacaltana:2011ze}. When we take the OPE of $p_1$ with $p_2$, we obtain the full puncture, $q= \includegraphics[width=61.5pt]{D4_11111111}$, coupled to $X=$ the ${(E_7)}_8$ SCFT, via $H=\Spin(7)$ (and similarly for the OPE of $p_3$ with $p_4$). However, there are three inequivalent embeddings of $\Spin(7)\hookrightarrow \Spin(8)$, depending on which of the three 8-dimensional irreducible representation of $\Spin(8)$ decomposes as $7+1$. We can indicate this choice by putting a subscript on $H$, or (in the notation of \cite{Chacaltana:2011ze}) by colouring the Young diagram corresponding to $q$:
\begin{displaymath}
\begin{split}
( \includegraphics[width=61.5pt]{D4_11111111_green}, \Spin(7)) &= ( \includegraphics[width=61.5pt]{D4_11111111},\Spin(7)_{8_v})\\
( \includegraphics[width=61.5pt]{D4_11111111_red}, \Spin(7)) &= ( \includegraphics[width=61.5pt]{D4_11111111},\Spin(7)_{8_s})\\
( \includegraphics[width=61.5pt]{D4_11111111_blue},\Spin(7)) &= ( \includegraphics[width=61.5pt]{D4_11111111},\Spin(7)_{8_c})\\
\end{split}
\end{displaymath}
In the notation of \eqref{degenerationwithtwosphere}, we have $H=\Spin(7)_{8_s}$ and $H'=\Spin(7)_{8_c}$, and the two-punctured sphere, \[
I( \includegraphics[width=61.5pt]{D4_11111111}, \includegraphics[width=61.5pt]{D4_11111111})
\] Higgses $H\times H'=\Spin(7)_{8_s}\times \Spin(7)_{8_c}$ down to the common subgroup, $G_2$. So the final 4D description of this limit is
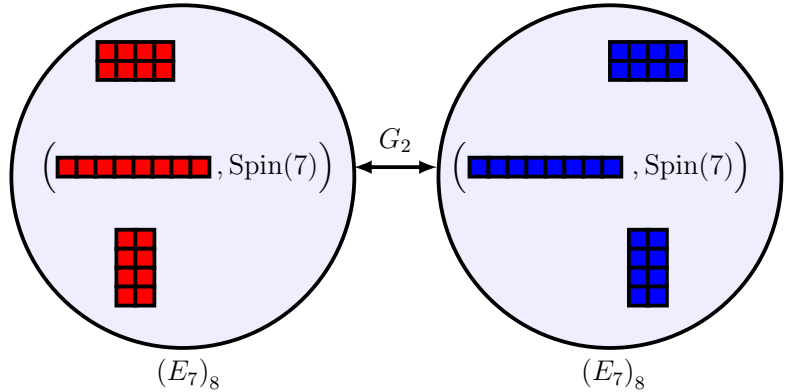
\begin{displaymath}
 \includegraphics[width=306.75pt]{irregD4example}
\end{displaymath}
where, as in Sec.~\ref{untwistedexample1}, we have a cylinder connecting two irregular punctures.

\hypertarget{higgs_ope_two_punctures}{}
\subsection{Determining the OPE via the Higgs field}\label{higgs_ope_two_punctures}

In light of Sec.~\ref{ope_of_punctures_on_a_plane_7} and \ref{degeneration_of_a_curve_and_irregular_puncture_8}, we would like to study the basic problem of two punctures $p_1$ and $p_2$ colliding on a plane. We have seen that in the collision limit, an irregular puncture $(q,H)$ arises, which is connected to a regular puncture $q$ by a cylinder with gauge group $H$. Let us discuss how to find $q$ and $H$.

To determine $q$, we construct a solution to the Higgs field on the plane that includes $p_1$ and $p_2$, and compute the residue that arises in the collision limit. This residue provides the Higgs-field boundary condition for $q$. Thus, one can determine the Nahm pole for $q$, e.g., by looking at the degeneracy of the mass deformations in the residue. Also, the number of independent mass deformations is equal to the rank of $H$.

To gather more information about $H$, we consider the $k$-differentials $\phi_k$, and take the limit where the punctures collide, which reveals the scaling dimensions of the Casimirs of $H$. Knowing these usually suffices to identify the gauge group. Only in a handful of cases, often to distinguish $\Sp(k)$ from $\SO(2k+1)$, must one do further consistency checks, such as computing the matter representation for the fixture that arises in the degeneration limit, and corroborating that it provides the right contribution to the beta function of $H$.

Because of these observations, in the next subsections we will study the Higgs field on a plane with two punctures, in the limit where these collide. Later, in Sec~\ref{degenerations_of_a_sphere_9}, we will do the same for $k$-differentials. But before doing this, let us briefly discuss a situation that will arise often.

Consider $C$ to be the complex plane or a sphere, with complex coordinate $z$, and put $k$ punctures, $p_1,\dots,p_k$, on $C$. Let the positions of the punctures be $\lambda z_1,\dots, \lambda z_m, z_{m+1},\dots z_k$, so that we can collide the first $m$ of them by taking the limit $\lambda\to 0$. Now consider a meromorphic $k$-differential on $C$ of the form
\begin{equation}
\frac{A\lambda^{\alpha}z^s}{\prod_{i=1}^{m}(z-\lambda z_i)^{r_i}\prod_{j=m+1}^{k}(z-z_{j})^{r_j}}(dz)^k,
\label{kdiffterm}\end{equation}
where $\alpha, s, r_1,\dots, r_k$ are rational numbers; $A$ is a coefficient. This is a typical term in a $k$-differential, including the case of the Higgs field ($k=1$). In the $\lambda\to 0$ limit, we get $C$ in the presence of the $k-m$ punctures $p_{m+1},\dots, p_k$, plus a new puncture, $q$, at $z=0$.

The $\lambda\to 0$ limit may also be represented by the conformally equivalent picture of a sphere $C'$ that bubbles off $C$, containing the $m$ punctures $p_1,\dots, p_m$, plus the irregular puncture $(q,H)$. Such picture is obtained through the change of variables $z=\lambda/z'$. Then, requiring that \eqref{kdiffterm} have a finite limit as $\lambda\to 0$ in the $z'$ coordinates puts a lower bound on $\alpha$,
\begin{equation}
\alpha\geq \text{max}(0,-s-k+\sum_{i=1}^m r_i).
\label{boundalpha}\end{equation}
We will use this result quite often. If the bound is not saturated, the $k$-differential simply vanishes, in the $\lambda\to 0$ limit, on both $C$ and $C'$. On the other hand, if the bound is saturated, we have three possibilities when $\lambda\to 0$:
\begin{enumerate}
\item If $\sum_{i=1}^m r_i>s+k$, \eqref{kdiffterm} vanishes on $C$, but not on $C'$.
\item If $\sum_{i=1}^m r_i<s+k$, \eqref{kdiffterm} vanishes on $C'$, but not on $C$.
\item If $\sum_{i=1}^m r_i=s+k$, \eqref{kdiffterm} does not vanish on $C$ nor $C'$.
\end{enumerate}
In most cases (such as in the following subsections), the coefficient $A$ in \eqref{kdiffterm} will represent a physical degree of freedom, and so it should not be lost in the $\lambda\to 0$ limit; therefore, it will be desirable that the bound be saturated. Case 1 (2) corresponds to $A$ being a degree of freedom for the theory on $C'$ ($C$), whereas case 3 corresponds to $A$ being a degree of freedom of the gauge group on the cylinder, which in the $\lambda\to 0$ limit looks like a mass deformation on both $C$ and $C'$. However, in a few cases where the coefficient $A$ carries redundant information (because of local constraints), consistency will require that the bound not be saturated. We will see such cases in Sec. \ref{atypical_degenerations_11}.

\subsubsection{Untwisted-untwisted}\label{higgs_untwisted_untwisted}

Consider now the Higgs field $\Phi$ on a plane with two untwisted punctures of type $p_1$ and $p_2$ at positions $z=0$ and $z=\lambda$, respectively, where $z$ is the coordinate on the plane. Let $A$ and $B$ be representatives of the (massless or mass-deformed) adjoint orbits in $\mathfrak{sl}(2N)$ corresponding to $p_1$ and $p_2$, respectively. Then, we can write an ansatz,
\begin{equation}
\Phi(z)=\frac{A(z-\lambda)+Bz+z(z-\lambda)P(z)}{z(z-\lambda )}\,dz,
\label{higgs-u-u}\end{equation}
where $P(z)$ is a power series in $z$ whose coefficients are generic elements of $\mathfrak{sl}(2N)$. $P(z)$ simply represents the infinite degrees of freedom contained in the plane.

At finite $\lambda$, the expansions of $\Phi(z)$ near $p_1$ and $p_2$ are, respectively,
\begin{equation}
\begin{aligned}
\Phi(z)&=\left(\frac{A}{z}+\text{generic in }\mathfrak{sl}(2N)\right)dz,\qquad
\Phi(z)&=\left(\frac{B}{z-\lambda}+\text{generic in }\mathfrak{sl}(2N)\right)dz,
\end{aligned}
\end{equation}
In the limit $\lambda\to 0$, a new untwisted puncture, $q$, arises at $z=0$. The expansion of $\Phi(z)$ near this point is
\begin{equation}
\Phi(z)=\left(\frac{A+B}{z}+\text{generic in }\mathfrak{sl}(2N)\right)dz.
\end{equation}
This is the expected expansion for an untwisted defect. Thus, the Higgs field boundary condition for the new puncture $q$ is given by $A+B$.

Notice that \eqref{higgs-u-u} saturates the bound of \eqref{boundalpha}. In the $\lambda\to 0$ limit, we have the complex plane in the presence of just the new puncture, $q$. The conformally-equivalent picture is that of a fixture that bubbles off the plane, containing the two original punctures, $p_1$ and $p_2$, plus the irregular puncture, $(q,H)$. The Higgs field for the fixture is
\begin{equation}
\Phi(z')=\frac{-A(z'-1)+B}{z'(z'-1)}\,dz',
\end{equation}
where we used $z=\lambda/z'$ in \eqref{higgs-u-u} and took the $\lambda\to  0$ limit. The punctures $q, p_1, p_2$ are at $z'=0, \infty, 1$, respectively. Notice that the Riemann-Roch theorem for a 3-punctured sphere requires only two coefficients, not three, for $\Phi(z')$. In other words, in a fixture, the choice of representatives, $A$ and $B$, for the adjoint orbits for two punctures, $p_1$ and $p_2$, completely determines the adjoint orbit (that is, its mass deformations), plus a representative of such orbit, for the third puncture, $q$.

\subsubsection{Twisted-twisted}\label{higgs_twisted_twisted}

Let us now take two twisted punctures of type $p_1$ and $p_2$ at positions $z=0$ and $z=\lambda$ on a complex plane with coordinate $z$. Let $A$ and $B$ be representatives of the (massless or mass-deformed) adjoint orbits in $\Sp(N)$ corresponding to $p_1$ and $p_2$, respectively. Recall the decomposition of $\mathfrak{sl}(2N)$ in eigenspaces of the $\mathbb{Z}_2$-outer automorphism, $\mathfrak{sl}(2N)\simeq \mathfrak{sp}(N)\oplus o_{-1}$. Then, we can write the following ansatz for the Higgs field,
\begin{equation}
\Phi(z)=
\frac{A(z-\lambda)+Bz+z(z-\lambda)P(z)}{z(z-\lambda )}+\frac{D+zQ(z)}{z^{1/2}(z-\lambda )^{1/2}},
\label{higgs-t-t}\end{equation}
where $D$ is a generic element of $o_{-1}$, and $P(z)$ and $Q(z)$ are power series in $z$ whose coefficients are, respectively, generic elements of $\mathfrak{sp}(N)$ and $o_{-1}$.

At finite $\lambda$, the expansions of $\Phi(z)$ near $p_1$ and $p_2$ are, respectively,
\begin{equation}
\Phi(z)=\frac{A}{z}+\frac{\text{generic in }o_{-1}}{z^{1/2}}+\text{generic in }\mathfrak{sp}(N),
\end{equation}
\begin{equation}
\Phi(z)=\frac{B}{z-\lambda}+\frac{\text{generic in }o_{-1}}{(z-\lambda)^{1/2}}+\text{generic in }\mathfrak{sp}(N),
\end{equation}
These expansions are of the expected form for twisted defects.

Now, in the collision limit $\lambda\to 0$, a new untwisted puncture, $q$, arises at $z=0$. The expansion of $\Phi$ near $q$ is
\begin{equation}
\Phi(z)=\frac{A+B+D}{z}+\text{generic in }\mathfrak{sl}(2N)
\end{equation}
This is, again, the expected expansion for an untwisted defect. Thus, $q$ has Higgs field residue $A+B+D$, with $D$ generic in $o_{-1}$.

For completeness, we show the Higgs field for the fixture in the conformally equivalent picture, as before,
\begin{equation}
\Phi(z')=\frac{-A(z'-1)+B}{z'(z'-1)}+\frac{iD}{z'(z'-1)^{1/2}}.
\end{equation}

\subsubsection{Twisted-untwisted}\label{higgs_twisted_untwisted}

Finally, consider a twisted and an untwisted puncture, of types $p_1$ and $p_2$, respectively, at positions $z=0$ and $z=\lambda$. Let $A$ and $B$ be representatives of the (massless or mass-deformed) adjoint orbits corresponding to $p_1$ and $p_2$, respectively. Notice that $A$ is in $\mathfrak{sp}(N)$, while $B$ is in $\mathfrak{sl}(2N)$. Let us decompose $B$ as $B=B_1+B_{-1}$, where $B_1$ is in $\mathfrak{sp}(N)$ and $B_{-1}$ is in $o_{-1}$. Then, we can write an ansatz for the Higgs field,
\begin{equation}
\Phi(z)=
\frac{A(z-\lambda)+B_{1}z+z(z-\lambda)P(z)}{z(z-\lambda )}+\frac{B_{-1}\lambda^{1/2}+(z-\lambda)Q(z)}{z^{1/2}(z-\lambda )},
\label{higgs-t-u}\end{equation}
where $P(z)$ and $Q(z)$ are power series in $z$ whose coefficients are, respectively, generic elements of $\mathfrak{sp}(N)$ and $o_{-1}$. The expansions of $\Phi(z)$ near $p_1$ and $p_2$ are, respectively,
\begin{equation}
\begin{aligned}
\Phi(z)&=\frac{A}{z}+\frac{\text{generic in }o_{-1}}{z^{1/2}}+\text{generic in }\mathfrak{sp}(N),\\
\Phi(z)&=\frac{B}{z-\lambda}+\text{generic in }\mathfrak{sl}(2N).
\end{aligned}
\end{equation}
Again, these expansions have the expected forms. Now, in the limit $\lambda\to 0$, a new twisted puncture, $q$, arises at $z=0$. The expansion of $\Phi(z)$ near $q$ is
\begin{equation}
\Phi(z)=\frac{A+B_1}{z}+\frac{\text{generic in }o_{-1}}{z^{1/2}}+\text{generic in }\mathfrak{sp}(N),
\end{equation}
This expansion has the correct form for a twisted defect. Thus, the Higgs field boundary condition for $q$ is given by $A+B_1$, with $B_1$ the projection of $B$ to $\mathfrak{sp}(N)$.

Finally, the Higgs field for the fixture in the conformally equivalent picture is
\begin{equation}
\Phi(z')=\frac{-A(z'-1)+B_1}{z'(z'-1)}+\frac{B_{-1}}{z'(z'-1)^{1/2}}.
\end{equation}

\subsubsection{{Degenerating $k$-differentials}}\label{degenerations_of_a_sphere_9}

Let us discuss how to find the scaling dimensions of the VEVs for the gauge group $H$ that arises when two punctures $p_1$ and $p_2$ on a plane collide. In most cases, these provide enough information to determine $H$. The natural way to find such VEVs is to use \eqref{spectralcurve} to compute the $k$-differentials from the Higgs field residue of the new puncture, $q$.  If $q$ is at $z=0$, we have, in principle, a gauge group VEV $u_k$ of scaling dimension $k$ if
\begin{equation}
\phi_k(z)=\frac{u_k}{z^{2k}}+\dots
\label{kdiffvevterm}\end{equation}
However, some of these $u_k$ may not be independent. If $p_1$ and $p_2$ have mass deformations, for instance, the $u_k$ might contain combinations of these masses. If a parameter $u_k$ vanishes when we turn off the the masses, then such $u_k$ is not an actual gauge group VEVs. Also, the $u_k$ might depend on each other. Hence, it is convenient, for the purpose of finding the gauge group, to consider the OPE with  massless $p_1$ and $p_2$. It might also be useful to study the $k$-differentials \emph{before} the collision, taking into account whichever c-constraints and a-constraints the two punctures have, and then take the degeneration limit. 
(For the concept of c- and a-constraints, see Sec.~\ref{constraints_4}.)
This makes evident what relations there might exist among the $u_k$ in a gauge-invariant way.

Let us also briefly discuss the basic problem of the degeneration of a sphere with $n_1+n_2$ massless punctures, in which we collide the first $n_1$ punctures. Each $k$-differential contains $1-2k+\sum_{i=1}^{n_1+n_2} p^{(i)}_k$ terms of the form \eqref{kdiffterm}. We want to find conditions for a gauge group VEV term \eqref{kdiffvevterm} to exist at every $k$. Note that if $k$ is odd, and we are colliding an odd number of twisted punctures, there cannot be a gauge group term because the power $s$ in \eqref{kdiffvevterm} must be an integer. So, if we collide an odd number of twisted punctures\footnote{Notice that one can always get a VEV with odd scaling dimension from a parameter with even scaling dimension, if the latter satisfies an a-constraint, e.g., $u_{2(2k+1)}=(a_{2k+1})^2$.}, the discussion below is restricted to even $k$. 

If none of the punctures satisfy constraints, then, at every $k$, there will exist a gauge group term \eqref{kdiffvevterm} if and only if $t_k\equiv -k+\sum_{i=1}^{n_1}p^{(i)}_k\geq 0$ and $t'_k\equiv -k+\sum_{j=n_1+1}^{n_1+n_2}p^{(j)}_k\geq 0$. If there are c-constraints and a-constraints, these will act on the parameters of the $k$-differential on the side of the degeneration where they exist, and so may constrain the gauge group VEV. 
However, if $t_k\geq n_c + n_a$, where $n_c$ and $n_a$ are, respectively, the number of c-constraints and a-constraints of dimension $k$ for the punctures on one side of the degeneration, and if a similar condition holds for the other side of the degeneration, then the gauge group term will not be affected by the constraints. But if these conditions do not hold, one needs to analyze the $k$-differential in detail to see how the gauge group VEVs are affected.

Now we wish to illustrate our techniques with some examples. To write residues explicitly, we need to use a basis. We use the embedding of $Sp(N)$ in $\mathfrak{sl}(2N)$ of \cite{CollingwoodMcGovern},
\begin{equation}
\begin{pmatrix}A & B\\C&-A^t\end{pmatrix},
\end{equation}
where $A, B$ and $C$ are $n\times n$ complex matrices, and $B$ and $C$ are furthermore symmetric. We also take from \cite{CollingwoodMcGovern} the nilpotent orbit representatives of $\mathfrak{sl}(2N)$ and $Sp(N)$.

\subsubsection{Example 1}

Let us consider again the untwisted example of Sec.~\ref{untwistedexample1}. The OPE of $[N^2]$ and $[N^2]$ on a plane is found to have diagonalized Higgs field residue of the form
\begin{equation}
\text{diag}(m_1,m_2,\dots, m_N;-m_1,-m_2,\dots,-m_N),
\end{equation}
with $\sum_{i=1}^N m_i =0$. One can compute the $k$-differentials for this residue to find that the cylinder has VEVs of scaling dimension $k=2,4,6,\dots, 2N$, which  is consistent with the interpretation that the gauge group is $Sp(N)$. (It could not be $SO(2N+1)$ because this cannot be embedded in $SU(2N)$.) Thus, we have an $Sp(N)$ cylinder and, since all masses are generically different, the new puncture is of type $[1^{2N}]$.

Similarly, the OPE of $[2,1^{N-2}]$ and $[2N-1,1]$ on a plane has diagonalized Higgs field residue of the form
\begin{equation}
\text{diag}(r_1,r_2,\dots, r_{2N-1},0),
\end{equation}
with $\sum_{j=1}^{2N-1} r_j=0$. This corresponds to an $SU(2N-1)$ cylinder and, since all masses are generically different, the new puncture is again of type $[1^{2N}]$.

Now, to get a 4-punctured sphere, we connect the $[1^{2N}]$ punctures. Up to permutation, there is only one way the forms of the two residues above can match, namely, by making $r_i=m_i$ and $r_{N+i}=-m_i$, for $i=1,\dots, N-1$, and $r_{N}=m_{N}=0$. This corresponds to an embedding of $Sp(N-1)$ in $SU(2N)$, as claimed in Sec.~\ref{untwistedexample1}.

\subsubsection{Example 2}

Let us study the case of a five minimal twisted punctures, $\BP{[2N+1]}$, on a plane, and study the consecutive collisions of two punctures. We do this to illustrate the rules for combining residues.

Colliding two minimal twisted punctures yields an untwisted puncture with residue equal to a generic element in  $o_{-1}$. This can be diagonalized to an element of the form
\begin{equation}
\text{diag}(m_1,m_2,\dots, m_N;m_1,m_2,\dots, m_N),
\label{OPEtwomintwistedpunctures}\end{equation}
with $\sum_{i=1}^N m_i=0$. Since the $2N$ eigenvalues are split into pairs of identical elements, this residue should be a mass-deformed representative of the untwisted $[2^N]$ puncture. This puncture has $\SU(N)$ global symmetry group. The fact that there are $N$ masses $m_i$ adding up to zero suggests  that the whole $\SU(N)$ group is gauged. This can be checked by computing the $k$-differentials. In fact,
to leading order we have, for $k=2,3,\dots, 2N$,
\begin{equation}
\phi_k = \frac{u_k}{z^{k/2}(z-\lambda)^{k/2}}+\dots,
\end{equation}
where the two colliding $\BP{[2N+1]}$ punctures are at $z=0, \lambda$. All of the $u_k$ survive the $\lambda\to 0$ limit, and so each should become a Casimir of the gauge group. However, only the $u_k$ for $k\leq N$ are independent because of the constraints from the minimal twisted punctures. (Actually, since there are \emph{two} $\BP{[2N+1]}$ punctures, the subleading coefficients in the expansion above for $k>N$ are also constrained.) Thus, the gauge group has only $N-1$ Casimirs of scaling dimensions $2,3,\dots, N$. Hence, it must be $\SU(N)$.

Now let us find the OPE of the $[2^N]$ puncture with the third $\BP{[2N+1]}$ puncture. Our prescription tells us that the residue of the new puncture will be the Sp-projection of the residue of the $[2^N]$ puncture\footnote{Notice that the Sp-projection of any element of $o_{-1}$ is trivially zero, so naive application of the prescription of Sec.~\ref{higgs_twisted_untwisted} for the OPE of a mass-deformed $[2^N]$ puncture and a $\BP{[2N+1]}$ puncture would not appear to reproduce \eqref{residue1}. This happens because the mass deformations of the new residue are encoded in the choice of representatives of the colliding punctures. So, we should first take of a more generic representative of $[2^N]$ (an element conjugate to the generic element of $o_{-1}$) to get the generic residue \eqref{residue1} for the OPE.\label{noticefootnote}}. The diagonalized form of the new residue is
\begin{equation}
\text{diag}(r_1,r_2,\dots, r_N;-r_1,-r_2,\dots, -r_N),
\label{residue1}\end{equation}
with $\sum_{i=1}^N  r_i=0$. For $N=2$, this residue takes the more particular form $\text{diag}(r,r,-r,-r)$. So, for $N=2$, this is the $\BP{[2^2,1]}$ puncture, and for $N\geq 3$, it is the $\BP{[1^{2N+1}]}$ puncture.

To find the gauge group, we can either use three colliding $\BP{[2N+1]}$ punctures, or the $[2^N]$ puncture colliding with a $\BP{[2N+1]}$ puncture. Let us use the first. A $\BP{[2N+1]}$ puncture has pole structure $p_k=k/2$, and satisfies a c-constraint for $k=N+1, N+2,\dots,2N$. Plus, for $k=2,\dots, N$, we have $t_k=\sum_{i=1}^3 p_k^{i}-k = k/2\geq 0$, so we have an unconstrained VEV for each even $k$ in this range. On the other hand, at each $k=N+1,\dots, 2N$, we have three c-constraints, so we have $t_k-n_c=\sum_{i=1}^3 p_k^{i}-k-3=k/2-3$, and we can be sure we have an unconstrained VEV for every even $k$ in this range such that $k\geq 6$. So, for $N\geq 5$, the gauge group is just $\Sp(N)$, while for $N=2,3,4$, we need to check by hand. For $N=2$, we have $k=4<6$, so we only have one VEV of dimension 2, and the gauge group should be $\SU(2)$. For $N=3$, we have $k=4<6$, but $k=6\geq 6$, so we have VEVs of dimensions 2 and 6, and the gauge group should be $G_2$. For $N=4$, we have $k=6\geq 6$ and $k=8\geq 6$, so we have an $\Sp(4)$ gauge group.

Colliding four $\BP{[2N+1]}$ punctures, we get a diagonalized residue of the form
\begin{equation}
\text{diag}(q_1,q_2,\dots,q_{2N}),
\end{equation}
with $\sum_{i=1}^{2N} q_i=0$. For $N=2$ this takes the more particular form $\text{diag}(q,s,-q,-s)$. Hence, for $N=2$, we find an irregular version of $[1^{4}]$, with $Sp(2)$ cylinder, while for $N\geq 3$, we find the regular $[1^{2N}]$, with gauge group $SU(2N)$.

Finally, for the collision of five $\BP{[2N+1]}$ punctures, for $k=2,\dots, N$, we have $t_k=\sum_{i=1}^5 p_k^{i}-k=3k/2\geq 0$, while for $k=N+1,\dots, 2N$, we have five c-constraints, and $t_k-n_c=3k/2-5$ is nonnegative for $k\geq 3$, and thus the gauge group is $\Sp(N)$ for any $N\geq 2$, and the new puncture is the regular $\BP{[1^{2N+1}]}$.

\hypertarget{atypical_degenerations_11}{}
\section{{Atypical Degenerations}}\label{atypical_degenerations_11}

The conventional understanding of Gaiotto duality is that one starts with a Riemann surface, $C$, with punctures, and, in any degeneration limit of $C$, a weakly-coupled gauge group, $G$, in the low-energy 4D theory arises. There is a specific connection between the plumbing fixture $q$ of the degenerating cylinder and the weak gauge coupling $\tau$ for $G$,
\begin{equation}
q=e^{2\pi i \tau}
\end{equation}
So, the pinch limit $q\to 0$ of the surface corresponds to the weakly-coupled limit $\tau\to i\infty$ for the gauge group. Besides, the gauge-invariant VEVs constructed from the scalars in the $G$-vector multiplets can be similarly assigned to the cylinder: upon degenerating the curve, the VEVs of $G$ become mass-deformations of the new punctures that appear on both sides of the degeneration. Finally, upon complete degeneration, the $G$-vector multiplets completely decouple; they are not present on either side of the degeneration.

In this section we want to study certain degenerations involving curves with certain combinations of twisted and untwisted punctures $A_{2N-1}$ series where the picture explained above relating cylinders and gauge groups does not hold.\footnote{Such ``{}atypical''{} degenerations also appear in the twisted $D_N$ series, as already seen in the analysis of linear quiver tails with $\SO(4)$ gauge groups and empty cylinders, in \cite{Tachikawa:2009rb}. A systematic treatment of the twisted $D_N$ series is work in progress.} Our goal is to understand to what extent the conventional picture of the previous paragraph is still correct, and how it should be modified when our curve contains these ``{}dangerous''{} combinations of punctures. Happily, ``{}atypical''{} degenerations are actually rare.

By ``{}atypical''{} degenerations in the twisted $A_{2N-1}$ series we refer to either of the following situations:
\begin{itemize}%
\item A degeneration brings a certain gauge coupling to a point in the interior of moduli space of couplings, instead of a weakly-coupled cusp. This interior point can either be a strongly-coupled point, or a point where two gauge group couplings become equal. 
\item A degeneration brings not only the corresponding gauge group to its weakly-coupled cusp, but also other gauge groups (adjacent to, but not directly localized at the degenerating cylinder) to weakly-coupled cusps.

\end{itemize}
Fortunately, these atypical degenerations seem to occur only when a \emph{sphere}, containing certain combinations of punctures, bubbles off a generic surface. Furthermore, for the bubbling sphere to be ``{}atypical''{}, there is a bound on the number of punctures it may contain, as well as a restriction on the types of punctures. (These can only be minimal twisted or minimal untwisted.) Thus, these ``{}atypical''{} spheres are easy to classify.
Let us study these ``{}atypical''{} spheres one by one.

\hypertarget{gauge_theories_on_a_fixture_12}{}
\subsection{{Gauge theories on a fixture}}\label{gauge_theories_on_a_fixture_12}
\subsubsection{Fixtures with $\BP{[2N-1,1^2]}$}
Consider a ``{}good''{} fixture (that is, one including only regular punctures, and whose virtual Coulomb branch dimensions are all non-negative) in the $A_{2N-1}$ series with one twisted puncture of type $\BP{[2N-1,1^2]}$, a twisted puncture of type $\BP{T}$, and an untwisted puncture of type $U$.
\begin{displaymath}
 \includegraphics[width=134.25pt]{fig12}
\end{displaymath}
Now, notice that the a-constraint $c^{(4)}_3=a^2$ in the $\BP{[2N-1,1^2]}$ puncture introduces a Coulomb branch VEV $a$ of scaling dimension two. We want to argue that such a fixture represents a gauge theory whose gauge coupling $\tau$ is locked at the $\mathbb{Z}_2$-symmetric point of its moduli space, $\tau=i$, and that the $\mathbb{Z}_2$ action can be identified with the disconnected part of the flavour symmetry $\mathrm{O}(2)$ of $\BP{[2N-1,1^2]}$.
A 3-punctured sphere has no moduli, so the fact that a gauge coupling is frozen at a point in coupling-constant moduli space is the only way that we could have a gauge theory on a 3-punctured sphere. Now, it is not obvious that such point in the moduli space should lie in its interior, or that it should take the value $\tau=i$. We will verify below these assertions by studying the Seiberg-Witten curve for the fixture.

Let the punctures $\BP{[2N-1,1^2]}$, $U$, and $\BP{T}$ be at the positions $z=0, 1,\infty$, respectively. Since we are just interested in seeing at which point in gauge-coupling moduli space the theory is, we remove all the unnecessary degrees of freedom, such as mass deformations and Coulomb branch VEVs of scaling dimension different from two. Then, the only surviving $k$-differential is $\phi_4$, which includes the square of the Coulomb branch parameter $a$. Notice that the 2-differential $\phi_2$ vanishes because we have only three massless punctures. So, the Seiberg-Witten curve for the reduced theory is:
\begin{equation}
y^{2N-4}\left(y^4-\frac{a^2}{z^3(z-1)^2}\right)=0
\end{equation}
The factor $y^{2N-4}$ tells us that the original $2N$-sheeted cover $\Sigma$ of the fixture splits into $2N-4$ unramified branches plus a four-sheeted cover, $\Sigma'$. Let us dispose of the unramified branches. The Seiberg-Witten curve factorizes:
\begin{equation}
\left(y^2-\frac{a}{z^{3/2}(z-1)}\right)\left(y^2+\frac{a}{z^{3/2}(z-1)}\right)=0
\label{swcurvetwofactors1}\end{equation}
This expression tells us that $\Sigma'$ globally splits into two double covers which differ only by the choice of sign in $a$.

Let us explore the first factor in \eqref{swcurvetwofactors1}. Consider the transformation $z=t^2$, $y=y'/2t$, which preserves the Seiberg-Witten differential, $\lambda=ydz=y'dt$. Then the first factor takes the form
\begin{equation}
y'^2-\frac{4a}{t(t-1)(t+1)}=0
\end{equation}
Now, this is the Seiberg-Witten for a four-punctured sphere in the untwisted $A_1$ theory, with punctures at $t=0,1,-1,\infty$, which represents the $\SU(2)$ $N_f=4$ gauge theory. If the curve above had really arisen from a four-punctured sphere, we would identify the marginal coupling $q$ of the $\SU(2)$ gauge group with the cross-ratio $x$ of the four punctures. But here, since we started with a fixture, we are not allowed to vary the cross-ratio; the curve we obtain is fixed at the cross ratio, $x=-1$. This value corresponds to the $\mathbb{Z}_2$-orbifold point, $\tau=i$, in gauge-coupling moduli space.

What about the second factor in \eqref{swcurvetwofactors1}? It clearly represents again the $\SU(2)$ gauge theory, but with the other choice of sign for $a$. The origin of this second factor is the a-constraint of the $\BP{[2N-1,1^2]}$ puncture, which does not fix the sign of $a$. This second copy is \emph{not} a second, independent, $\SU(2)$ gauge theory, since its degrees of freedom simply mirror those of the first factor up to a sign. So, the original fixture we started with represents a \emph{single} copy of the $\SU(2)$ gauge theory at the $\tau=i$ point in moduli space. This suggests that the original, good $A_{2N-1}$ fixture can be interpreted to contain a gauge group $G$ with gauge coupling $\tau=i$. This interpretation is confirmed by computing the total number of hypers and vectors for this fixture, as well as the representations for the matter, and is consistent with S-duality in all examples we studied.

\subsubsection{Separating $\BP{[2N-1,1^2]}$ to a pair of $\BP{[2N+1]}$ and $[2N-1,1]$ }
Is there a way to ``{}turn on''{} the frozen gauge coupling of a gauge theory fixture, so that we are able to move in the moduli space of the gauge theory? The answer lies in coupling the $\BP{[2N-1,1^2]}$ to another fixture so that we have a cross-ratio. The $\BP{[2N-1,1^2]}$ puncture has flavour symmetry group $\mathrm{O}(2)$, so a cylinder involving this puncture cannot support a non-abelian gauge group. Still, an irregular version of this puncture, $(\BP{[2N-1,1^2]},\varnothing)$, does exist, and it arises in the OPE of a minimal twisted puncture, $\BP{[2N+1]}$, and a minimal untwisted puncture, $[2N-1,1]$. This irregular puncture connects to the regular $\BP{[2N-1,1^2]}$ via an ``{}empty''{} cylinder, to form the following four-punctured sphere:
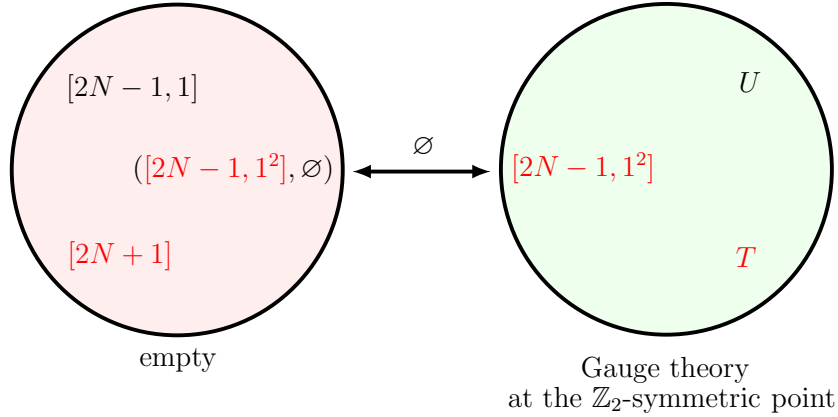
\begin{displaymath}
 \includegraphics[width=323.25pt]{fig13}
\end{displaymath}
In this degeneration, the cross-ratio $x$ of this four-punctured sphere controls the perturbation of the gauge coupling $\tau$ away from the $\mathbb{Z}_2$ symmetric point, $\tau=i$. Locating the punctures $\BP{[2N+1]}, \BP{T}, [2N-1,1], U$ at $z_1,z_2,z_3,z_4$ respectively, let
$$
   w^2 = x \equiv \frac{z_{13} z_{24}}{z_{14}z_{23}}
$$
We will find that the relation between the cross-ratio of our original twisted 4-punctured sphere
and the $\SU(2)$ gauge coupling $q$ is given by 
\begin{equation}
q=-\frac{1+w}{1-w}.
\label{relationQX}\end{equation}
At $x=0$, the gauge coupling becomes $q=-1$, i.e., the $\mathbb{Z}_{2}$ orbifold point, which is consistent with what we found for the gauge-theory fixture alone. 
Note that before the degeneration, the flavour symmetry comes from the puncture $[2N-1,1]$ which has $\mathrm{U}(1)$ symmetry.
After the degeneration, the flavour symmetry comes from $\BP{[2N-1,1^2]}$ which has $\mathrm{O}(2)$ symmetry. This contains the original $\mathrm{U}(1)$ together with its outer automorphism.

The appearance of a square root of $x$ in \eqref{relationQX} means that the marginal-coupling moduli space of the theory, which we will denote by $X_4$, is not the complex structure moduli space of the punctured sphere, $\mathcal{M}_{0,4}$, but a double cover of it. $X_4$ is parametrized by $w$, rather than the cross-ratio, $x$. This feature recurs later in the discussions in Sec.~\ref{dshaped_quivers_15}. The $\mathbb{Z}_2$ deck transformation for this cover, $w\to -w$, implements the $\mathbb{Z}_2$ S-duality transformation
\begin{equation}
q\to {1}/{q}.
\end{equation}
on the family of gauge theories, of which the theory at $q=-1$ is a fixed-point. Moreover, this $\mathbb{Z}_{2}$ S-duality transformation is accompanied by a $\mathbb{Z}_2$ outer automorphism that acts on the global symmetry group of the theory. 
In particular, it acts as a $\mathbb{Z}_2$ automorphism of the SCFT at $q=-1$. 

A generic $\mathcal{N}=2$ gauge theory with simple gauge group and vanishing $\beta$-function need not possess such a $\mathbb{Z}_2$-symmetric point. But, for example, all the $A_3$ gauge-theory fixtures of \S\ref{gaugetheory_fixtures_44} do:
\begin{itemize}
\item $\SU(2)+4(\mathbf{2})$
\item $\SU(2)\times \SU(2)+4(\mathbf2,\mathbf1)+4(\mathbf1,\mathbf2)$
\item $\SU(3)+6(\mathbf{3})$
\item $\SU(4)+4(\mathbf{4})+2(\mathbf{6})$
\item $\Sp(2)+4(\mathbf{4})+1(\mathbf{5})$
\end{itemize}

For instance, $\SU(4)+4(\mathbf{4})+2(\mathbf{6})$ (as reviewed in \S\ref{id50_25}) has an $\SU(4)_8\times \UU(1)\times Sp(2)_6$ global symmetry. The outer automorphism corresponds to flipping the sign of the $\UU(1)$ charge (or, alternatively, complex-conjugating the $\SU(4)$ representation). As the $\UU(1)$ in question is the global symmetry group of the $[2N-1,1]$ puncture, this action is implemented in the mass-deformed theory as $m\to -m$, where $m$ is the mass-deformation parameter of the $[2N-1,1]$ puncture. The change in sign of $m$, as $[2N-1,1]$ circles the $\BP{[2N+1]}$ puncture is, thus, seen as the deck transformation $w\to -w$ on $X_4$.

\subsubsection{Derivation}\label{derivation1}
Using the prescription of Sec.~\ref{higgs_twisted_untwisted}, one can check that the OPE of the massless punctures of types $\BP{[2N+1]}$ and $[2N-1,1]$ yields the \emph{massless} $\BP{[2N-1,1^2]}$ puncture. This is a remarkable property. Usually, the OPE of two massless punctures yields a \emph{mass-deformed} new puncture; these masses give rise to the VEVs for the gauge group on the cylinder. In general, the masses of the new puncture are encoded in the choice of representatives of the orbits for the two original punctures, as emphasized in footnote \ref{noticefootnote}. In our case, the $\BP{[2N+1]}$ puncture has only one possible representative, the zero element. However, one still has a choice of representative for the $[2N-1,1]$ orbit. In fact, the OPE of $\BP{[2N+1]}$ with any other puncture does give a mass-deformed new puncture. So, the property we just described is very particular to the pair $\BP{[2N+1]}$ and $[2N-1,1]$. One immediate consequence is that the gauge group, if any, cannot have any VEVs supported on the cylinder. This is what we mean by an ``empty'' cylinder, which we denote by $\varnothing$.

Next, computing the OPE of the \emph{mass-deformed} $[2N-1,1]$ puncture with the $[2N+1]$ puncture, one obtains the mass-deformed $\BP{[2N-1,1^2]}$. We verify that the $\mathrm{U}(1)$ flavour group of the $\BP{[2N-1,1^2]}$ puncture can be identified with the $\mathrm{U}(1)$ of the $[2N-1,1]$ puncture.

Let us now derive the relation \eqref{relationQX} explicitly by studying the Seiberg-Witten curve of the 4-punctured sphere. We put the punctures of types $\BP{[2N+1]}$, $[2N-1,1]$, $U$, $\BP{T}$, at the positions $z=0$, $x$, $1$, $\infty,$ respectively, and consider the $x\to 0$ limit. As before, we discard all parameters irrelevant to the problem, that is, all mass deformations and Coulomb branch VEVs of scaling dimension different from two. So, the only non-zero $k$-differentials will be $\phi_2$ and $\phi_4$. In eliminating Coulomb branch VEVs, we have solved the constraints of the  $\BP{[2N+1]}$ puncture by imposing the relation $c_{2}^{(4)}=\frac{1}{4}\left(c_{1}^{(2)}\right)^{2}$ (which is itself a constraint for $N=2$) and setting to zero any other parameters. Thus, effectively, we have reduced a problem in the $A_{2N-1}$ theory to one in $A_{3}$. The $k$-differentials are:
\begin{equation}
\begin{aligned}
	\phi_{2}&=\frac{2x^\gamma u_{2}}{z(z-x)(z-1)}, &
	\phi_{4}&=\frac{x^\beta u_{4}}{z^{2}(z-x)(z-1)^{2}}.
\label{kdifferentials1}\end{aligned}
\end{equation}
The bound in \eqref{boundalpha} implies that $\gamma\geq 0$ and $\beta\geq 0$, while the constraint requires $u_4=-(u_{2})^2$ and $\beta=2\gamma-1$. Hence, the bounds are refined to $\gamma\geq1/2$ and $\beta\geq0$. But we cannot have $\gamma\gt 1/2$, $\beta\gt 0$ because then both $u_{2}$ and $u_{4}$ would disappear from either side of the degeneration when $x\to 0$, and we would `{}lose'{} a physical VEV. Thus, we must have $\gamma=1/2$, $\beta=0$. Hence, $u_{ 2}$ vanishes on both sides of the degeneration when $x\to 0$, and $u_{4}$ survives only on the gauge-theory fixture side. There are no gauge group VEVs supported on the cylinder, as we expected. Also, the parameter $u_{4}$ is a square, which is consistent with the a-constraint $c_{3}^{(4)}=(a^{(2)})^{2}$ of the new puncture, $\BP{[2N-1,1^{2}]}$.

Using our newly found values, we write the Seiberg-Witten curve:
\begin{equation}
y^{4}+\frac{2x^{1/2}u_{2}}{z(z-x)(z-1)}y^{2}-\frac{(u_{2})^{2}}{z^{2}(z-x)(z-1)^{2}}=0.
\end{equation}
This expression can be written, as before, as the product of two global factors:
\begin{equation}
\left(y^2+\frac{u_2}{z\left(z^{1/2}-x^{1/2}\right)(z-1)}\right)\left(y^2-\frac{u_2}{z\left(z^{1/2}+x^{1/2}\right)(z-1)}\right) = 0.
\label{swcurvetwofactors2}\end{equation}
Let us pick the first factor in this expression, and use again the transformation $z=t^{2}$, $y=y'/2t$. We get:
\begin{equation}
y'^{2}+\frac{4u_{2}}{\left(t-w\right)(t-1)(t+1)}=0,
\label{swcurve1}\end{equation}
where $w^2=x$. This is again the Seiberg-Witten curve for the $A_1$ four-punctured sphere representation of the $\SU(2)$ $N_{f}=4$ theory, at gauge coupling \eqref{relationQX}.

What about the second factor in \eqref{swcurvetwofactors2}? One arrives at a result similar to \eqref{swcurve1}, but with $u_2$ and $w$ traded for $-u_2$ and $-w$, respectively. So, this also represents the $\SU(2)$ gauge theory, with gauge coupling controlled by a cross ratio $q'=\frac{-1+w}{1+w}$. Notice that $q'=1/q$, and since the points at $q$ and $1/q$ are related by S-duality, both factors in \eqref{swcurvetwofactors2} represent the gauge theory at the same point in gauge-coupling moduli space. Again, they differ only by the choice of sign in $u_2$, which is left unfixed by the a-constraint $c^{(4)}_3=a^2$ in the $\BP{[2N-1,1^2]}$ puncture.

\hypertarget{_cylinder_13}{}\subsection{{$\SU(N)\times \SU(N)$ cylinder}}\label{_cylinder_13}
\subsubsection{How it arises}
Let us study a sphere with one minimal untwisted and two minimal twisted punctures bubbling off a plane:
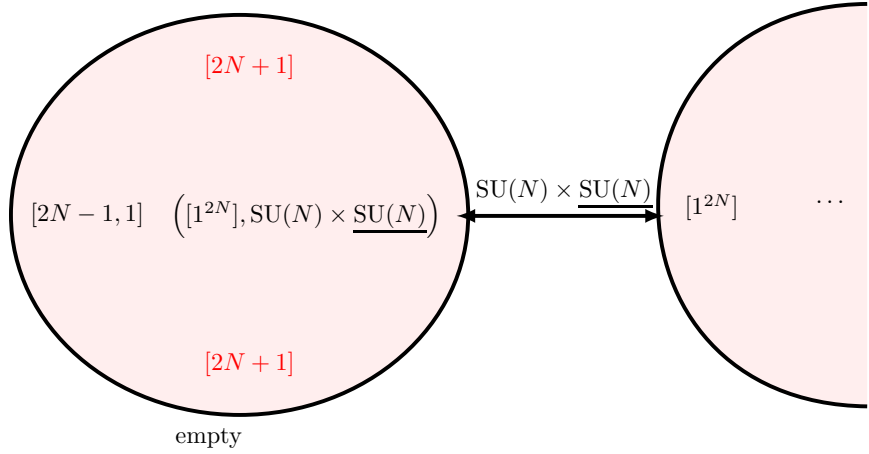
\begin{displaymath}
\footnotesize \includegraphics[width=331.5pt]{fig14}
\end{displaymath}
Here the maximal untwisted puncture, $[1^{2N}]$, at the right end of the cylinder has $\SU(2N)$ flavour symmetry, but only an $\SU(N)\times \SU(N)$ subgroup is gauged. Counting hypers reveals that the sphere to the left must be empty. Both $\SU(N)$ factors become weakly coupled when the cylinder degenerates. The second $\SU(N)$ factor is underlined to indicate, as we will see, that there is a specific degeneration of the empty sphere that takes the $\underline{\SU(N)}$ gauge coupling to zero, but does not decouple the non-underlined $\SU(N)$ factor.

Let us look at the degeneration where two $\BP{[2N+1]}$ collide:
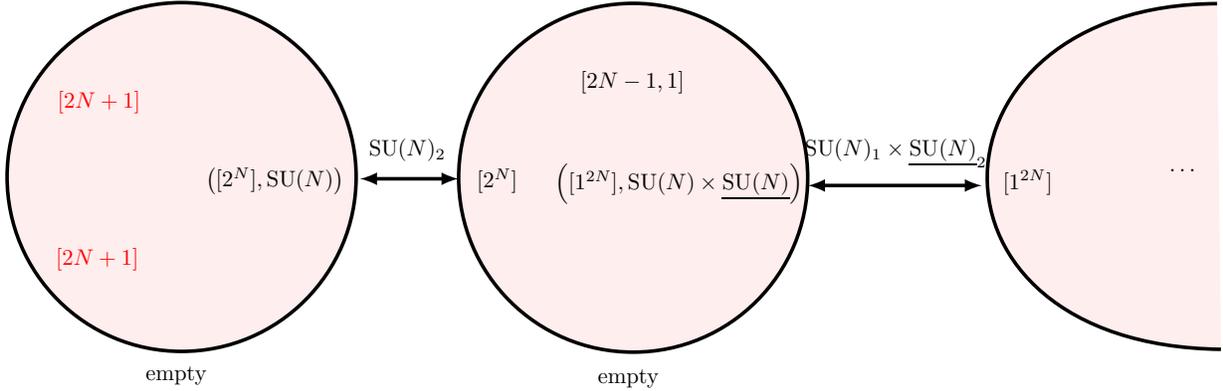
\begin{displaymath}
\footnotesize \includegraphics[width=\textwidth]{fig15}
\end{displaymath}
The $\bigl([1^{2N}],\SU(N)\times\underline{\SU(N)}\bigr)$ irregular puncture is more-or-less equivalent to $\bigl([1^{2N}],\SU(N)\bigr)$; it makes the same contribution to $n_h$, graded Coulomb branch dimensions, \emph{etc}. It differs only in how the global symmetries of the fixture are realized: the additional $\underline{\SU(N)}$ is identfied with the $\SU(N)$ symmetry of the $[2^N]$ puncture.

As one would expect, counting hypers tells us that the empty sphere decomposes into two empty fixtures. The underlined $\SU(N)$ gauge group is \emph{identified} with the $\SU(N)_2$ on the cylinder on the left. One can make the $\SU(N)_2$ gauge group weakly coupled by degenerating \emph{either} of the two shown cylinders. Degenerating completely the cylinder to the left turns off the $\SU(N)_2$ gauge coupling, and leaves just the $\SU(N)_1$ gauge group supported on the the cylinder on the right.

Instead, if we degenerate the cylinder to the right, both $\SU(N)_1$ and $\SU(N)_2$ factors decouple, and we are left with an empty four-punctured sphere on the left. Notice that the four-punctured sphere cannot be a conformal $\SU(N)$ gauge theory because it contains no hypers.

Let us look at this degeneration in more detail. We already saw that the collision of two minimal twisted punctures, $\BP{[2N+1]}$, yields a mass-deformed $[2^N]$ puncture with residue \eqref{OPEtwomintwistedpunctures} and an $SU(N)$ gauge group. As in \eqref{OPEtwomintwistedpunctures}, we take the $m_i$ (with $\sum_{i=1}^{N}m_i=0$) to be the mass deformations of the $[2^N]$ puncture.

Similarly, we can consider the OPE of the $[2^N]$ puncture with the massless ${[2N-1,1]}$ puncture, using the prescription of Sec.~\ref{higgs_untwisted_untwisted}. The resulting residue can be diagonalized to the form
\begin{equation}
\text{diag}(m_1,m_2,\dots,m_N;r_1,r_2,\dots,r_N),
\label{first-sun-sun-diagonal-form}\end{equation}
with $\sum_{i=1}^{N}r_i=0$. The $r_i$ are related to the $m_i$, but do not vanish if the $m_i$ are set to zero. Since all terms in \eqref{first-sun-sun-diagonal-form} are generically different, this boundary condition must correspond to an irregular version of the $[1^{2N}]$ puncture. In this case, we have two independent $\mathfrak{sl}(N)$ Lie algebras, with mass deformations $m_i$ and $r_i$, embedded in the $\mathfrak{sl}(2N)$ global symmetry group of the $[1^{2N}]$ puncture. Clearly, the $\mathfrak{sl}(N)$ factor with masses $m_i$ is identified with the $\mathfrak{sl}(N)$ global symmetry group of the $[2^N]$ puncture. In particular, if we turn off the $m_i$, we are still left with a $\mathfrak{sl}(N)$ factor, with Coulomb branch VEVs $r_i$. This is how the $\SU(N)\times \underline{\SU(N)}$ cylinder arises.

Consider also the degeneration where $\BP{[2N+1]}$ and $[2N-1,1]$ collide:
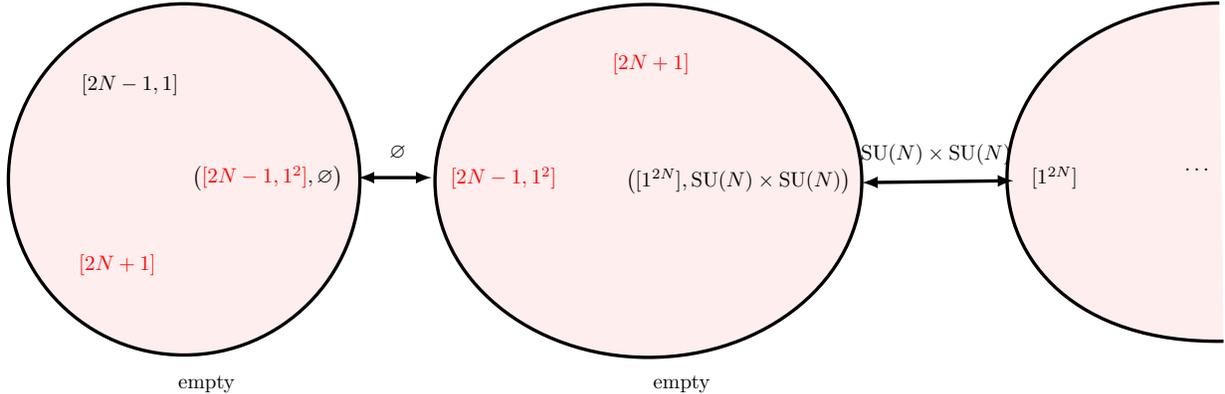
\begin{displaymath}
\footnotesize \includegraphics[width=\textwidth]{fig18}
\end{displaymath}
Here, again, as one would expect, both fixtures are empty. Degenerating the empty cylinder does not decouple either of the two $\SU(N)$ factors. On the other hand, if the $\SU(N)\times \SU(N)$ cylinder degenerates, both $\SU(N)$ factors become decoupled. We represent this behaviour by not underlining any of the $\SU(N)$ factors. Note that the $\mathbb{Z}_2$ action of the $\mathrm{O}(2)$ flavour symmetry now exchanges the two $\SU(N)$ gauge groups.

Let us now study the residues. We have already seen in Sec.~\ref{derivation1} that the OPE of a $\BP{[2N+1]}$ and a $[2N-1,1]$ puncture yields the $\BP{[2N-1,1^2]}$ puncture. So, we move on to the OPE of the $\BP{[2N-1,1^2]}$ puncture with a $\BP{[2N+1]}$. As one would expect, this yields a residue that can be diagonalized to the form \eqref{first-sun-sun-diagonal-form}. Notice that, in the present degeneration,  the $m_i$ do not have the interpretation of mass deformations of a puncture.

We leave the discussion of the dependence of the gauge couplings on the cross-sections to Sec.~\ref{dshaped_quivers_15}, which is an example that covers all the cases of atypical degenerations.

\hypertarget{_cylinder_14}{}\subsection{{$\SU(2)\times \SU(2)$ cylinder}}\label{_cylinder_14}

Let us now look at a sphere with two minimal untwisted and one minimal twisted punctures, bubbling off a plane. This example is similar to the previous one, but it involves an $\SU(2)\times \SU(2)$ cylinder.
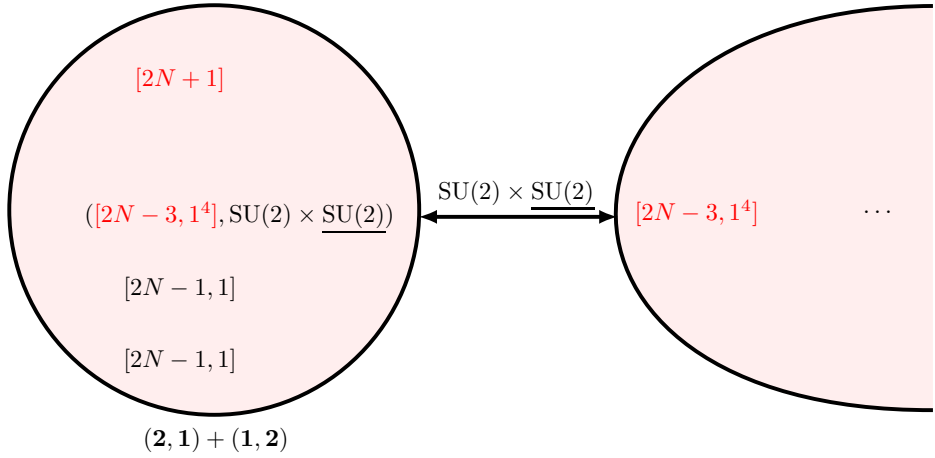
\begin{displaymath}
\footnotesize \includegraphics[width=357pt]{fig20}
\end{displaymath}
Here, the full $\SU(2)\times \SU(2)$ flavour symmetry of the $\BP{[2N-3,1^4]}$ puncture is gauged. (For $N=2$, the puncture is $[1^5]$, and the global symmetry group is enhanced to $\Sp(2)$, but only an $\SU(2)\times \SU(2)$ subgroup is gauged.) The sphere to the left contains four free hypers. Degenerating the cylinder decouples both $\SU(2)$ factors. As in the previous example, a certain degeneration of the four-punctured sphere, turns off only the underlined $\SU(2)$ factor. Such a degeneration is:
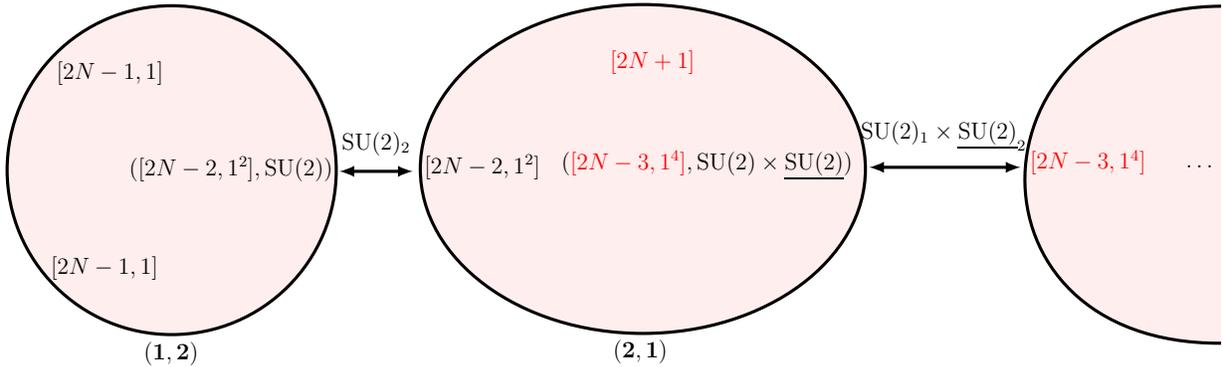
\begin{displaymath}
 \includegraphics[width=\textwidth]{fig21}
\end{displaymath}
As before, the underlined $\SU(2)$ is identified with the $\SU(2)$ on the left-hand cylinder. Notice that each fixture contains two hypers charged under one of the two $\SU(2)$ gauge group factors. The hypers in the left fixture are charged under $\SU(2)_2$. When the left cylinder decouples, the two hypers in the left fixture become free. If we instead degenerate the right cylinder, both $\SU(2)$ gauge groups decouple, and we get a 4-punctured sphere with four free hypers.

Let us look at the OPEs for this degeneration. Consider two punctures of type $[2N-1,1]$, with respective diagonalized residues
\begin{equation}
\text{diag}(m,m,\dots,m,-(2N-1)m),\qquad\quad\text{diag}(n,n,\dots,n,-(2N-1)n).
\end{equation}
The OPE of these punctures yields an untwisted puncture with diagonalized residue
\begin{equation}
\text{diag}(m+n,m+n,\dots,m+n,-(N-1)(m+n)+t,-(N-1)(m+n)-t).
\label{higgs-2Nminus21square}\end{equation}
This is a mass-deformed form of the $[2N-2,1^2]$ puncture. Here, $t$ depends on $m$ and $n$, but does not vanish if $m$ and $n$ are set to zero. So, $t$ supports the $\SU(2)_2$ cylinder that arises when the two $[2N-1,1]$ punctures get close to each other.

Now consider the OPE of a $[2N-2,1^2]$ with residue \eqref{higgs-2Nminus21square} and a $\BP{[2N+1]}$ puncture. This yields a twisted puncture with diagonalized residue
\begin{equation}
\text{diag}(-r-\frac{1}{2}t,-r+\frac{1}{2}t,0,0,\dots,0;r+\frac{1}{2}t,r-\frac{1}{2}t,0,0,\dots,0)
\end{equation}
This is a mass-deformed form of the $\BP{[2N-3,1^4]}$ puncture. Here $r$ depends on $t$, $m$ and $n$, but does not vanish if these parameters are set to zero. Thus, $r$ and $t$ parametrize the Coulomb branches of the $\SU(2)_1$ and $\SU(2)_2$ gauge groups, respectively.

The second degeneration is:
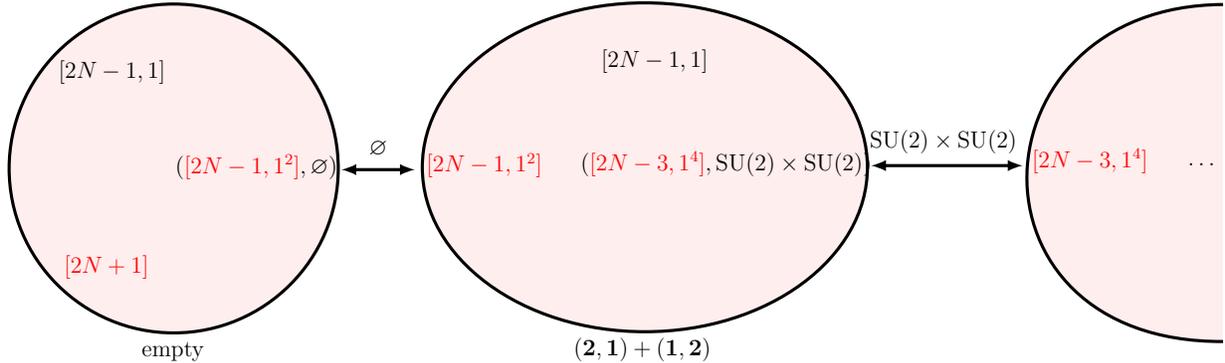
\begin{displaymath}
 \includegraphics[width=\textwidth]{fig24}
\end{displaymath}
Here the middle fixture contains the four hypers, while the left fixture is empty. Degenerating the left cylinder does not completely decouple either of the $\SU(2)$ factors, but degenerating the left cylinder decouples both. In this latter degeneration, the resulting four-punctured sphere again contains four free hypers.

We have already seen that the OPE of the $[2N-1,1]$ and the $\BP{[2N+1]}$ puncture is the $\BP{[2N-1,1^2]}$ puncture, so we may look at the OPE of the $\BP{[2N-1,1^2]}$ and the $[2N-1,1]$ puncture. This yields the twisted puncture with diagonalized residue
\begin{equation}
\text{diag}(w,s,0,0,\dots,0;-w,-s,0,0,\dots,0)
\end{equation}
This is a mass-deformed form of the $\BP{[2N-3,1^4]}$ puncture. It is clear that there is an $\SU(2)\times \SU(2)$ embedding in the $\SU(4)\times U(1)$ global symmetry group.

As we did for the case of the $\SU(N)\times \SU(N)$ cylinder, we leave the discussion of the dependence of the gauge couplings on the cross-sections to Sec.~\ref{dshaped_quivers_15}.

\hypertarget{dshaped_quivers_15}{}
\section{{D-Shaped Quivers}}\label{dshaped_quivers_15}

\begin{figure}
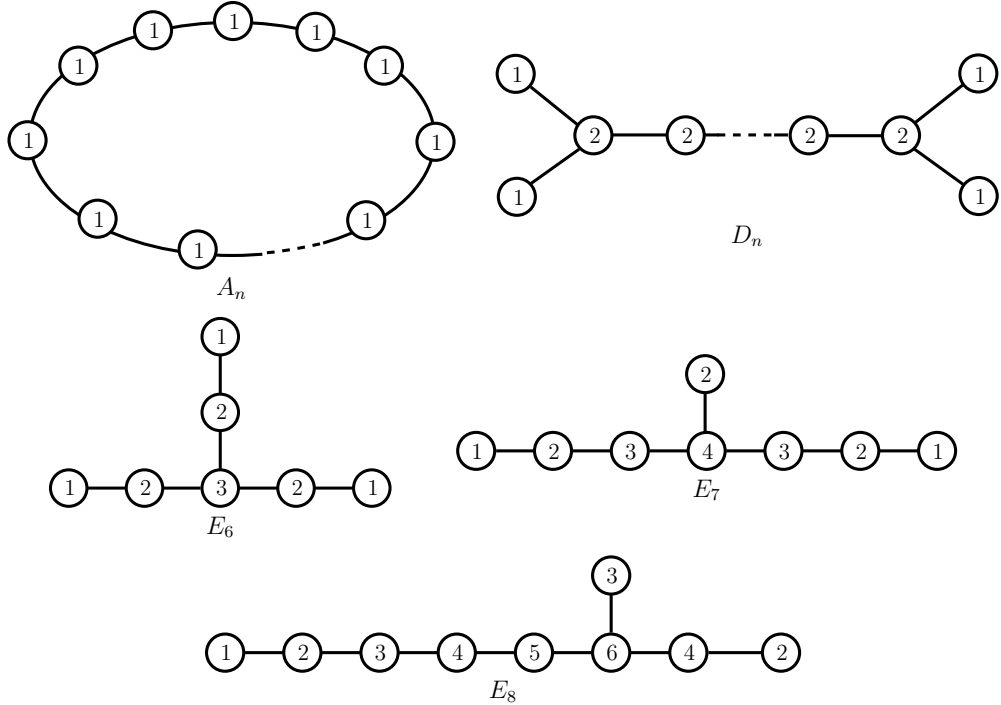

\[
\scalebox{.8}{$
\begin{gathered}
\begin{matrix} \includegraphics[width=219pt]{affineAn}\\ A_n\end{matrix}\quad
\begin{matrix} \includegraphics[width=245pt]{affineDn}\\ D_n\end{matrix}
\\
\begin{matrix} \includegraphics[width=169pt]{affineE6}\\ E_6 \end{matrix}\qquad
\begin{matrix} \includegraphics[width=244pt]{affinE7}\\ E_7\end{matrix}
\\
\begin{matrix} \includegraphics[width=289pt]{affineE8}\\ E_8\end{matrix}
\end{gathered}
$}
\]
\caption{Extended Dynkin diagrams and Dynkin labels. \label{dynkin}}
\end{figure}

Consider the extended Dynkin diagrams for the simply-laced Lie algebras, given in Fig.~\ref{dynkin}, where we have indicated the Dynkin label of each node.
It is well known that one obtains a conformally-invariant quiver gauge theory by assigning an $\SU(l_i N)$ gauge theory to the $i$-th node (whose Dynkin label is $l_i$), and a hypermultiplet, in the bi-fundamental, to each link. It has not, however, been known whether all of these affine quiver gauge theories can be realized as compactifications of the (2,0) theory. The realization of the affine $A_n$ quivers \emph{is} well-known: compactify the $A_{N-1}$ theory on a torus with $n$ simple punctures.
In this section, we present the analogous six-dimensional realization of the affine $D_n$ quivers in the twisted $A_{2N-1}$ theory. This was first found by Kapustin \cite{Kapustin:1998fa} using a chain of string dualities. We will first present our construction using twisted punctures, and then compare it with Kapustin'{}s. 

On the other hand, it is also known that any quiver gauge theory with $\SU$ gauge groups which is semiclassically conformal has its gauge groups arranged in the form of a non-affine Dynkin diagram, with $\SU(N_i)$ gauge groups on the $i$-th node and bifundamentals associated to the edges, together with some fundamental flavours at each of the nodes. The realization of the non-affine $A_n$-shaped quivers is known: compactify the $A_{N-1}$ theory on an untwisted sphere with two regular punctures and a number of simple punctures. At the end of this section, we show how an arbitrary non-affine $D_n$-shaped quiver can be analogously realized in the twisted $A_{2N-1}$ theory.

\hypertarget{AffineDnQuivers}{}\subsection{{Affine $D_n$-shaped quivers}}\label{AffineDnQuivers}

The affine $D_n$-shaped quiver arises from the compactification of the $A_{2N-1}$ theory on a sphere, with four copies of the minimal twisted puncture, $\BP{[2N+1]}$, and $n$ copies of the minimal untwisted puncture, $[2N-1,1]$. A partial degeneration of this curve is
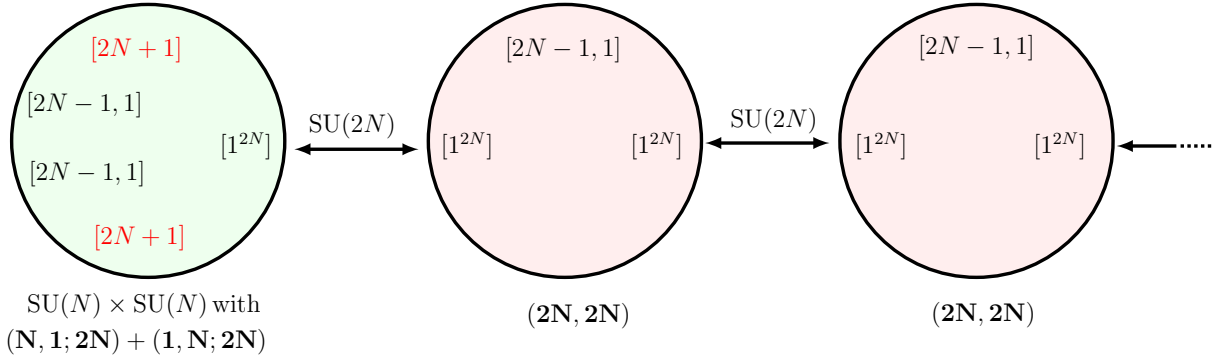
\begin{displaymath}
 \includegraphics[width=\textwidth]{fig30}
\end{displaymath}
Here we show only one of the two ends of the affine $D_n$ quiver, since the other end is identical. The $\SU(2N)$ cylinders here represent the nodes with a label ``{}2''{} in the affine D-series Dynkin diagram in the figure above, and the bifundamental fixtures represents the links. The non-trivial piece is the nodes with a ``{}1''{} at each end of the Dynkin diagram, which correspond to $\SU(N)$ gauge groups. In the sphere above, this piece is represented by the 5-punctured sphere at the left end of the figure. We have deliberately not degenerated the 5-punctured sphere at the end of quiver, since the punctures there are in combinations that lead to atypical degenerations, as studied in the previous section. Let us then examine the degenerations of this 5-punctured sphere in detail.

\hypertarget{Degenerations}{}\subsubsection{{Degenerations of the 5-punctured sphere}}\label{Degenerations}

The 5-punctured sphere at the end of the affine $D_n$ quiver has six inequivalent degenerations. When both $\SU(N)$s are weakly-coupled, we have a Lagrangian field theory, with matter in the $(N,1;2N)+(1,N;2N)$ of the $\SU(N)\times \SU(N)\times \SU(2N)$. The $\SU(2N)_{4N}$ global symmetry of the $[1^{2N}]$ puncture (which is gauged, when we attach this 5-punctured sphere to the rest of the surface) is realized as the diagonal embedding in the $\mathrm{S}[\UU(2N)\times \UU(2N)]$ flavour symmetry of the gauge theory. When one or the other of the $\SU(N)$s approaches its strongly-coupled cusp point, that part of the theory is better described by an $\SU(2)$ gauging of the $R_{0,N}$ SCFT (with an additional hypermultiplet in the fundamental of $\SU(2)$). 
The Lagrangian field theory arises only in an ``{}atypical''{} degeneration (in the nomenclature of the previous section).

\paragraph{Degeneration A:} The only degeneration that can be understood in the usual, non-atypical sense is:
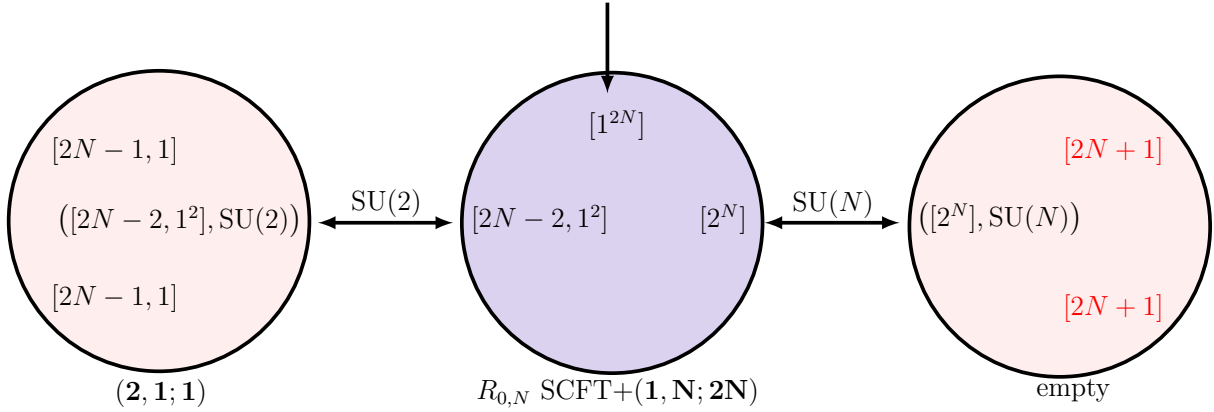
\begin{displaymath}
 \includegraphics[width=\textwidth]{fig31}
\end{displaymath}
The $\SU(2)$ and the $\SU(N)$ gauge groups shown here become weakly coupled as their respective cylinders degenerate. The representations for the matter above are those for the product $\SU(2)\times \SU(N)\times \SU(2N)$.

The next five degenerations are all atypical.

\paragraph{Degeneration B:} In this degeneration all of the dynamics is supported on the middle fixture, which is a gauge theory fixture:
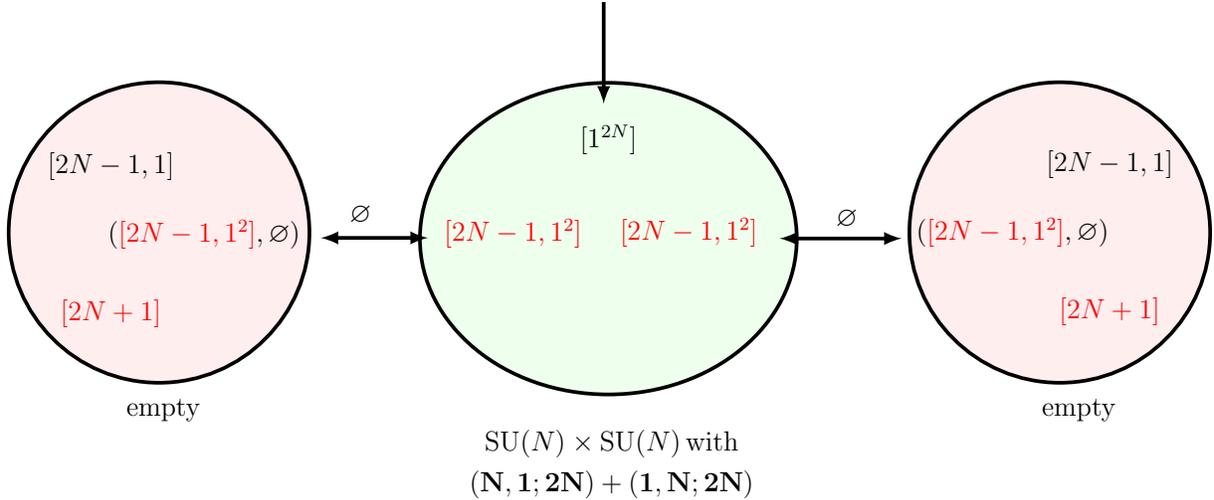
\begin{displaymath}
 \includegraphics[width=\textwidth]{fig32}
\end{displaymath}
When the two cylinders pinch off, the middle fixture gives two copies of the $\SU(N)$ $N_f=2N$ SCFT at the $\mathbb{Z}_2$-symmetric point of each theory'{}s marginal coupling moduli space. Turning on, say, the plumbing fixture for the cylinder to the right perturbs one of the gauge couplings away from the $\mathbb{Z}_2$-symmetric point.

\paragraph{Degeneration C:} 
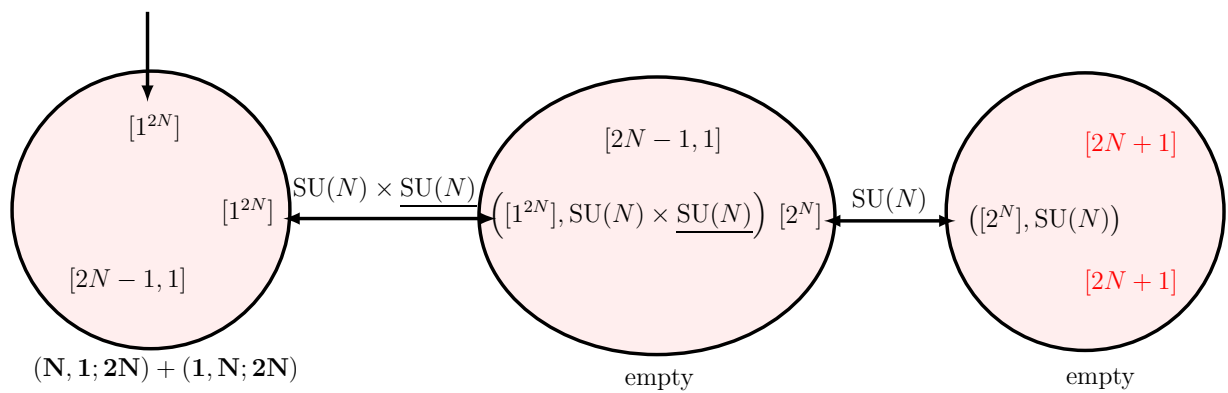
\begin{displaymath}
 \includegraphics[width=\textwidth]{fig33}
\end{displaymath}
Here, the underlined $\underline{\SU(N)}$ is identified with the $\SU(N)$ cylinder on the right hand side. It goes to zero gauge coupling when \emph{either} cylinder pinches off.

\paragraph{Degeneration D:} 
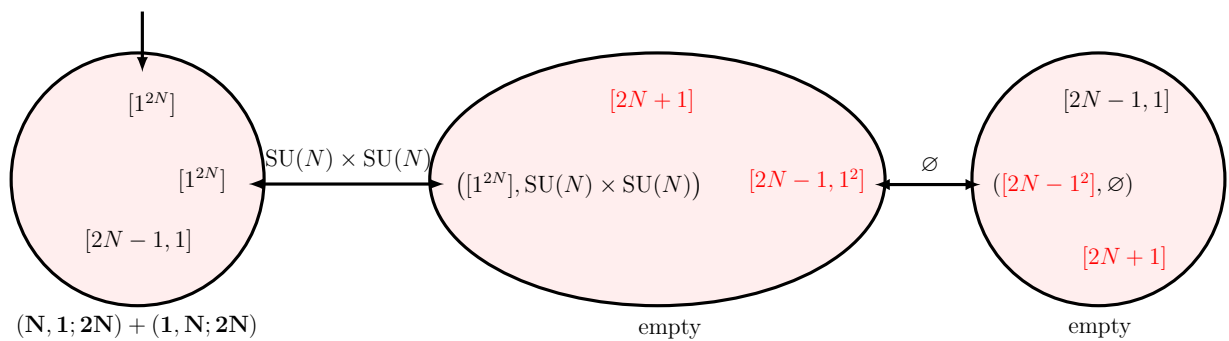
\begin{displaymath}
 \includegraphics[width=\textwidth]{fig34}
\end{displaymath}
In this degeneration, the two $\SU(N)$ gauge couplings become \emph{equal} when the cylinder on the right pinches off.

\paragraph{Degeneration E:} 
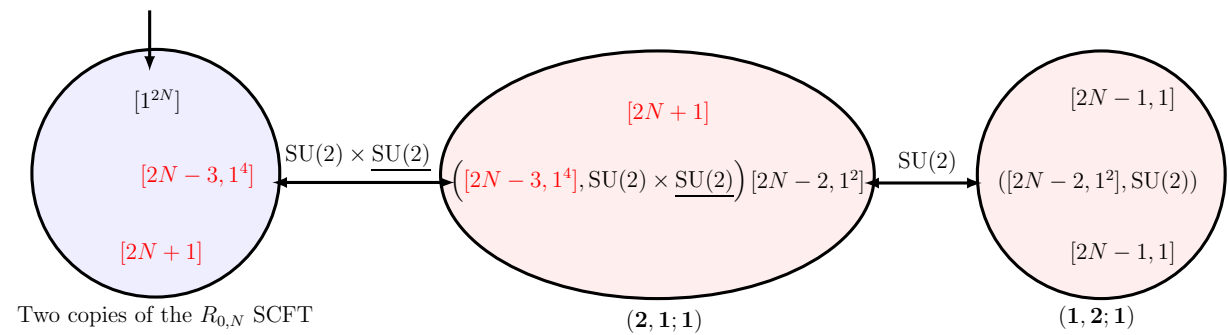
\begin{displaymath}
 \includegraphics[width=\textwidth]{fig35}
\end{displaymath}
Here, the ${\SU(2N)}_{4N}$ symmetry of the $[1^{2N}]$ puncture is the diagonal embedding in the ${\left(\SU(N)_{2N}\times \SU(2)_6\right)}^2$ global symmetry of two copies of the $R_{0,N}$ SCFT. The underlined $\underline{\SU(2)}$ is identified with the $\SU(2)$ on the righthand cylinder, and goes to zero coupling when \emph{either} cylinder pinches off.

\paragraph{Degeneration F:} 
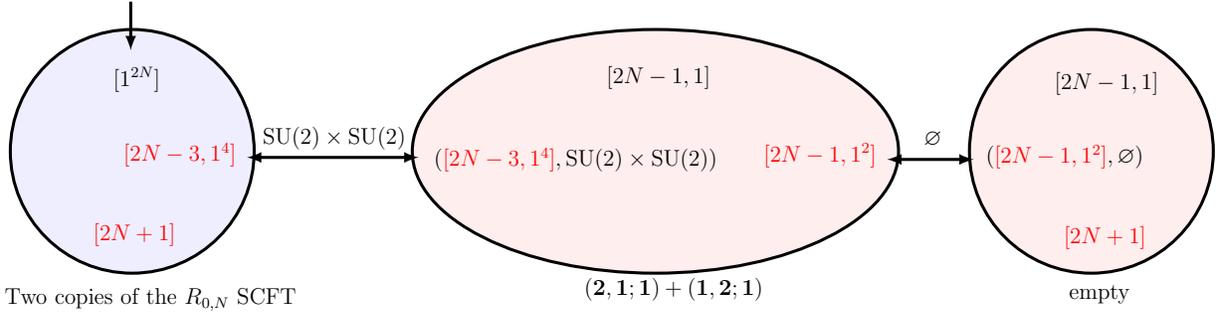
\begin{displaymath}
 \includegraphics[width=\textwidth]{fig36}
\end{displaymath}
In this last degeneration, the two $\SU(2)$ gauge couplings become equal when the righthand cylinder pinches off.

\hypertarget{true}{}\subsubsection{{The  moduli space of coupling constants vs.~the complex structure moduli space}}\label{true}

Let us locate the punctures at
\begin{equation}
\begin{matrix}
z_1&z_2&z_3&z_4&z_5\\
\BP{[2N+1]}&\BP{[2N+1]}&[2N-1,1]&[2N-1,1]&[1^{2N}]
\end{matrix}.
\end{equation}
The ring of meromorphic functions on $\bar{\mathcal{M}}_{0,5}$ consists of all rational functions of the independent cross-ratios
\begin{equation}
s_1 = \frac{z_{1 3} z_{2 5} }{z_{1 5} z_{2 3}},\qquad
   s_2 = \frac{z_{1 4} z_{2 5} }{z_{1 5} z_{2 4}}.
\label{indepxratios}\end{equation}
Then the compactified moduli space of the 5-punctured sphere, $\bar{\mathcal{M}}_{0,5}$ is
obtained by blowing up $\mathbb{CP}^1\times \mathbb{CP}^1$ described by $s_1$ and $s_2$ at 3 points; the result is a del Pezzo surface, $dP_4$. 
The boundary divisor consists of 10 rational curves, $D_{i j},\, 1\leq i\lt j\leq 5$, corresponding to the locus where the points $z_i$ and $z_j$ collide. Each $D_{i j}$ has self-intersection number $=-1$, and intersects precisely three others
\begin{equation}
D_{i j} \cap D_{k l} = +1\quad \text{for}\, i,j,k,l\, \text{all distinct}.
\end{equation}

The moduli space of the coupling constants, however, is \emph{not} $\bar{\mathcal{M}}_{0,5}$. Instead, it is a 4-sheeted branched cover $X_5\to \bar{\mathcal{M}}_{0,5}$, branched over the compactification divisor. We will be more precise about the nature of the ramification, below, but $X_5$ is most effectively described as being the rational surface whose ring of meromorphic functions consists of all rational functions of $y,w$ where
\begin{equation}
y^2 =s_1,\quad w^2 = s_2.
\label{goodcoords}\end{equation}
The UV gauge couplings of our two decoupled gauge theories are
\begin{equation}
\begin{aligned}
   q_1 &= \frac{y-1}{y+1}\,\frac{w+1}{w-1}, &
   q_2 &= \frac{y-1}{y+1}\,\frac{w-1}{w+1}
\end{aligned}
\label{gaugecouplings}\end{equation}
where $q=0,\infty$ correspond to a weakly-coupled $\SU(N)$ gauge group and $q=1$ is the point where the dual $\SU(2)$ gauge group is weakly-coupled.

\begin{table}
\centering
\begin{tabular}{|c|c|c|}
\hline
Divisor&$\#$ sheets&$(q_1,q_2)$\\
\hline 
\hline 
$D_{1 2}$&4&$\begin{gathered}(0,q_2)\\ (\infty,q_2)\\ (q_1,0)\\(q_1,\infty)\end{gathered}$\\
\hline 
$D_{3 4}$&4&$\begin{gathered}(1,q_2)\\ (q_1,1)\end{gathered}$\\
\hline 
$D_{3 5}$&4&$\begin{gathered}(0,0)\\ (\infty,\infty)\end{gathered}$\\
\hline 
$D_{4 5}$&4&$\begin{gathered}(0,0)\\ (\infty,\infty)\end{gathered}$\\
\hline 
$D_{1 3}$&2&$(q_1,1/q_1)$\\
\hline 
$D_{1 4}$&2&$(q_1,q_1)$\\
\hline 
$D_{2 3}$&2&$(q_1,1/q_1)$\\
\hline 
$D_{2 4}$&2&$(q_1,q_1)$\\
\hline 
$D_{1 5}$&1&$(1,1)$\\
\hline 
$D_{2 5}$&1&$(1,1)$\\
\hline
\end{tabular}
\qquad
\begin{tabular}{|c|c|c|}
\hline
Degeneration&Intersection&$(q_1,q_2)$\\
\hline 
\hline
\textbf{A} &$D_{1 2}\cap D_{3 4}$&$\begin{gathered}(1,0)\\ (1,\infty)\\ (0,1)\\ (\infty,1)\end{gathered}$\\
\hline 
\textbf{B} &$\begin{gathered}D_{1 3}\cap D_{2 4}\\ D_{1 4}\cap D_{2 3}\end{gathered}$&$(-1,-1)$\\
\hline 
\textbf{C} &$\begin{gathered}D_{1 2}\cap D_{3 5}\\ D_{1 2}\cap D_{4 5}\end{gathered}$&$\begin{gathered}(0,0)\\ (0,\infty)\\ (\infty,0)\\ (\infty,\infty)\end{gathered}$\\
\hline 
\textbf{D} &$\begin{gathered}D_{1 3}\cap D_{4 5}\\ D_{1 4}\cap D_{3 5}\\ D_{2 3}\cap D_{4 5}\\ D_{2 4}\cap D_{3 5}\end{gathered}$&$\begin{gathered} (0,\infty)\\ (\infty,0)\end{gathered}$\\
\hline 
\textbf{E} &$\begin{gathered}D_{1 5}\cap D_{3 4}\\ D_{2 5}\cap D_{3 4}\end{gathered}$&$(1,1)$\\
\hline 
\textbf{F} &$\begin{gathered}D_{1 5}\cap D_{2 3}\\ D_{1 5}\cap D_{2 4}\\ D_{2 5}\cap D_{1 3}\\ D_{2 5}\cap D_{1 4}\end{gathered}$&$(1,1)$\\
\hline
\end{tabular}
\caption{Behaviour of the couplings at each of the degenerations and double degenerations.\label{ramification}}
\end{table}

There is a natural action of the dihedral group, $D_4$, on our family of gauge theories, generated by
\begin{equation}
\begin{aligned}
\alpha:\ q_1&\to 1/q_1,&  q_2&\to q_2,\\
\beta:\    q_1&\to q_1,& q_2&\to 1/q_2,\\
\gamma:\  q_1&\leftrightarrow q_2.
\end{aligned}
\end{equation}
This $D_4$ action is implemented on $X_5$ by
\begin{equation}
\begin{aligned}
\alpha:\ (y,w)&\to(y,-w),\\
\beta:\ (y,w)&\to(w,y),\\
\gamma:\  (y,w)&\to (-w,-y).
\end{aligned}
\end{equation}
While they are easy to compute from \eqref{indepxratios}, \eqref{goodcoords} and \eqref{gaugecouplings}, we indicate, in Table~\ref{ramification}, the number of sheets over a generic point on the divisor and the behaviour of the gauge couplings \eqref{gaugecouplings} on the pre-images of each of the components of the compactification divisor. E.g., on one of the pre-images of $D_{3 4}$, $q_1\equiv 1$, while $q_2$ varies. On the other pre-image, $q_1$ varies, while $q_2\equiv 1$. From these, we easily see that the behaviour of the gauge theory, at each of the degenerations discussed in the previous subsection, is as we claimed. For instance, on $D_{1 3}\cap D_{24}$ (or $D_{1 4}\cap D_{2 3}$), we have $y=0,\, w=\infty$ ($y=\infty,\, w=0$) and hence $q_1=q_2=-1$.

\hypertarget{Kapustin}{}
\subsection{{Comparison to Kapustin'{}s work}}\label{Kapustin}

In \cite{Kapustin:1998fa}, Kapustin realized the affine D-shaped quiver in Type IIA string theory. As always, consider $2N$ D4-branes extending along directions 01236, and $k$ NS5-branes extending along directions 012345. Furthermore, introduce a suitable orbifold whose action includes $(-1)^{F_L}$ and flips the directions 6789. Let the orientifold action fix $x^6=0,L$, and let the $x^6$-coordinate of the NS5-branes be $L_{1,2,\ldots,k}$. Then, each segment $[L_i,L_{i+1}]$, for $i=1,\ldots,{k-1}$ gives an $\SU(2N)$ gauge group, each NS5-brane gives a bifundamental, and the segments $[0,L_1]$ and $[L_k,L]$ give each an $\SU(N)\times \SU(N)$ gauge group. Hence, this construction realizes the affine D-shaped quiver. In this construction, the inverse square of each 4D gauge coupling is given by the length of the corresponding segment; in particular, the two coupling constants of the $\SU(N)\times \SU(N)$ gauge group at the leftmost end are fixed to be at the same value.

This particular orbifold is known to be magnetically charged under the B-field, like an NS5-brane. The M-theory lift of the configuration is then given by $2N$ M5-branes wrapped on a torus parametrized by $z$, together with the M-theory orientifold action $z\to -z$, which has four fixed points. Each of the two orbifold planes lifts to a pair of fixed points, each pair having an M5-brane on top of it. We also have $k$ M5-branes intersecting the torus. We can move the two M5-branes away from the fixed points, and the final configuration becomes $2N$ M5-branes on a torus with the orientifold action $z\to -z$, plus $k+2$ M5-branes.

We can take the decoupling limit. Each of the M-theory orientifold fixed points becomes a twisted simple puncture of type $\BP{[2N+1]}$; the torus divided by $z\to -z$ is a sphere; and the intersection with an M5-brane is a untwisted simple puncture of type $[2N-1,1]$. Thus we have the 6D theory of type $A_{2N-1}$ on a sphere with four twisted punctures of type $\BP{[2N+1]}$ and $k+2$ untwisted simple punctures of type $[2N-1,1]$, which reproduces our previous analysis.

We saw above that the degeneration limit where an untwisted simple puncture collides with a twisted simple puncture corresponds to the point where the two gauge couplings of an $\SU(N)\times \SU(N)$ gauge group become equal. In the M-theory construction, this corresponds to the fact that to take the IIA limit, two of the fixed points need to be paired, with an M5-brane on top of them.

\hypertarget{nonAffine}{}
\subsection{{Non-affine $D_n$-shaped quivers}}\label{nonAffine}

Let us next consider semiclassically-conformal non-affine D-shaped quivers:
\begin{equation}
\SU(n_1)\times \SU(n_2)\times \SU(N_1)\times \SU(N_2)\times \cdots \SU(N_l)
\label{nonaffinequiver}\end{equation}
with bifundamentals for $\SU(n_1)\times \SU(N_1)$, $\SU(n_2)\times \SU(N_1)$, and $\SU(N_i)\times \SU(N_{i+1})$, $k_{1,2}$ fundamental flavours for $\SU(n_{1,2})$, and $K_{1,2,\ldots,l}$ fundamental flavours for $\SU(N_{1,2,\ldots,l})$, respectively.

Without sacrificing generality, we can assume $n_1\ge n_2$. We must also have $n_1+n_2\ge N_1$; otherwise, $n_2<N_1/2$, which would render the $\SU(n_2)$ gauge group non-asymptotically-free. This forces $N_1\ge N_2\ge N_3\cdots \ge N_l$.

Now, consider the $A_{2n_1-1}$ theory compactified on a sphere with a full untwisted puncture, a regular untwisted puncture, and $l+1$ simple untwisted punctures; this sphere realizes the 4D linear quiver tail:
\begin{equation}
[\SU(n_1+n_2)]\times \SU(N_1)\times \SU(N_2)\times \cdots \SU(N_l),
\label{quivertail}\end{equation}
with $K_i$ additional flavours for $\SU(N_i)$. Here, the original $SU(2n_1)$ global symmetry group of the full puncture spontaneously breaks to an $\SU(n_1+n_2)$ subgroup, which we represent by $[\SU(n_1+n_2)]$ at the left end of the quiver.

On the other hand, from Sec. \ref{_cylinder_13}, we know that in a 4-punctured sphere in the $A_{2n_1-1}$ theory with two $\BP{[2n_1+1]}$ punctures, one $[2n_1-1,1]$ puncture, and a $[1^{2n_1}]$ puncture, the $\SU(2n_1)$ flavour symmetry of the full puncture spontaneously breaks to $\SU(n_1)\times \SU(n_1)$.

Now, let us combine these two sets of punctures by connecting the two full punctures. Specifically, consider a single sphere with
\begin{itemize}%
\item two $\BP{[2n_1+1]}$ punctures,
\item $(l+2)$ $[2n_1-1,1]$ punctures,
\item a regular untwisted puncture, which characterizes the linear quiver tail \eqref{quivertail}.

\end{itemize}
The cylinder connecting the two full punctures has one side spontaneously broken to $\SU(n_1)\times \SU(n_1)$, and the other to $\SU(n_1+n_2)$. So, in the 4D limit, it supports an $\SU(n_1)\times \SU(n_2)$ gauge group.
The combined system hence realizes the non-affine D-shaped quiver \eqref{nonaffinequiver}.

\hypertarget{_and__gauge_theories_21}{}
\section{{$\SU(4)$ and $\Sp(2)$ Gauge Theories}}\label{_and__gauge_theories_21}

As an application of the $A_3$ twisted theory, we study below S-duality of the $\SU(4)$ and $\Sp(2)$ superconformal gauge theories with matter in all allowed combinations of the fundamental and antisymmetric representations. The full tables of twisted and untwisted punctures, cylinders and fixtures of the $A_3$ theory are shown in Appendix~\ref{_twisted_sector_36}.

\hypertarget{_gauge_theory_22}{}
\subsection{{$\SU(4)$ gauge theory}}\label{_gauge_theory_22}

Including the twisted sector of the $A_3$ theory allows us to write down Argyres-Seiberg duals for $\SU(4)$ gauge theory, with matter in the $n(\mathbf{6}) +(8-2n)(\mathbf{4})$, for $n=0,1,2,3,4$. The cases $n=0,1,2$ were already accessible in the untwisted sector. We will review those, first, before describing $n=3,4$, where we will have recourse to punctures from the twisted sector.

\hypertarget{id48_23}{}
\subsubsection{{$8(\mathbf{4})$}}\label{id48_23}

$\SU(4)$, with 8 hypermultiplets in the $\mathbf{4}$, can be realized by
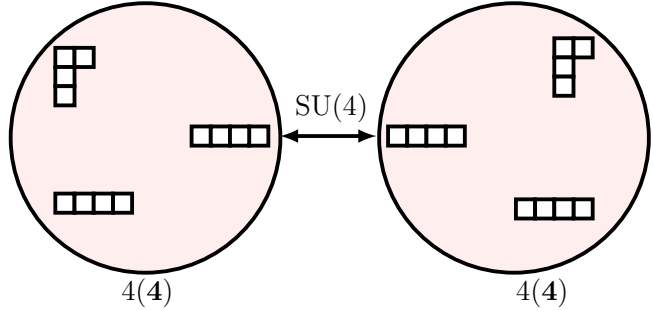
\begin{displaymath}
 \includegraphics[width=253.5pt]{SU4_8fund}
\end{displaymath}
The strong coupling dual is an $\SU(2)$ gauge theory, with a hypermultiplet in the $\mathbf{2}$, gauging the ${\SU(2)}_6$ of the ${\SU(2)}_6\times \SU(8)_8$ SCFT
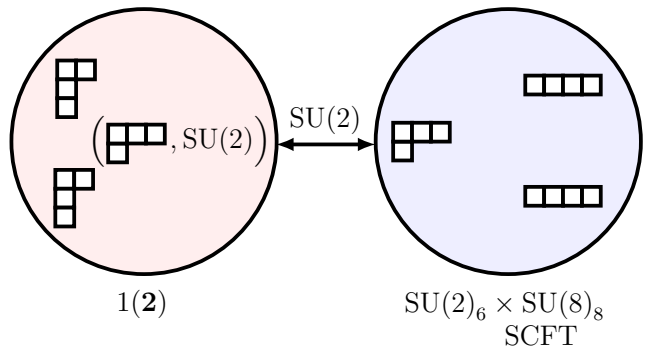
\begin{displaymath}
 \includegraphics[width=253.5pt]{SU4_8fund_dual}
\end{displaymath}
\hypertarget{id49_24}{}\subsubsection{{$1(\mathbf{6})+6(\mathbf{4})$}}\label{id49_24}

Replacing one of the free-field fixtures by one that yields matter in the $1(\mathbf{6})+2(\mathbf{4})$, we obtain
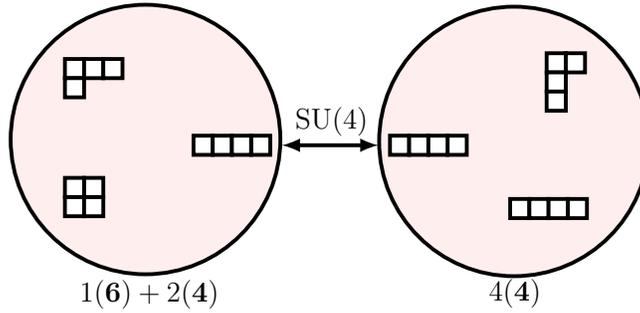
\begin{displaymath}
 \includegraphics[width=253.5pt]{SU4_1six_6fund}
\end{displaymath}
which yields an $\SU(4)$ gauge theory with matter in the $1(\mathbf{6})+6(\mathbf{4})$.
Now there are two distinct strong coupling limits. One is
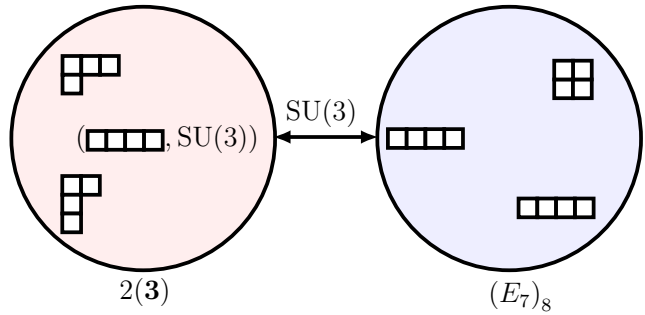
\begin{displaymath}
 \includegraphics[width=253.5pt]{SU4_1six_6fund_dual1}
\end{displaymath}
which is an $\SU(3)$ gauge theory with matter in the $2(\mathbf{3})$, gauging an $\SU(3)$ subgroup of the ${(E_7)}_8$ SCFT. The other is
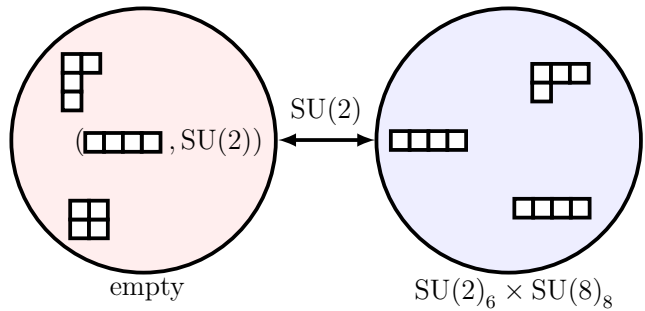
\begin{displaymath}
 \includegraphics[width=253.5pt]{SU4_1six_6fund_dual2}
\end{displaymath}
which is a gauging of an $\SU(2)\subset {\SU(8)}_8$ of the ${\SU(2)}_6\times{\SU(8)}_8$ SCFT.

\hypertarget{id50_25}{}\subsubsection{{$2(\mathbf{6})+4(\mathbf{4})$}}\label{id50_25}

Now, we take a configuration where both fixtures yield matter in the $1(\mathbf{6})+2(\mathbf{4})$.
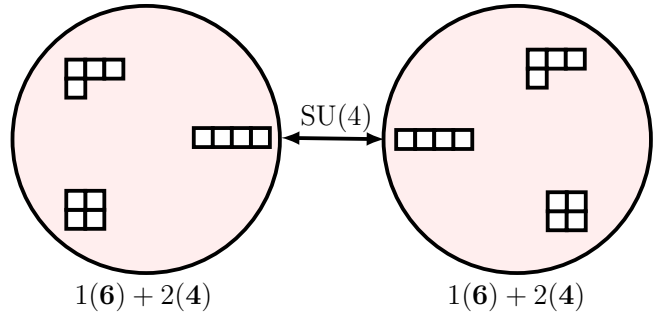
\begin{displaymath}
 \includegraphics[width=253.5pt]{SU4_2six_4fund}
\end{displaymath}
This yields $\SU(4)$ gauge theory with matter in the $2(\mathbf{6})+4(\mathbf{4})$.
Taking the S-dual, we obtain
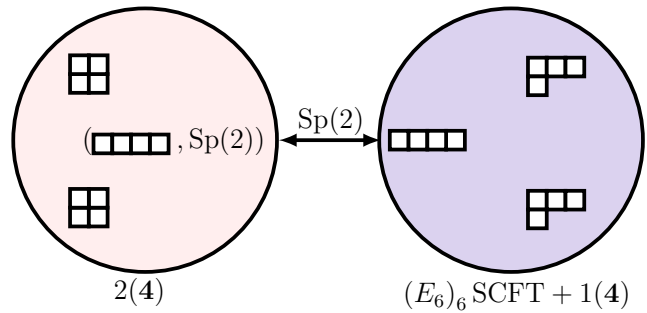
\begin{displaymath}
 \includegraphics[width=253.5pt]{SU4_2six_4fund_dual}
\end{displaymath}
The fixture on the right is a mixed fixture, consisting of the ${(E_6)}_6$ SCFT and additional hypermultiplets, transforming as the $1(\mathbf{4})$. All in all, the S-dual is an $\Sp(2)$ gauge theory, coupled to the ${(E_6)}_6$ SCFT, with matter in the $3(\mathbf{4})$. The examples so far were already discussed in \cite{Chacaltana:2010ks}.

\hypertarget{id51_26}{}\subsubsection{{$3(\mathbf{6})+2(\mathbf{4})$}}\label{id51_26}

To proceed further, we replace one of the previous fixtures with one from the twisted sector of the $A_3$ theory
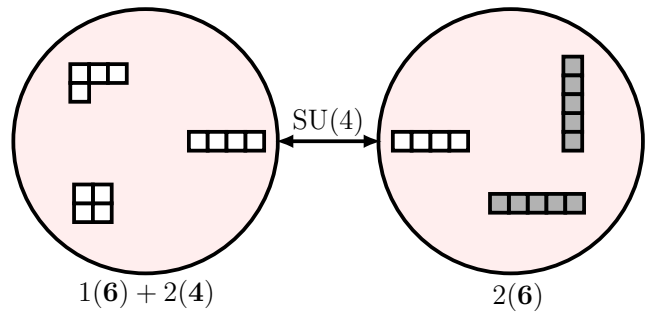
\begin{displaymath}
 \includegraphics[width=253.5pt]{SU4_3six_2fund}
\end{displaymath}
This yields a realization of $\SU(4)$ gauge theory with matter in the $3(\mathbf{6})+2(\mathbf{4})$.

There are two distinct strong coupling limits.
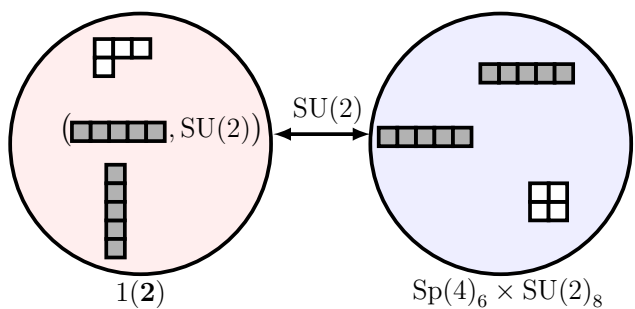
\begin{displaymath}
 \includegraphics[width=253.5pt]{SU4_3six_2fund_dual1}
\end{displaymath}
is an $\SU(2)$ gauge theory, with matter in the $1(\mathbf{2})$, coupled to the ${\Sp(4)}_6\times {\SU(2)}_8$ SCFT (gauging an $\SU(2)\subset {\Sp(4)}_6$).
The other limit,
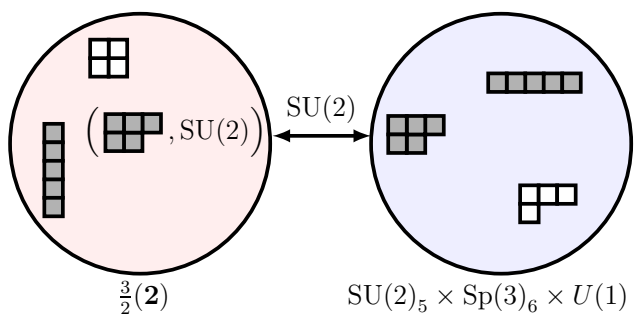
\begin{displaymath}
 \includegraphics[width=253.5pt]{SU4_3six_2fund_dual2}
\end{displaymath}
is an $\SU(2)$ gauge theory, with \emph{three} half-hypermultiplets in the $\mathbf{2}$, coupled to the ${\SU(2)}_5\times {\Sp(3)}_6\times \UU(1)$ SCFT.

\hypertarget{id52_27}{}\subsubsection{{$4(\mathbf{6})$}}\label{id52_27}

Finally, we can take both fixtures from the twisted sector.
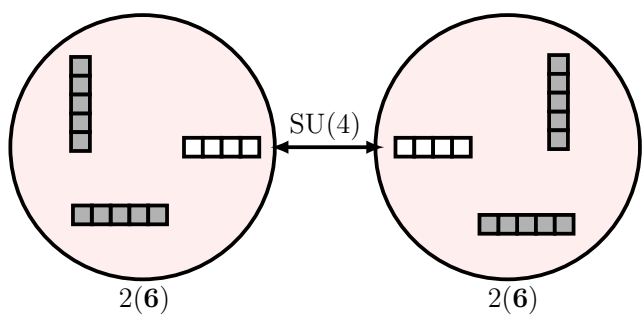
\begin{displaymath}
 \includegraphics[width=253.5pt]{SU4_4six}
\end{displaymath}
The S-dual theory,
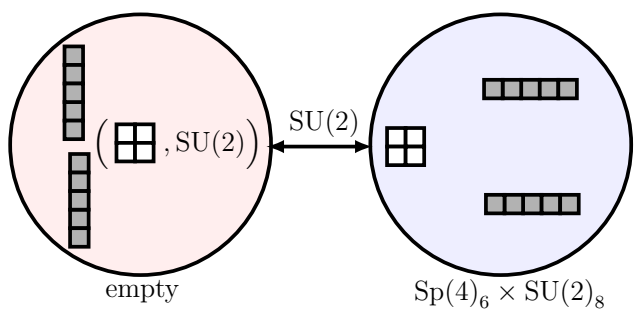
\begin{displaymath}
 \includegraphics[width=253.5pt]{SU4_4six_dual}
\end{displaymath}
is an $\SU(2)$ gauging of the ${\Sp(4)}_6\times {\SU(2)}_8$ SCFT (this time, gauging the ${\SU(2)}_8$).

\hypertarget{_gauge_theory_28}{}
\subsection{{$\Sp(2)$ gauge theory}}\label{_gauge_theory_28}

Including the twisted sector of the $A_3$ theory allows us to write down Argyres-Seiberg duals for $\Sp(2)$ gauge theory, with matter in the $n(\mathbf{5}) +(6-2n)(\mathbf{4})$, for $n=0,1,2,3$.

\hypertarget{id53_29}{}\subsubsection{{$6(\mathbf{4})$}}\label{id53_29}

This classic example of Argyres-Seiberg duality is realized in the untwisted sector of the $A_3$ theory.
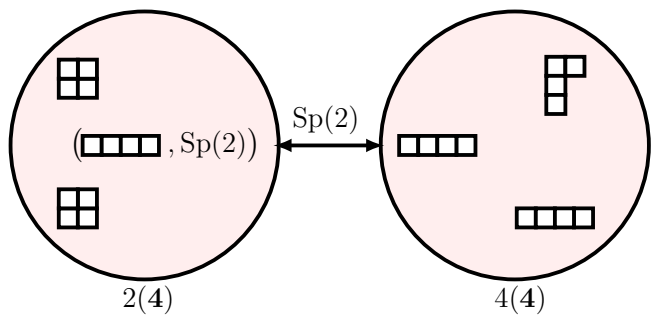
\begin{displaymath}
 \includegraphics[width=253.5pt]{Sp2_6four}
\end{displaymath}
is an $\Sp(2)$ gauge theory with matter in the $6(\mathbf{4})$.
The S-dual theory,
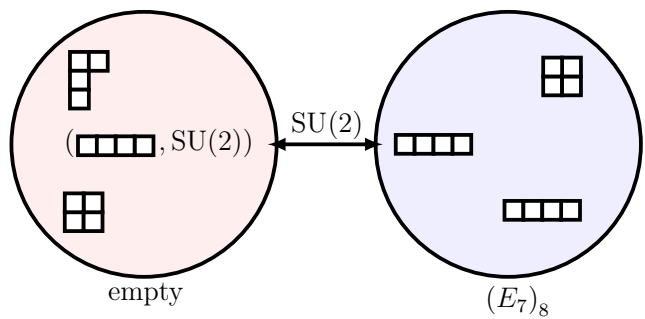
\begin{displaymath}
 \includegraphics[width=253.5pt]{Sp2_6four_dual}
\end{displaymath}
is an $\SU(2)$ gauging of the ${(E_7)}_8$ SCFT.

\hypertarget{Sp21544}{}\subsubsection{{$1(\mathbf{5})+4(\mathbf{4})$}}\label{Sp21544}

The 4-punctured sphere, of interest, has \emph{two} degeneration limits which correspond to $\Sp(2)$ gauge with matter in the $1(\mathbf{5})+4(\mathbf{4})$
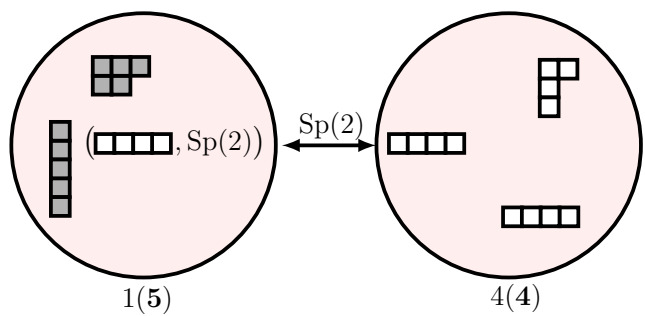
\begin{displaymath}
 \includegraphics[width=253.5pt]{Sp2_1five_4four}
\end{displaymath}
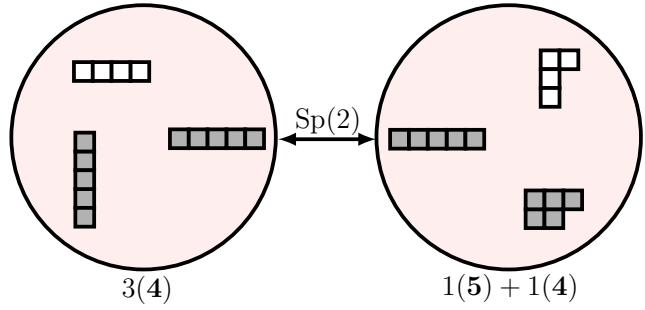
\begin{displaymath}
 \includegraphics[width=253.5pt]{Sp2_1five_4four_dual1}
\end{displaymath}
The remaining degeneration limit,
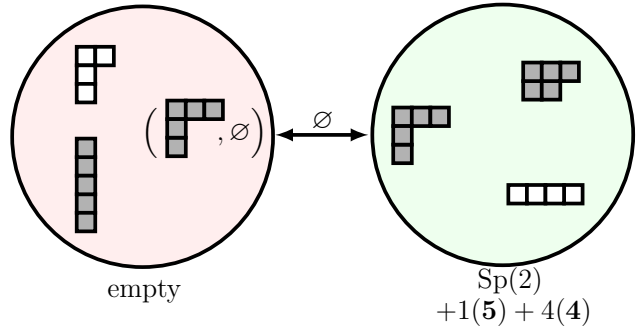
\begin{displaymath}
 \includegraphics[width=253.5pt]{Sp2_1five_4six_dual2}
\end{displaymath}
involves a gauge theory fixture.

\hypertarget{id55_31}{}\subsubsection{{$2(\mathbf{5})+2(\mathbf{4})$}}\label{id55_31}

$\Sp(2)$ gauge theory, with matter in the $2(\mathbf{5})+2(\mathbf{4})$ is realized by
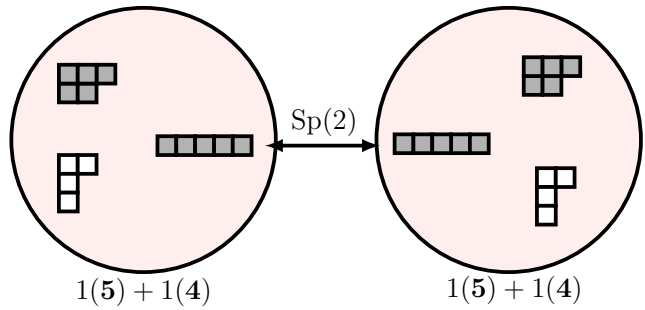
\begin{displaymath}
 \includegraphics[width=253.5pt]{Sp2_2five_2six}
\end{displaymath}
The S-dual theory
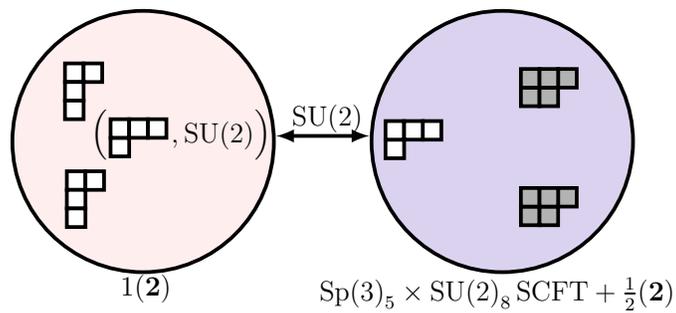
\begin{displaymath}
 \includegraphics[width=268.5pt]{Sp2_2five_2six_dual}
\end{displaymath}
is an $\SU(2)$ gauge theory, with \emph{three} half-hypermultiplets in the $\mathbf{2}$, coupled to the ${\Sp(3)}_5\times {\SU(2)}_8$ SCFT.
An alternative realization of the same family of theories (with the addition of one \emph{free} hypermultiplet) is given by
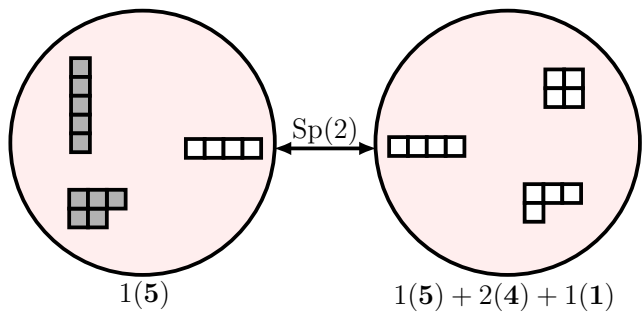
\begin{displaymath}
 \includegraphics[width=253.5pt]{Sp2_2five_2six_alt}
\end{displaymath}
The two other degenerations both give rise to an $\SU(2)$ gauge theory, with matter in the three half-hypermultiplets in $\mathbf{2}$, coupled to the ${\Sp(3)}_5\times{\SU(2)}_8$ SCFT.
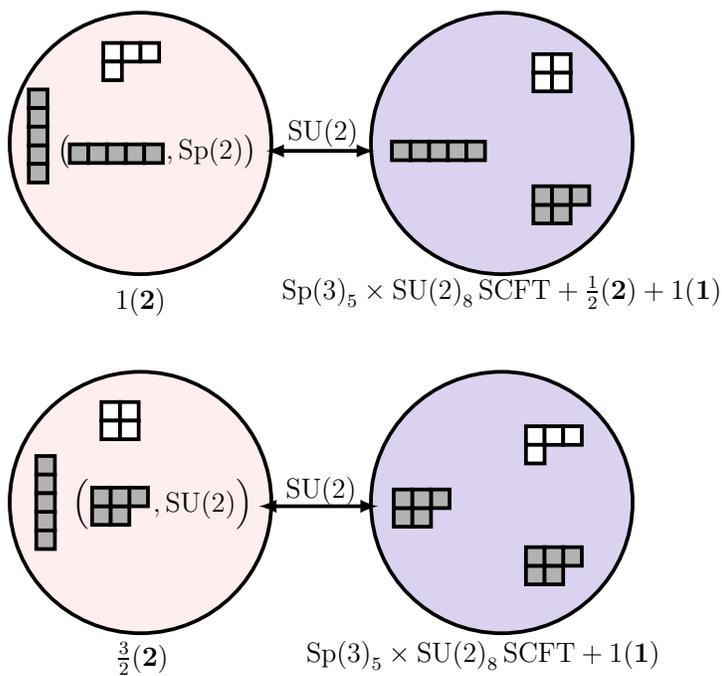
\begin{displaymath}
 \includegraphics[width=287.25pt]{Sp2_2five_2six_alt_dual}
\end{displaymath}

Note that this S-duality is the Example 13 of Argyres-Wittig \cite{Argyres:2007tq} and the ${\Sp(3)}_5\times{\SU(2)}_8$ SCFT is one of the \emph{new} rank-1 SCFT, to which we will come back in Sec.~\ref{rank1_scfts_33}.

\hypertarget{id56_32}{}\subsubsection{{$3(\mathbf{5})$}}\label{id56_32}

The best we can do, to capture $\Sp(2)$ with matter in the $3(\mathbf{5})$, is
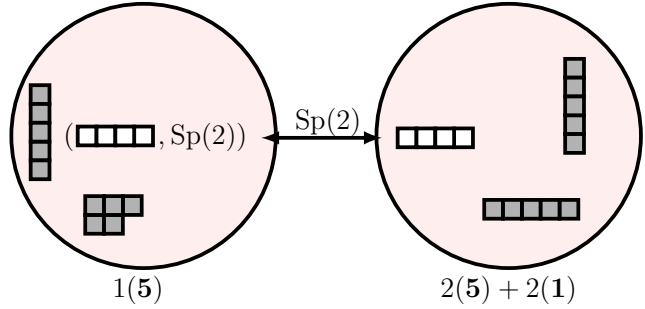
\begin{displaymath}
 \includegraphics[width=253.5pt]{Sp2_3five}
\end{displaymath}
which yields the desired gauge theory, with the addition of two free hypermultiplets.
The S-dual
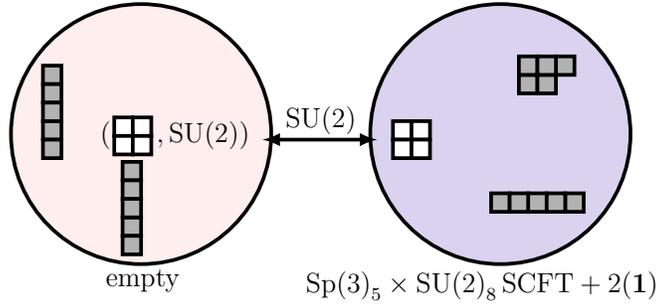
\begin{displaymath}
 \includegraphics[width=262.5pt]{Sp2_3five_dual}
\end{displaymath}
is an $\SU(2)$ gauging of the ${\Sp(3)}_5\times {\SU(2)}_8$ SCFT plus two free hypermultiplets.
Note that this S-duality is the Example 12 of Argyres-Wittig \cite{Argyres:2007tq}.

\subsection{A family of $\SU(2)\times \Sp(2)$ gauge theories}\label{curious}

In this subsection, we wish to discuss a 5-punctured sphere which ``{}interpolates''{} between two of the gauge theories discussed in the previous section, namely

\begin{itemize}%
\item $\Sp(2)$ with $6(\mathbf4)$,
\item $\Sp(2)$ with $1(\mathbf5)+4(\mathbf4)+3(\mathbf1)$.

\end{itemize}
We achieve this by gauging an $\SU(2)$ subgroup of the flavour-symmetry group of one of these theories. The first has flavour symmetry $\Spin(12)_{k=8}$; the second has flavour symmetry group ${\SU(2)}_5\times {\Spin(8)}_8\times$ 3 free hypers. In each case, gauging an $\SU(2)$ (where, in the latter, we treat the 3 free hypermultiplets as 3 half-hypers in the fundamental of $\SU(2)$) breaks the flavour-symmetry group to $F={\SU(2)}_8\times{\Spin(8)}_8$.

Taking the $\SU(2)$ from weak coupling to strong coupling, it again decouples, leaving us with the \emph{other} $\Sp(2)$ gauge theory.
Along the way, a crucial role is played by the gauge theory fixture,

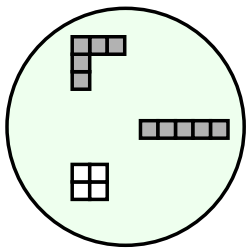
\begin{displaymath}
 \includegraphics[width=104pt]{A3gaugeth2}
\end{displaymath}
which is an $\SU(2)$ gauge theory, with matter in the $4(\mathbf2)+4(\mathbf1)$, at its $\mathbb{Z}_2$ symmetric point. Only an $\SU(2)\times \Sp(2)$ subgroup of the $\Spin(8)$ global symmetry of the gauge theory is manifestly realized by the punctures. Being at the $\mathbb{Z}_2$ symmetric point means that a $\mathbb{Z}_2$ subgroup of the outer-automorphisms of $\Spin(8)$ acts an automorphism of the SCFT. In particular, it exchanges two 8-dimensional representations of $\Spin(8)$ which transform, respectively, as $(\mathbf3,\mathbf1)+(\mathbf1,\mathbf5)$ and as $(\mathbf2,\mathbf4)$ under $\SU(2)\times \Sp(2)\subset \Spin(8)$. Thus, the matter can be interpreted \emph{either} as transforming as the $\tfrac{1}{2}(\mathbf3,\mathbf1;\mathbf2)+\tfrac{1}{2}(\mathbf1,\mathbf5;\mathbf2)$ \emph{or} as the $(\mathbf4,\mathbf2;\mathbf2)$ of $\SU(2)\times \Sp(2)\times {\SU(2)}_{\text{gauge}}$.

There are nine distinct degenerations of the 5-punctured sphere. The first three correspond to weakly-coupled descriptions with an $\SU(2)\times \Sp(2)$ gauge group and matter in the $\tfrac{3}{2}(\mathbf2,\mathbf1)+\tfrac{1}{2}(\mathbf2,\mathbf5)+4(\mathbf1,\mathbf4)$.

\hypertarget{degeneration_a_1}{}\paragraph*{{Degeneration A:}}
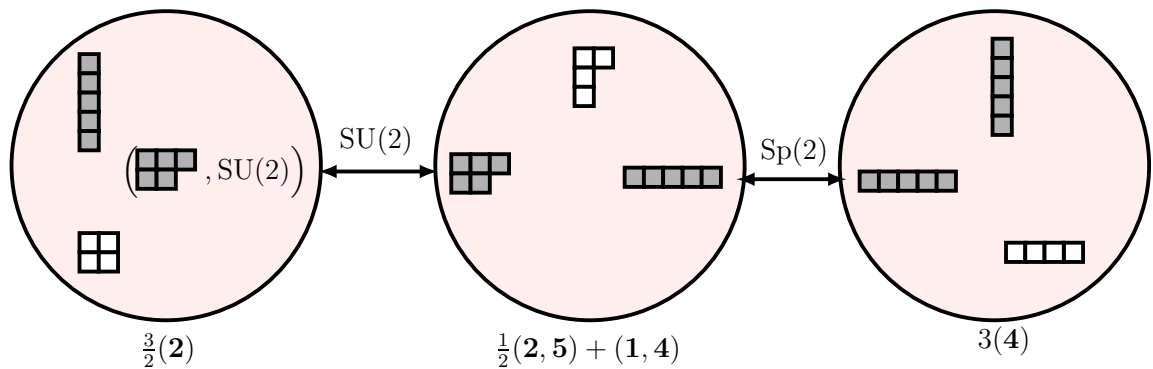
\label{degeneration_a_1}

\begin{displaymath}
 \includegraphics[width=446pt]{SU2Sp2deg1}
\end{displaymath}
\hypertarget{degeneration_b_2}{}\paragraph*{{Degeneration B:}}
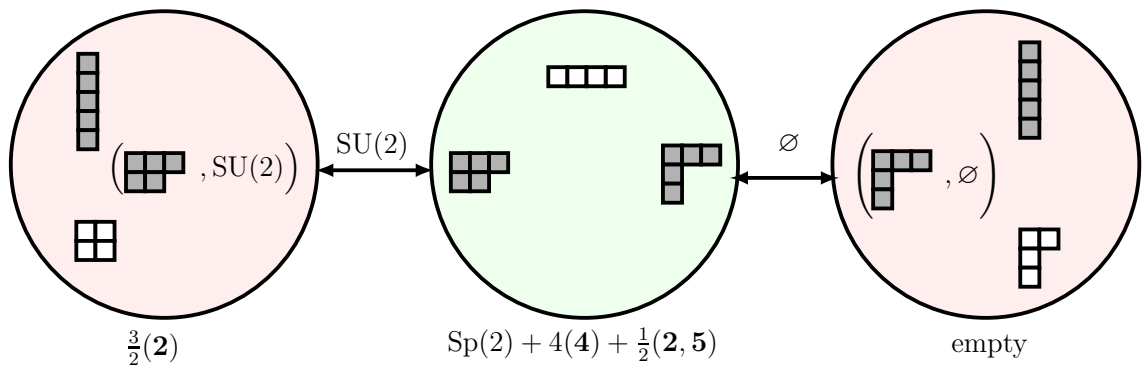
\label{degeneration_b_2}

\begin{displaymath}
 \includegraphics[width=446pt]{SU2Sp2deg2}
\end{displaymath}
\hypertarget{degeneration_c_3}{}\paragraph*{{Degeneration C:}}
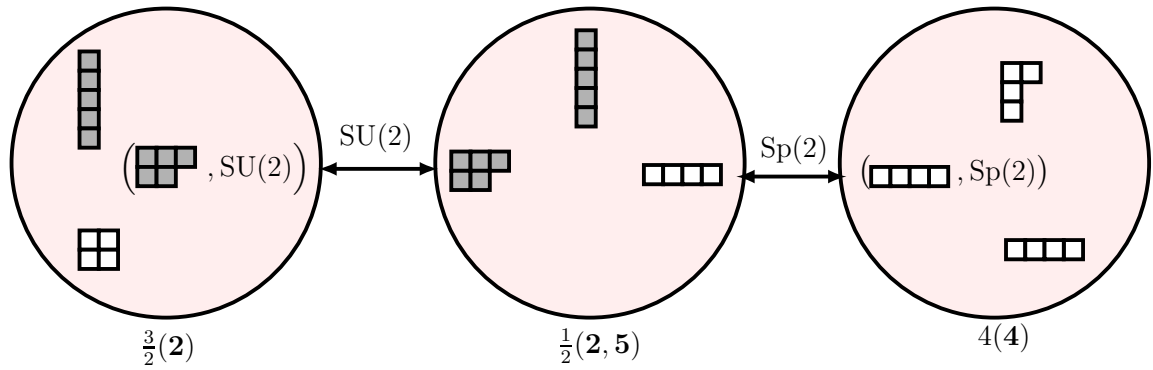
\label{degeneration_c_3}

\begin{displaymath}
 \includegraphics[width=446pt]{SU2Sp2deg3}
\end{displaymath}
The next two correspond to weakly-coupled descriptions with an $\SU(2)\times \Sp(2)$ gauge group and matter in the $(\mathbf2,\mathbf4)+4(\mathbf1,\mathbf4)$.

\hypertarget{degeneration_d_4}{}\paragraph*{{Degeneration D:}}
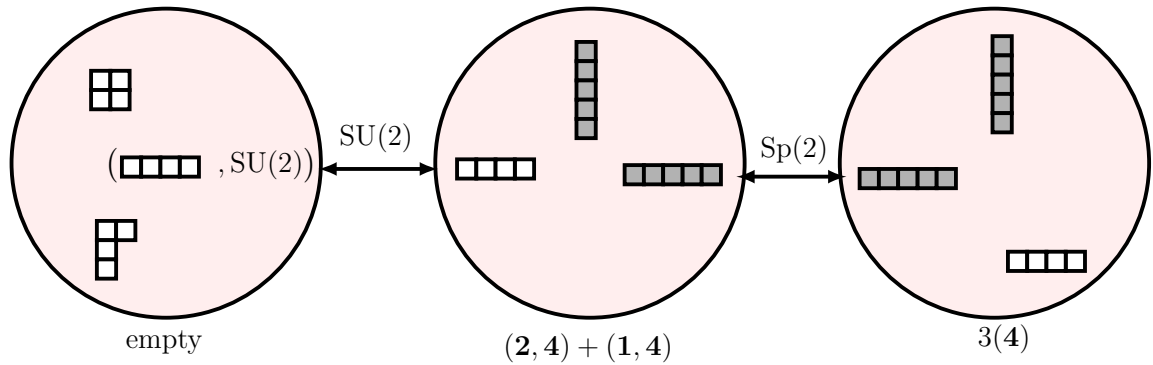
\label{degeneration_d_4}

\begin{displaymath}
 \includegraphics[width=446pt]{SU2Sp2deg5}
\end{displaymath}
\hypertarget{degeneration_e_5}{}\paragraph*{{Degeneration E:}}
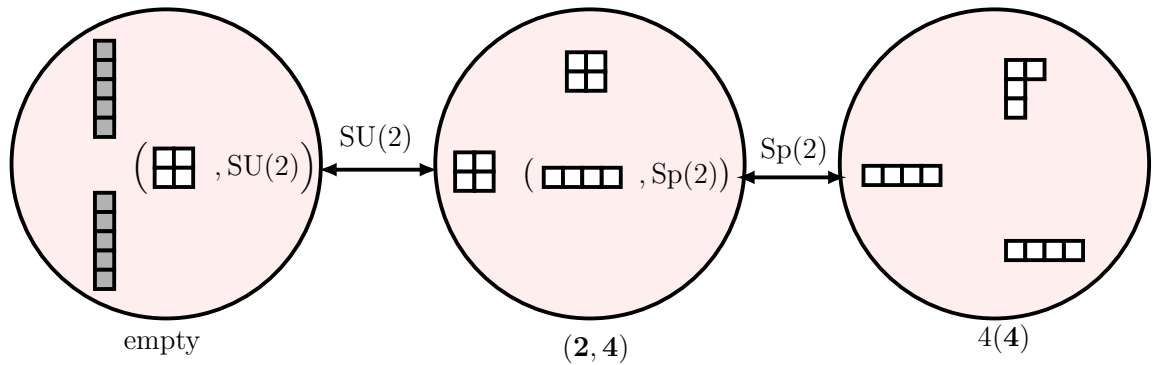
\label{degeneration_e_5}

\begin{displaymath}
 \includegraphics[width=446pt]{SU2Sp2deg6}
\end{displaymath}
Three more degenerations correspond to the S-dual description when the latter $\Sp(2)$ gauge theory is strongly-coupled: an $\SU(2)\times \SU(2)$ gauging of the ${(E_7)}_8$ SCFT.

\hypertarget{degeneration_f_6}{}\paragraph*{{Degeneration F:}}
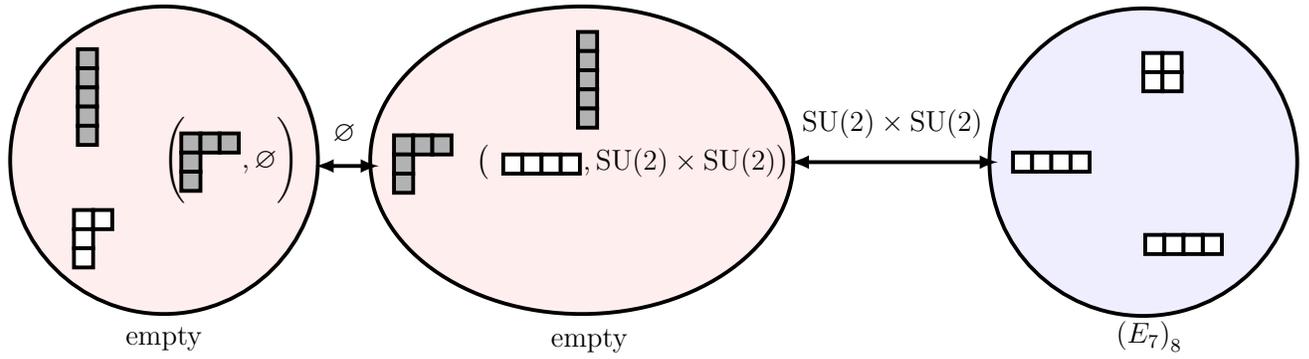
\label{degeneration_f_6}

\begin{displaymath}
 \includegraphics[width=506pt]{SU2Sp2deg7}
\end{displaymath}
\hypertarget{degeneration_g_7}{}\paragraph*{{Degeneration G:}}
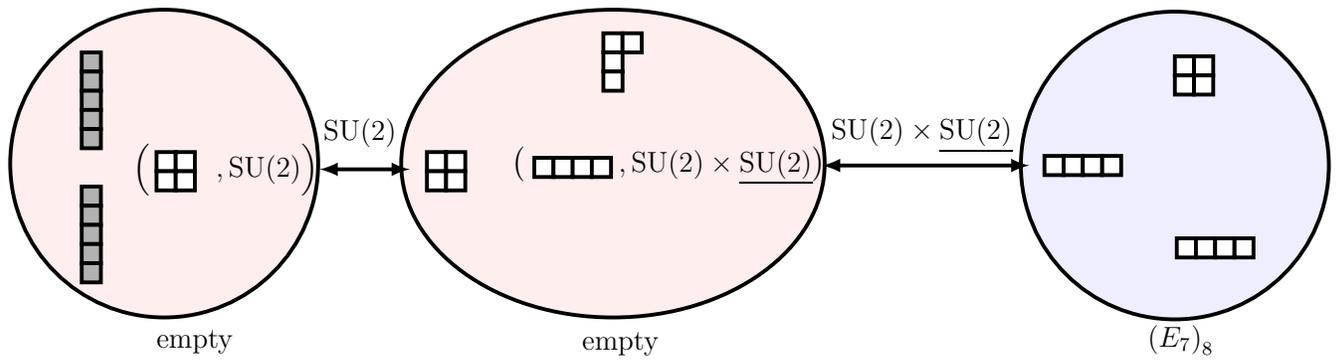
\label{degeneration_g_7}

\begin{displaymath}
 \includegraphics[width=518pt]{SU2Sp2deg8}
\end{displaymath}
\hypertarget{degeneration_h_8}{}\paragraph*{{Degeneration H:}}
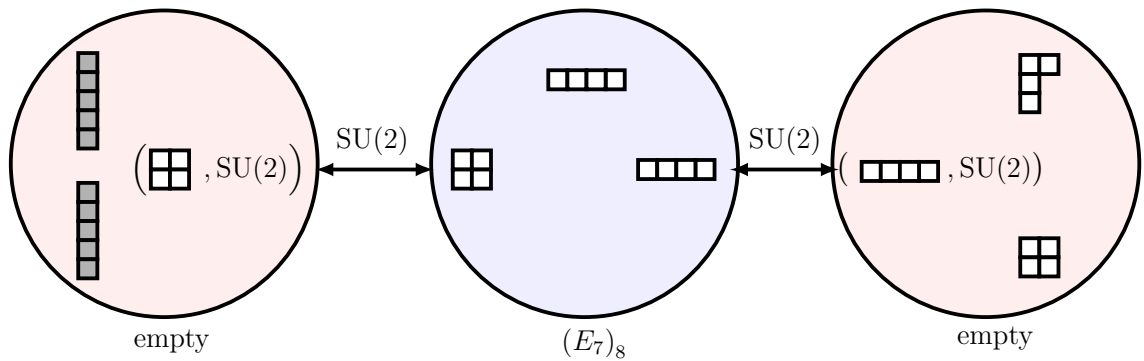
\label{degeneration_h_8}

\begin{displaymath}
 \includegraphics[width=446pt]{SU2Sp2deg9}
\end{displaymath}
Finally, comes the degeneration which ``{}interpolates''{} between these gauge theory descriptions

\hypertarget{degeneration_i_9}{}\paragraph*{{Degeneration I:}}
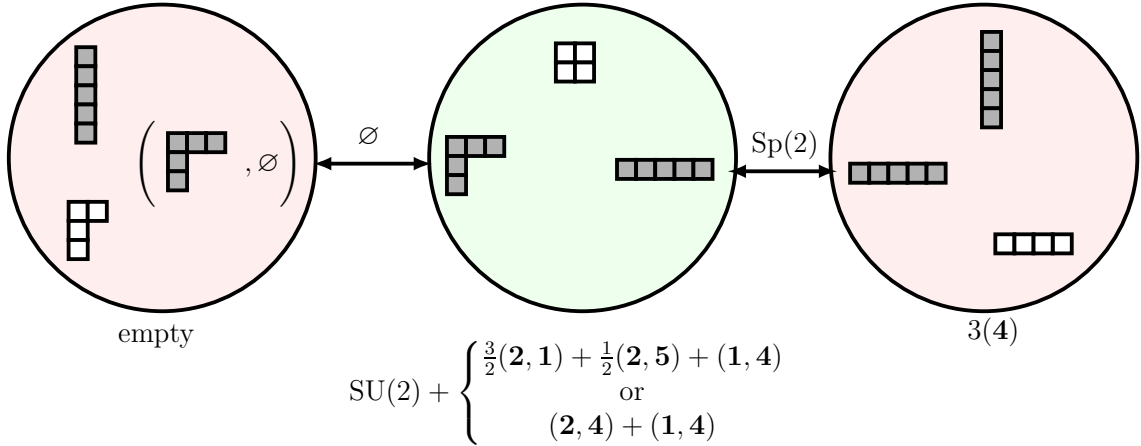
\label{degeneration_i_9}

\begin{displaymath}
 \includegraphics[width=446pt]{SU2Sp2deg4}
\end{displaymath}

\hypertarget{rank1_scfts_33}{}
\section{{Rank-1 SCFTs}}\label{rank1_scfts_33}
In this section, we summarize our current knowledge of rank-1 theories and their realizations, including constructions of new such theories using the twisted $A$ series.
\subsection{Summary of rank-1 SCFTs}
Recall that a rank-1 SCFT has, by definition, only one Coulomb branch operator, $u$. When all mass deformations are turned off, the scaling dimension $\Delta(u)$ of $u$ may only take the values $\tfrac{6}{5}$, $\tfrac{4}{3}$, $\tfrac{3}{2}$, $2$, $3$, $4$, or $6$, as reviewed in, e.g.,~\cite{Argyres:2005pp}. Examples of such SCFTs can be constructed from a D3-brane probing an F-theory 7-brane:
\begin{itemize}%
\item The SCFTs with $\Delta(u)=\tfrac{6}{5}$, $\tfrac{4}{3}$, or $\tfrac{3}{2}$, obtained in this way, can also be realized as the superconformal points of Argyres-Douglas type \cite{Argyres:1995jj,Argyres:1995xn}.
\item The SCFT with $\Delta(u)=2$ is the $\SU(2)$ gauge theory with four flavours.
\item The SCFTs with $\Delta(u)=3$, $4$, or $6$ are the interacting theories found in \cite{Minahan:1996fg,Minahan:1996cj}, with flavour symmetries $E_{6,7,8}$, respectively.

\end{itemize}
We call these SCFTs the \emph{old} rank-1 theories. For some time, these were thought to exhaust the list of rank-1 SCFTs. However, three new ones were found in \cite{Argyres:2007tq}:
\begin{itemize}%
\item An interacting SCFT with $\Delta(u)=6$, with $\Sp(5)_7$ flavour symmetry,
\item An interacting SCFT with $\Delta(u)=4$, with $\Sp(3)_5\times \SU(2)_8$ flavour symmetry, and
\item An interacting SCFT with $\Delta(u)=3$ theory, with unknown flavour symmetry.

\end{itemize}
We call these the \emph{new} rank-1 SCFTs.

Let us now see that both old and new rank-1 theories can be realized by compactifying a 6D $\mathcal{N}=(2,0)$ theory on a punctured curve. The constructions below are not necessarily unique; there are often distinct configurations that yield the same isolated SCFTs (possibly plus free hypermultiplets) in the 4D limit.
\begin{itemize}%
\item The old SCFTs with $\Delta(u)=\tfrac{6}{5}$, $\tfrac{4}{3}$, or $\tfrac{3}{2}$ can be obtained from the untwisted $A_1$ theory on a sphere with an irregular puncture and, possibly, a regular puncture; see, e.g., \cite{Bonelli:2011aa,Gaiotto:2012sf,Xie:2012hs}. (Here, we mean ``irregular'' in the sense of these papers, not in ours.)
\item The old SCFT with $\Delta(u)=2$ is obtained from the untwisted $A_1$ theory on a sphere with four full punctures.
\item The old SCFT with $\Delta(u)=3$ is obtained from the untwisted $A_2$ theory on a sphere with three full punctures, \cite{Gaiotto:2009we}.
\item The old SCFT with $\Delta(u)=4$ is obtained from the untwisted $A_3$ theory on a sphere with two full punctures and a puncture of type $[2^2]$ \cite{Benini:2009gi}.
\item The old SCFT with $\Delta(u)=6$ is obtained from the untwisted $A_5$ theory on a sphere with three punctures of types $[1^6]$, $[2^3]$ and $[3^2]$ \cite{Benini:2009gi}.
\item The new SCFT with $\Delta(u)=6$ is obtained from the untwisted $D_4$ theory on a sphere with two punctures of type $[3,2^2,1]$ and one puncture of type $[2^2,1^4]$. This realization includes three free hypers \cite{Chacaltana:2011ze}.
\item The new SCFT with $\Delta(u)=4$ is obtained from the $A_3$ theory on a sphere with an untwisted puncture of type $[2,1^2]$ and two twisted punctures of type $\BP{[2^2,1]}$. This realization comes with a free half-hypermultiplet in the fundamental of the $\SU(2)$ flavour symmetry. This is the example we studied in Sec.~\ref{id55_31}.

\end{itemize}

\subsection{On a new rank-1 SCFT with $\Delta(u)=3$}

To obtain the new SCFT with $\Delta(u)=3$, we need to extend our analysis to the twisted $A_{2n}$ theory. The $\mathbb{Z}_2$ twist of the $A_{2n}$ theory is particularly subtle, as emphasized in \cite{Tachikawa:2011ch}. Hence, we prefer to postpone a systematic analysis of the twisted $A_{2n}$ theory. Still, it is possible to show how to obtain the missing new SCFT from a 6D construction.

In \cite{Argyres:2007tq}, the new theory with $\Delta(u)=3$ is introduced in the following way. Consider the $\SU(3)$ gauge theory with one hyper in the fundamental and one hyper in the symmetric tensor representation. The S-dual theory is an $\SU(2)$ gauging of the new SCFT, coupled to $n$ half-hypermultiplets in the doublet. The field-theory arguments in \cite{Argyres:2010py} constrain $n$ to be 0 or 2, and require the flavour symmetry $\mathfrak{h}$ of the SCFT to satisfy $k_{\mathfrak{h}}=(8-n)/I$, where $I$ is the index of the embedding of $\mathfrak{su}(2)$ in $\mathfrak{h}$.

Now, the tensor product of two fundamentals of $\SU(3)$ decomposes as the direct sum of a fundamental plus a symmetric representation. The tensor product can in turn be obtained from the bifundamental of a product group $\SU(3)_1\times \SU(3)_2$, by taking a diagonal subgroup $\SU(3)_{diag}$, such that $\SU(3)_{diag}$ is embedded in $\SU(3)_1$ in the standard way, but embedded in $\SU(3)_2$ with the action of complex conjugation, i.e., the nontrivial outer automorphism.

So, consider the $A_2$ theory on a fixture with a simple puncture and two full punctures, which by itself simply gives rise to a bifundamental. Then, the $\SU(3)$ gauge theory with matter in the $1(\mathbf{3})+1(\mathbf{6})$ can be realized by connecting the full punctures in the fixture by a cylinder with a $\mathbb{Z}_2$ twist line looping around it. In other words, we have the $A_2$ theory on a torus with one simple puncture and a $\mathbb{Z}_2$ twist loop. See the left side of the figure below.
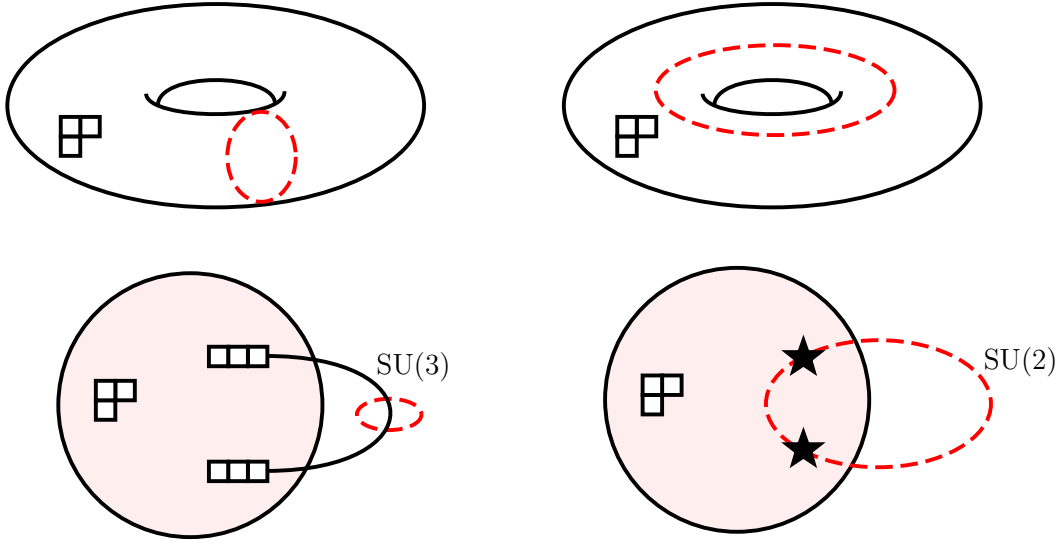
\begin{displaymath}
 \includegraphics[width=412.5pt]{A2toruswithatwist}
\end{displaymath}
In the S-dual frame, shown on the right side of the figure, we have a fixture with an untwisted simple puncture and two \emph{twisted} full punctures (denoted by $\includegraphics[width=11.25pt]{fullstar}$), and the full punctures are connected by a cylinder with a $\mathbb{Z}_2$-twist line \emph{along} the cylinder. Clearly, this gives a weakly-coupled $\SU(2)$ gauge field coupled to the fixture. Also, the flavour symmetry group of the fixture must contain the explicit $\SU(2)^2\times \UU(1)$ as a subgroup.

Thus, based on our findings, the fixture may be:
\begin{itemize}%
\item an interacting SCFT with flavour symmetry of rank 4 (if $n=0$),
\item an interacting SCFT with flavour symmetry of rank 3, plus free hypers in the $(2,1)+(\mathbf{1},\mathbf{2})$ of $\SU(2)^2$. The half-hypers are neutral under $\UU(1)$ (if $n=2$).
\end{itemize}
To see which of these two possibilities is the right one, recall \cite{Gaiotto:2011xs} that when the Riemann surface has two twisted (or untwisted) full punctures with flavour symmetry $G$, the holomorphic moment maps $\mu_1$, $\mu_2$ of the two $G$-actions on the Higgs branch must be equal,
\begin{equation}
\mathrm{tr}\,\mu_1{}^2 = \mathrm{tr}\,\mu_2{}^2
\label{holomapscondition}\end{equation}
Let us see what happens if $n=2$. In this case, the Higgs branch is $X\times \mathbb{H}_1 \times \mathbb{H}_2$, where $X$ is the Higgs branch of the interacting SCFT with an action of $\SU(2)_1\times \SU(2)_2$, and $\mathbb{H}_{1,2}$ is the Higgs branch for the $\SU(2)_{1,2}$ free hypers, respectively. Then we have
\begin{equation}
\mathrm{tr}\,\mu_i{}^2 = \mathrm{tr}\,\mu_{X,i}{}^2 + \mathrm{tr}\,\mu_{\mathbb{H}_i}^2
\end{equation}
for $i=1,2$. However, $\mathrm{tr}\,\mu_1{}^2$ depends on the position on $\mathbb{H}_1$, but not on the position on $\mathbb{H}_{2}$, while for $\mathrm{tr}\,\mu_2{}^2$ the opposite is true. Hence, $n=2$ does not satisfy the condition \eqref{holomapscondition}. 
Then, we conclude that $n=0$, and so the $A_2$ fixture with one untwisted simple puncture and two twisted full punctures contains just the new interacting rank-1 SCFT with $\Delta(u)=3$.

\section*{Acknowledgements}
The authors would like to thank Andrew Neitzke for very helpful conversations on the treatment of Hitchin fields. The authors would also like to thank Philip Argyres and Alfred Shapere for suggesting to YT the possibility of introduction of the twist fields in the type A theories in the first place, during the conference Quantum Theory and Symmetries held in July 2009.
The work of O.~C.~is supported in part by the INCT-Matem\'atica and the
ICTP-SAIFR in Brazil through a Capes postdoctoral fellowship.
The research of J.~D.~is based on work supported by the National Science Foundation under Grant No. PHY-0969020, and by the United States-Israel Binational Science Foundation under Grant \#2006157. The  work of Y.~T.~is  supported in part by World Premier International Research Center Initiative
(WPI Initiative),  MEXT, Japan through the Institute for the Physics and Mathematics
of the Universe, the University of Tokyo.

\appendix

\hypertarget{tables_34}{}

\hypertarget{list_of_twisted_punctures_35}{}
\section{{Tables of Properties of Twisted Sectors}}\label{list_of_twisted_punctures_35}

\hypertarget{_twisted_sector_36}{}\subsection{{$A_{3}$ twisted sector}}\label{_twisted_sector_36}
\subsubsection{Punctures}
\begin{tabular}{|c|c|c|c|c|c|}
\hline
\begin{tabular}{c}Flavour\\B-partition\end{tabular}&\begin{tabular}{c}Hitchin\\C-partition\end{tabular}&Pole structure&Constraints&Flavour group&$(\delta n_{h},\delta n_{v})$\\
\hline
\hline
$ \includegraphics[width=39pt]{A3twisted11111}$&$[4]$&$\{1,\frac{5}{2},3\}$&$-$&${\Sp(2)}_6$&$(40,\tfrac{73}{2})$\\
\hline
$ \includegraphics[width=24pt]{A3twisted311}$&$[2^{2}]$&$\{1,\frac{3}{2},3\}$&$c^{(4)}_{3}=(a^{(2)})^{2}$&$\UU(1)$&$(28,\tfrac{55}{2})$\\
\hline
$ \includegraphics[width=24pt]{A3twisted221}$&$([2^2],\mathbb{Z}_2)$&$\{1,\frac{3}{2},3\}$&$-$&${\SU(2)}_5$&$(33,\tfrac{63}{2})$\\
\hline
$ \includegraphics[width=9pt]{A3twisted5}$&$[1^{4}]$&$\{1,\frac{1}{2},2\}$&$c^{(4)}_{2}=\frac{1}{4}\left(c^{(2)}_{1}\right)^{2}$&none&$(12,\tfrac{25}{2})$\\
\hline 
\end{tabular}

\bigskip

Since the $A_3$ and $D_3$ (2,0) theories are isomorphic, the defects have labels in both descriptions. In $D_3$ notation, an untwisted puncture is labeled by a D-partition of 6, whereas a twisted one is labeled by a C-partition of 4. To facilitate comparison with the tables in \cite{Tachikawa:2009rb}, we list below the labels for the punctures in both descriptions.
\[
{\hbox{\begin{tabular}{|c|c|}
\multicolumn{2}{c}{\bfseries untwisted}\\
\hline 
Partition of 4 &D-partition of 6\\
($A_3$)& ($D_3$) \\
\hline 
$[1^4]$&$[1^6]$\\
$[2,1^2]$&$[2^2,1^2]$\\
$[2^2]$&$[3,1^3]$\\
$[3,1]$&$[3^2]$\\
$[4]$&$[5,1]$\\
\hline
\end{tabular}}}
\quad
{\hbox{\begin{tabular}{|c|c|}
\multicolumn{2}{c}{\bfseries  twisted}\\
\hline 
B-partition of 5 &C-partition of 4 \\
($A_3$)&($D_3$)\\
\hline 
$[1^5]$&$[1^4]$\\
$[3,1^2]$&$[2,1^2]$\\
$[2^2,1]$&$[2^2]$\\
$[5]$&$[4]$\\
\hline 
\end{tabular}}}
\]

\hypertarget{cylinders_38}{}\subsubsection{{Cylinders}}\label{cylinders_38}
\hypertarget{untwisted_39}{}\paragraph{{Untwisted}}
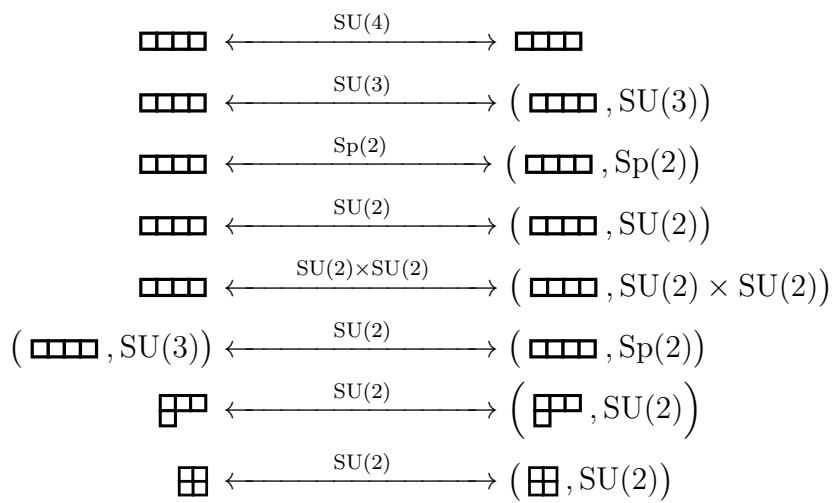
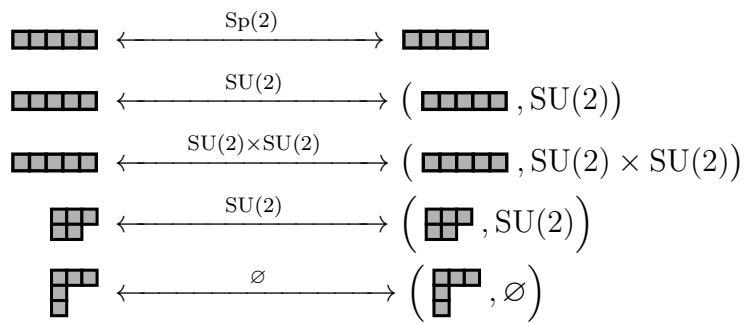
\label{untwisted_39}
\begin{displaymath}
\begin{aligned}
\begin{matrix} \includegraphics[width=31.5pt]{A31111}\end{matrix}
&\xleftrightarrow{\qquad\;\;\;\;\;\; \SU(4)\qquad\;\;\;\;\;\;}
\begin{matrix} \includegraphics[width=31.5pt]{A31111}\end{matrix}\\
\begin{matrix} \includegraphics[width=31.5pt]{A31111}\end{matrix}
&\xleftrightarrow{\qquad\;\;\;\;\;\; \SU(3)\qquad\;\;\;\;\;\;}
\left(\begin{matrix} \includegraphics[width=31.5pt]{A31111}\end{matrix},\SU(3)\right)\\
\begin{matrix} \includegraphics[width=31.5pt]{A31111}\end{matrix}
&\xleftrightarrow{\qquad\;\;\;\;\;\; \Sp(2)\qquad\;\;\;\;\;\;}
\left(\begin{matrix} \includegraphics[width=31.5pt]{A31111}\end{matrix},\Sp(2)\right)\\
\begin{matrix} \includegraphics[width=31.5pt]{A31111}\end{matrix}
&\xleftrightarrow{\qquad\;\;\;\;\;\; \SU(2)\qquad\;\;\;\;\;\;}
\left(\begin{matrix} \includegraphics[width=31.5pt]{A31111}\end{matrix},\SU(2)\right)\\
\begin{matrix} \includegraphics[width=31.5pt]{A31111}\end{matrix}
&\xleftrightarrow{\qquad \SU(2)\times \SU(2)\qquad}\left(\begin{matrix} \includegraphics[width=31.5pt]{A31111}\end{matrix},\SU(2)\times \SU(2)\right)\\
\left(\begin{matrix} \includegraphics[width=31.5pt]{A31111}\end{matrix},\SU(3)\right)
&\xleftrightarrow{\qquad\;\;\;\;\;\; \SU(2)\qquad\;\;\;\;\;\;}
\left(\begin{matrix} \includegraphics[width=31.5pt]{A31111}\end{matrix},\Sp(2)\right)\\
\begin{matrix} \includegraphics[width=24pt]{A3211}\end{matrix}
&\xleftrightarrow{\qquad\;\;\;\;\;\; \SU(2)\qquad\;\;\;\;\;\;}
\left(\begin{matrix} \includegraphics[width=24pt]{A3211}\end{matrix},\SU(2)\right)\\
\begin{matrix} \includegraphics[width=16.5pt]{A322}\end{matrix}
&\xleftrightarrow{\qquad\;\;\;\;\;\; \SU(2)\qquad\;\;\;\;\;\;}
\left(\begin{matrix} \includegraphics[width=16.5pt]{A322}\end{matrix},\SU(2)\right)\\
\end{aligned}
\end{displaymath}

\hypertarget{twisted_40}{}\paragraph{{Twisted}}\label{twisted_40}
\begin{displaymath}
\begin{aligned}
\begin{matrix} \includegraphics[width=39pt]{A3twisted11111}\end{matrix}
&\xleftrightarrow{\qquad\;\;\;\;\;\; \Sp(2)\qquad\;\;\;\;\;\;}
\begin{matrix} \includegraphics[width=39pt]{A3twisted11111}\end{matrix}\\
\begin{matrix} \includegraphics[width=39pt]{A3twisted11111}\end{matrix}
&\xleftrightarrow{\qquad\;\;\;\;\;\; \SU(2)\qquad\;\;\;\;\;\;}
\left(\begin{matrix} \includegraphics[width=39pt]{A3twisted11111}\end{matrix},\SU(2)\right)\\
\begin{matrix} \includegraphics[width=39pt]{A3twisted11111}\end{matrix}
&\xleftrightarrow{\qquad \SU(2)\times \SU(2)\qquad}
\left(\begin{matrix} \includegraphics[width=39pt]{A3twisted11111}\end{matrix},\SU(2)\times \SU(2)\right)\\
\begin{matrix} \includegraphics[width=24pt]{A3twisted221}\end{matrix}
&\xleftrightarrow{\qquad\;\;\;\;\;\; \SU(2)\qquad\;\;\;\;\;\;}
\left(\begin{matrix} \includegraphics[width=24pt]{A3twisted221}\end{matrix},\SU(2)\right)\\
\begin{matrix} \includegraphics[width=24pt]{A3twisted311}\end{matrix}
&\xleftrightarrow{\qquad\qquad \varnothing\qquad\qquad}
\left(\begin{matrix} \includegraphics[width=24pt]{A3twisted311}\end{matrix},\varnothing\right)\\
\end{aligned}
\end{displaymath}

\hypertarget{freefield_fixtures_41}{}\subsubsection{{Free-field fixtures}}\label{freefield_fixtures_41}

\begin{longtable}{|c|c|c|}
\hline 
Untwisted fixture&Number of Hypers&Representation\\
\hline 
\hline 
\endhead
$ \includegraphics[width=104.25pt]{A3freehypers1}$&2&$\mathbf{2}$ of $\SU(2)$\\
\hline 
$ \includegraphics[width=104.25pt]{A3freehypers2}$&0&empty\\
\hline 
$ \includegraphics[width=104.25pt]{A3freehypers3}$&6&$(\mathbf{2},\mathbf{3})$ of $\SU(2)\times \SU(3)$\\
\hline 
$ \includegraphics[width=104.25pt]{A3freehypers4}$&16&$(\mathbf{4},\mathbf{4})$ of $\SU(4)\times \SU(4)$\\
\hline
$ \includegraphics[width=104.25pt]{A3freehypers5}$&8&$\tfrac{1}{2}(\mathbf{2},\mathbf{2},\mathbf{4})$ of $\SU(2)\times \SU(2)\times \Sp(2)$\\
\hline
$ \includegraphics[width=104.25pt]{A3freehypers6}$&14&$(\mathbf2,\mathbf1,\mathbf4)+\tfrac{1}{2}(\mathbf{1},\mathbf{2},\mathbf{6})$ of $\SU(2)\times \SU(2)\times \SU(4)$\\
\hline 
\end{longtable}

\begin{longtable}{|c|c|c|}
\hline
Twisted fixture&Number of Hypers&Representation\\
\hline
\hline 
\endhead
$ \includegraphics[width=104.25pt]{A3twistedfreehypers1}$&0&empty\\
\hline 
$ \includegraphics[width=104.25pt]{A3twistedfreehypers2}$&3&$\tfrac{1}{2}(\mathbf{3},\mathbf{2})$ of $\SU(2)\times \SU(2)$\\
\hline 
$ \includegraphics[width=104.25pt]{A3twistedfreehypers3}$&2&$(\mathbf{1},\mathbf{2})$ of $\SU(2)\times \SU(2)$\\
\hline 
$ \includegraphics[width=104.25pt]{A3twistedfreehypers4}$&12&$\tfrac{1}{2}(\mathbf{6},\mathbf{4})$ of $\SU(4)\times \Sp(2)$\\
\hline 
$ \includegraphics[width=104.25pt]{A3twistedfreehypers5}$&0&empty\\
\hline 
$ \includegraphics[width=104.25pt]{A3twistedfreehypers6}$&5&$\tfrac{1}{2}(\mathbf{2},\mathbf{5})$ of $\SU(2)\times \Sp(2)$\\
\hline 
$ \includegraphics[width=104.25pt]{A3twistedfreehypers7}$&9&$\tfrac{1}{2}(\mathbf2,\mathbf5)+(\mathbf{1},\mathbf{4})$ of $\SU(2)\times \Sp(2)$\\
\hline 
$ \includegraphics[width=114.75pt]{A3twistedfreehypers8}$&0&empty\\
\hline 
$ \includegraphics[width=122.25pt]{A3twistedfreehypers9}$&4&$(\mathbf1,\mathbf2)+(\mathbf{2},\mathbf{1})$ of ${\SU(2)}\times \SU(2)$\\
\hline 
\end{longtable}

\hypertarget{gaugetheory_fixtures_44}{}\subsubsection{{Gauge-theory fixtures}}\label{gaugetheory_fixtures_44}

Gauge-theory fixtures, which represent a gauge theory with gauge group $G$ and coupling $\tau$ fixed at the $\mathbb{Z}_2$-symmetric point, are always twisted. These fixtures may also include free hypermultiplets.

\begin{longtable}{|c|c|c|c|}
\hline
Fixture&$G$&Number of Hypers&Representation\\
\hline 
\hline
\endhead
$ \includegraphics[width=104.25pt]{A3gaugeth1}$&$\SU(2)$&11&\begin{tabular}{c}$\tfrac{1}{2}(\mathbf2,\mathbf1;\mathbf2)+\tfrac{1}{2}(\mathbf2,\mathbf2;\mathbf1)+\tfrac{1}{2}(\mathbf2,\mathbf1;\mathbf1)$\\$+(\mathbf1,\mathbf2;\mathbf2)+(\mathbf{1},\mathbf{1};\mathbf{2})$ of\\${\SU(2)}^2\times G$\end{tabular}\\
\hline
$ \includegraphics[width=104.25pt]{A3gaugeth2}$&$\SU(2)$&12&\begin{tabular}{c}$\tfrac{1}{2}(\mathbf3,\mathbf1;\mathbf2)+\tfrac{1}{2}(\mathbf1,\mathbf5;\mathbf2)+(\mathbf1,\mathbf4;\mathbf1)$\\ \emph{or} $\tfrac{1}{2}(\mathbf2,\mathbf4;\mathbf2)+(\mathbf{1},\mathbf{4};\mathbf{1})$ of\\$\SU(2)\times \Sp(2)\times G$\end{tabular}\\
\hline
$ \includegraphics[width=104.25pt]{A3gaugeth3}$&$\SU(2)\times \SU(2)$&16&\begin{tabular}{c}$(\mathbf4;\mathbf2,\mathbf1)+(\mathbf4;\mathbf1,\mathbf2)$ of\\$\SU(4)\times G$\end{tabular}\\
\hline
$ \includegraphics[width=104.25pt]{A3gaugeth4}$&$\SU(3)$&18&\begin{tabular}{c}$(\mathbf4,\mathbf1;\mathbf3)+(\mathbf{1},\mathbf{2};\mathbf{3})$ of\\$\Sp(2)\times \SU(2)\times G$\end{tabular}\\
\hline
$ \includegraphics[width=104.25pt]{A3gaugeth5}$&$\Sp(2)$&21&\begin{tabular}{c}$(\mathbf4,\mathbf1;\mathbf4)+\tfrac{1}{2}(\mathbf{1},\mathbf{2};\mathbf{5})$ of\\$\SU(4)\times \SU(2)\times G$\end{tabular}\\
\hline
$ \includegraphics[width=104.25pt]{A3gaugeth6}$&$\SU(4)$&28&\begin{tabular}{c}$(\mathbf4,\mathbf1;\mathbf4)+\tfrac{1}{2}(\mathbf{1},\mathbf{4};\mathbf{6})$ of\\$\SU(4)\times \Sp(2)\times G$\end{tabular}\\
\hline
\end{longtable}

\hypertarget{interacting_scfts_46}{}\subsubsection{{Interacting SCFTs}}\label{interacting_scfts_46}

\begin{longtable}{|c|c|c|c|}
\hline
Untwisted fixture&$(d_2,d_3,d_4)$&$(n_h,n_v)$&${(G_{\text{global}})}_k$\\
\hline 
\hline
\endhead
$ \includegraphics[width=104.25pt]{A3interacting1}$&$(0,0,1)$&$(24,7)$&${(E_7)}_8$\\
\hline
$ \includegraphics[width=104.25pt]{A3interacting2}$&$(0,1,1)$&$(30,12)$&${\SU(2)}_6\times {\SU(8)}_8$\\
\hline
$ \includegraphics[width=104.25pt]{A3interacting3}$&$(0,1,2)$&$(40,19)$&${\SU(4)}_8^{3}$\\
\hline
\end{longtable}

\begin{longtable}{|c|c|c|c|}
\hline
Twisted fixture&$(d_2,d_3,d_4)$&$(n_h,n_v)$&${(G_{\text{global}})}_k$\\
\hline 
\hline
\endhead
$ \includegraphics[width=104.25pt]{A3twistedinteracting1}$&$(0,1,1)$&$(23,12)$&${\SU(2)}_5\times{\Sp(3)}_6 \times \UU(1)$\\
\hline
$ \includegraphics[width=104.25pt]{A3twistedinteracting2}$&$(0,1,2)$&$(33,19)$&${\SU(2)}_5\times {\SU(4)}_8\times {\Sp(2)}_6$\\
\hline
$ \includegraphics[width=104.25pt]{A3twistedinteracting3}$&$(0,0,2)$&$(26,14)$&${\SU(2)}_5^2\times {\SO(7)}_8$\\
\hline
$ \includegraphics[width=104.25pt]{A3twistedinteracting4}$&$(0,1,0)$&$(16,5)$&${(E_6)}_6$\\
\hline
$ \includegraphics[width=104.25pt]{A3twistedinteracting5}$&$(0,1,1)$&$(24,12)$&${\Sp(4)}_6\times {\SU(2)}_8$\\
\hline
$ \includegraphics[width=104.25pt]{A3twistedinteracting6}$&$(0,2,1)$&$(30,17)$&${\Sp(2)}_6^2\times {\SU(2)}_6\times \UU(1)$\\
\hline
$ \includegraphics[width=104.25pt]{A3twistedinteracting7}$&$(0,2,2)$&$(40,24)$&${\Sp(4)}_6\times {\SU(4)}_8$\\
\hline
\end{longtable}

\hypertarget{mixed_fixtures_49}{}\subsubsection{{Mixed Fixtures}}\label{mixed_fixtures_49}

\begin{center}
\begin{tabular}{|c|c|}
\hline 
Untwisted fixture&Theory\\
\hline 
$ \includegraphics[width=104.25pt]{A3mixed1}$&${(E_{6})_6}$ ${}+(\mathbf{1},\mathbf{1},\mathbf{4})$ of $\SU(2)\times \SU(2)\times \SU(4)$\\
\hline 
\end{tabular}
\end{center}

There is one \emph{new} SCFT, that appears as part of a mixed fixture in the twisted sector of the $A_3$ theory. It is the ${\Sp(3)}_5\times {\SU(2)}_8$ SCFT, which has $(d_2,d_3,d_4)=(0,0,1)$ and $(n_h,n_v)=(15,7)$.

\begin{longtable}{|c|c|}
\hline
Twisted fixture&Theory\\
\hline 
\hline
\endhead
$ \includegraphics[width=104.25pt]{A3twistedmixed1}$&${\Sp(3)}_5\times {\SU(2)}_8$ SCFT ${}+ \tfrac{1}{2}(\mathbf{1},\mathbf{1},\mathbf{4})$ of $\SU(2)\times \SU(2)\times \Sp(2)$\\
\hline
$ \includegraphics[width=104.25pt]{A3twistedmixed2}$&${\Sp(3)}_5\times {\SU(2)}_8$ SCFT ${}+ \tfrac{1}{2}(\mathbf{1},\mathbf{2},\mathbf{1})$ of $\SU(2)\times \SU(2)\times \SU(2)$\\
\hline
\end{longtable}

\hypertarget{_twisted_sector_52}{}\subsection{{$A_{5}$ twisted sector}}\label{_twisted_sector_52}

\begin{longtable}{|c|c|c|c|c|c|}
\hline
\begin{tabular}{c}Flavour\\B-partition\end{tabular}&\begin{tabular}{c}Hitchin\\C-partition\end{tabular}&Pole structure&Constraints&Flavour group&$(\delta n_{h},\delta n_{v})$\\
\hline 
\hline
\endhead
$[1^{7}]$&$[6]$&$\{1,\tfrac{5}{2},3,\tfrac{9}{2},5\}$&$-$&$\SO(7)$&$(140,132)$\\
\hline
$[3,1^{4}]$&$[4,2]$&$\{1,\tfrac{5}{2},3,\tfrac{7}{2},5\}$&$c^{(6)}_{5}=\left(a^{(3)}_{5/2}\right)^{2}$&$\SU(2)^{2}$&$(120,117)$\\
\hline
$[2^{2},1^{3}]$&$([4,2],\mathbb{Z}_2)$&$\{1,\tfrac{5}{2},3,\tfrac{7}{2},5\}$&$-$&$\SU(2)^{2}$&$(127,123)$\\
\hline
$[3^{2},1]$&$[2^{3}]$&$\{1,\tfrac{3}{2},3,\tfrac{7}{2},4\}$&$-$&$\UU(1)$&$(108,107)$\\
\hline
$[3,2^{2}]$&$[3^{2}]$&$\{1,\tfrac{5}{2},3,\tfrac{7}{2},5\}$&$c^{(6)}_{5}=\tfrac{1}{4}\left(c^{(3)}_{5/2}\right)^{2}$&$\SU(2)$&$(114,112)$\\
\hline
$[5,1^{2}]$&$[2^{2},1^{2}]$&$\{1,\tfrac{3}{2},3,\tfrac{7}{2},4\}$&$\begin{aligned}c^{(4)}_{3}&=\left(a^{(2)}_{3/2}\right)^{2}\\c^{(5)}_{7/2}&=2a^{(2)}_{3/2}a^{(3)}_{2}\\c^{(6)}_{4}&=\left(a^{(3)}_{2}\right)^{2}\end{aligned}$&$\UU(1)$&$(88,88)$\\
\hline
$[7]$&$[1^{6}]$&$\{1,\tfrac{3}{2},2,\tfrac{5}{2},3\}$&$\begin{aligned}c^{(4)}_{2}&=\tfrac{1}{4}\left(c^{(2)}_{1}\right)^{2}\\c^{(5)}_{5/2}&=\tfrac{1}{2}c^{(2)}_{1}c^{(3)}_{3/2}\\c^{(6)}_{3}&=\tfrac{1}{4}\left(c^{(3)}_{3/2}\right)^{2}\end{aligned}$&none&$(52,53)$\\
\hline
\end{longtable}

\hypertarget{_twisted_sector_53}{}\subsection{{$A_{7}$ twisted sector}}\label{_twisted_sector_53}

\small
\begin{longtable}{|c|c|c|c|c|c|}
\hline
\begin{tabular}{c}Flavour\\B-partition\end{tabular}&\begin{tabular}{c}Hitchin\\C-partition\end{tabular}&Pole structure&Constraints&
\begin{tabular}{c}Flavour\\group\end{tabular}&$(\delta n_{h},\delta n_{v})$\\
\hline 
\hline
\endhead
$[1^{9}]$&$[8]$&$\{1,\tfrac{5}{2},3,\tfrac{9}{2},5,\tfrac{13}{2},7\}$ &$-$&$\SO(9)$&$(336,\tfrac{643}{2})$\\
\hline
$[3,1^{6}]$&$[6,2]$&$\{1,\tfrac{5}{2},3,\tfrac{9}{2},5,\tfrac{11}{2},7\}$&$c^{(8)}_{7}=\left(a^{(4)}_{7/2}\right)^{2}$&$\SU(4)$&$(308,\tfrac{601}{2})$\\
\hline
$[2^{2},1^{5}]$&$([6,2],\mathbb{Z}_2)$&$\{1,\tfrac{5}{2},3,\tfrac{9}{2},5,\tfrac{11}{2},7\}$&$-$&\footnotesize $\Sp(2)\times \SU(2)$  &$(317,\tfrac{617}{2})$\\
\hline
$[3,2^{2},1^{2}]$&$[4^{2}]$&$\{1,\tfrac{5}{2},3,\tfrac{7}{2},5,\tfrac{11}{2},7\}$&$c^{(8)}_{7}=\left(a^{(4)}_{7/2}\right)^{2}$&\footnotesize $\SU(2)\times \UU(1)$ &$(296,\tfrac{583}{2})$\\
\hline
$[2^{4},1]$&$([4^2],\mathbb{Z}_2)$&$\{1,\tfrac{5}{2},3,\tfrac{7}{2},5,\tfrac{11}{2},7\}$&$-$&$\Sp(2)$&$(306,\tfrac{599}{2})$\\
\hline
$[3^{2},1^{3}]$&$[4,2^{2}]$&$\{1,\tfrac{5}{2},3,\tfrac{7}{2},5,\tfrac{11}{2},6\}$&$-$&\footnotesize $\SU(2)\times \UU(1)$ &$(288,\tfrac{569}{2})$\\
\hline
$[3^{3}]$&$[3^{2},2]$&$\{1,\tfrac{5}{2},3,\tfrac{7}{2},5,\tfrac{11}{2},6\}$&$c^{(6)}_{5}=\tfrac{1}{4}\left(c^{(3)}_{5/2}\right)^{2}$&$\SU(2)$&$(276,\tfrac{547}{2})$\\
\hline
$[5,1^{4}]$&$[4,2,1^{2}]$&$\{1,\tfrac{5}{2},3,\tfrac{7}{2},5,\tfrac{11}{2},6\}$&$\begin{aligned}c^{(6)}_{5}&=\left(a^{(3)}_{5/2}\right)^{2}\\c^{(7)}_{11/2}&=2a^{(3)}_{5/2} a^{(4)}_{3}\\c^{(8)}_{6}&=\left(a^{(4)}_{3}\right)^{2}\end{aligned}$&$\SU(2)^{2}$&$(260,\tfrac{515}{2})$\\
\hline
$[5,2^{2}]$&$[3^{2},1^{2}]$&$\{1,\tfrac{5}{2},3,\tfrac{7}{2},5,\tfrac{11}{2},6\}$&$\begin{aligned}c^{(6)}_{5}&=\tfrac{1}{4}\left(c^{(3)}_{5/2}\right)^{2}\\c^{(7)}_{11/2}&=c^{(3)}_{5/2} a^{(4)}_{3}\\c^{(8)}_{6}&=\left(a^{(4)}_{3}\right)^{2}\end{aligned}$&$\SU(2)$&$(254,\tfrac{505}{2})$\\
\hline
$[5,3,1]$&$[2^{4}]$&$\{1,\tfrac{3}{2},3,\tfrac{7}{2},4,\tfrac{9}{2},6\}$&$c^{(8)}_{6}=\left(a^{(4)}_{3}\right)^{2}$&none&$(248,\tfrac{495}{2})$\\
\hline
$[4^{2},1]$&$([2^4],\mathbb{Z}_2)$&$\{1,\tfrac{3}{2},3,\tfrac{7}{2},4,\tfrac{9}{2},6\}$&$-$&$\Sp(2)$&$(257,\tfrac{511}{2})$\\
\hline
$[7,1^{2}]$&$[2^{2},1^{4}]$&$\{1,\tfrac{3}{2},3,\tfrac{7}{2},4,\tfrac{9}{2},5\}$&
\footnotesize $\begin{aligned}
c^{(4)}_{3}&=\left(a^{(2)}_{3/2}\right)^{2}\\
c^{(5)}_{7/2}&=2a^{(2)}_{3/2} a^{(3)}_{2}\\
c^{(6)}_{4}&=2a^{(2)}_{3/2}a^{(4)}_{5/2}\\
&+\left(a^{(3)}_{2}\right)^{2}\\
c^{(7)}_{9/2}&=2a^{(3)}_{2} a^{(4)}_{5/2}\\
c^{(8)}_{5}&=\left(a^{(4)}_{5/2}\right)^{2}
\end{aligned}$
\normalsize
&$\UU(1)$&$(200,\tfrac{401}{2})$\\
\hline
$[9]$&$[1^{8}]$&$\{1,\tfrac{3}{2},2,\tfrac{5}{2},3,\tfrac{7}{2},4\}$&
\footnotesize $\begin{aligned}
c'^{(4)}_{2}&\equiv c^{(4)}_{2}-\tfrac{1}{4}\left(c^{(2)}_{1}\right)^{2}\\
c^{(5)}_{5/2}&=c^{(2)}_{1} c^{(3)}_{3/2}\\
c^{(6)}_{3}&=\tfrac{1}{2}c^{(2)}_{1} c'^{(4)}_{2}\\
&+\tfrac{1}{4}\left(c^{(3)}_{3/2}\right)^{2}\\
c^{(7)}_{7/2}&=\tfrac{1}{2}c^{(3)}_{3/2} c'^{(4)}_{2}\\
c^{(8)}_{4}&=\tfrac{1}{4}\left(c'^{(4)}_{2}\right)^2
\end{aligned}$
&none&$(136,\tfrac{275}{2})$\\
\hline
\end{longtable}
\normalsize

\hypertarget{_twisted_sector_54}{}\subsection{{$A_{9}$ twisted sector}}\label{_twisted_sector_54}

\small
\begin{longtable}{|c|c|c|c|c|c|}
\hline
\begin{tabular}{c}Flavour\\B-partition\end{tabular}&\begin{tabular}{c}Hitchin\\C-partition\end{tabular}&Pole structure&Constraints&\begin{tabular}{c}Flavour\\group\end{tabular}&$(\delta n_{h},\delta n_{v})$\\
\hline 
\hline
\endhead
$[1^{11}]$&$[10]$&$\{1,\tfrac{5}{2},3,\tfrac{9}{2},5,\tfrac{13}{2},7,\tfrac{17}{2},9\}$&-&$\SO(11)$&(660,637)\\
\hline
$[3,1^{8}]$&$[8,2]$&$\{1,\tfrac{5}{2},3,\tfrac{9}{2},5,\tfrac{13}{2},7,\tfrac{15}{2},9\}$&$c_{9}^{(10)}=(a_{9/2}^{(5)})^{2}$&$\SO(8)$&(624,610)\\
\hline
$[2^{2},1^{7}]$&$([8,2],\mathbb{Z}_2)$&$\{1,\tfrac{5}{2},3,\tfrac{9}{2},5,\tfrac{13}{2},7,\tfrac{15}{2},9\}$&-&\footnotesize $\SO(7)\times \SU(2)$&(635,620)\\
\hline
$[3,2^{2},1^{4}]$&$[6,4]$&$\{1,\tfrac{5}{2},3,\tfrac{9}{2},5,\tfrac{11}{2},7,\tfrac{15}{2},9\}$&$c_{9}^{(10)}=(a_{9/2}^{(5)})^{2}$&$\SU(2)^{3}$&(606,597)\\
\hline
$[2^{4},1^{3}]$&$([6,4],\mathbb{Z}_2)$&$\{1,\tfrac{5}{2},3,\tfrac{9}{2},5,\tfrac{11}{2},7,\tfrac{15}{2},9\}$&-&\footnotesize $\Sp(2)\times \SU(2)$&(618,607)\\
\hline
$[3,2^{4}]$&$[5^{2}]$&$\{1,\tfrac{5}{2},3,\tfrac{9}{2},5,\tfrac{11}{2},7,\tfrac{15}{2},9\}$&$c_{9}^{(10)}=\tfrac{1}{4}(c_{9/2}^{(5)})^{2}$&$\Sp(2)$&(596,588)\\
\hline
$[3^{2},1^{5}]$&$[6,2^{2}]$&$\{1,\tfrac{5}{2},3,\tfrac{9}{2},5,\tfrac{11}{2},7,\tfrac{15}{2},8\}$&-&\footnotesize $\Sp(2)\times \UU(1)$&(596,588)\\
\hline
$[3^{3},1^{2}]$&$[4^{2},2]$&$\{1,\tfrac{5}{2},3,\tfrac{7}{2},5,\tfrac{11}{2},7,\tfrac{15}{2},8\}$&$c_{7}^{(8)}=(a_{7/2}^{(4)})^{2}$&\footnotesize $\SU(2)\times \UU(1)$&(576,571)\\
\hline
$[3^{2},2^{2},1]$&$([4^{2},2],\mathbb{Z}_2)$&$\{1,\tfrac{5}{2},3,\tfrac{7}{2},5,\tfrac{11}{2},7,\tfrac{15}{2},8\}$&-&\footnotesize $\SU(2)\times \UU(1)$&(585,579)\\
\hline
$[4^{2},1^{3}]$&$[4,2^{3}]$&$\{1,\tfrac{5}{2},3,\tfrac{7}{2},5,\tfrac{11}{2},6,\tfrac{13}{2},8\}$&-&$\SU(2)^{2}$&(551,547)\\
\hline
$[5,3^{2}]$&$[3^{2},2^{2}]$&$\{1,\tfrac{5}{2},3,\tfrac{7}{2},5,\tfrac{11}{2},6,\tfrac{13}{2},8\}$&$\begin{aligned}c_{5}^{(6)}&=\tfrac{1}{4}(c_{5/2}^{(3)})^{2}\\c_{8}^{(10)}&=(a_{4}^{(5)})^{2}\\\end{aligned}$&$\UU(1)$&(528,526)\\
\hline
$[4^{2},3]$&$([3^{2},2^{2}],\mathbb{Z}_2)$&$\{1,\tfrac{5}{2},3,\tfrac{7}{2},5,\tfrac{11}{2},6,\tfrac{13}{2},8\}$&$c_{5}^{(6)}=\tfrac{1}{4}(c_{5/2}^{(3)})^{2}$&$\SU(2)$&(539,536)\\
\hline
$[5,1^{6}]$&$[6,2,1^{2}]$&$\{1,\tfrac{5}{2},3,\tfrac{9}{2},5,\tfrac{11}{2},7,\tfrac{15}{2},8\}$&$\begin{aligned}c_{7}^{(8)}&=(a_{7/2}^{(4)})^{2}\\c^{(9)}_{15/2}&=2a^{(4)}_{7/2}a^{(5)}_{4}\\c_{8}^{(10)}&=(a_{4}^{(5)})^{2}\end{aligned}$&$\SO(6)$&(560,553)\\
\hline
$[5,2^{2},1^{2}]$&$[4^{2},1^{2}]$&$\{1,\tfrac{5}{2},3,\tfrac{7}{2},5,\tfrac{11}{2},7,\tfrac{15}{2},8\}$&$\begin{aligned}c_{7}^{(8)}&=(a_{7/2}^{(4)})^{2}\\c_{15/2}^{(9)}&=2a_{7/2}^{(4)}a_{4}^{(5)}\\c_{8}^{(10)}&=(a_{4}^{(5)})^{2}\end{aligned}$&\footnotesize $\SU(2)\times \UU(1)$&(548,544)\\
\hline
$[5,3,1^{3}]$&$[4,2^{3}]$&$\{1,\tfrac{5}{2},3,\tfrac{7}{2},5,\tfrac{11}{2},6,\tfrac{13}{2},8\}$&$c_{8}^{(10)}=(a_{4}^{(5)})^{2}$&$\SU(2)$&(540,537)\\
\hline
$[5^{2},1]$&$[2^{5}]$&$\{1,\tfrac{3}{2},3,\tfrac{7}{2},4,\tfrac{9}{2},6,\tfrac{13}{2},7\}$&-&$\UU(1)$&(500,499)\\
\hline
$[7,1^{4}]$&$[4,2,1^{4}]$&$\{1,\tfrac{5}{2},3,\tfrac{7}{2},5,\tfrac{11}{2},6,\tfrac{13}{2},7\}$&$
\begin{aligned}c_{5}^{(6)}&=(a_{5/2}^{(3)})^{2}\\
c_{11/2}^{(7)}&=2a_{5/2}^{(3)}a_{3}^{(4)}\\
c_{6}^{(8)}&=2a_{5/2}^{(3)}a_{7/2}^{(5)}\\
&+(a_{3}^{(4)})^{2}\\
c_{13/2}^{(9)}&=2a_{3}^{(4)}a_{7/2}^{(5)}\\c_{7}^{(10)}&=(a_{7/2}^{(5)})^{2}\end{aligned}$&$\SU(2)^{2}$&(476,474)\\
\hline
$[7,2^{2}]$&$[3^{2},1^{4}]$&$\{1,\tfrac{5}{2},3,\tfrac{7}{2},5,\tfrac{11}{2},6,\tfrac{13}{2},7\}$&$
\begin{aligned}
c_{5}^{(6)}&=\tfrac{1}{4}(c_{5/2}^{(3)})^{2}\\
c_{11/2}^{(7)}&=c_{5/2}^{(3)}a_{3}^{(4)}\\
c_{6}^{(8)}&=c_{5/2}^{(3)}a_{7/2}^{(5)}\\
&+(a_{3}^{(4)})^{2}\\
c_{13/2}^{(9)}&=2a_{3}^{(4)}a_{7/2}^{(5)}\\c_{7}^{(10)}&=(a_{7/2}^{(5)})^{2}\end{aligned}$&$\SU(2)$&(470,469)\\
\hline
$[7,3,1]$&$[2^{4},1^{2}]$&$\{1,\tfrac{3}{2},3,\tfrac{7}{2},4,\tfrac{9}{2},6,\tfrac{13}{2},7\}$&$\begin{aligned}c_{6}^{(8)}&=(a_{3}^{(4)})^{2}\\c_{13/2}^{(9)}&=2a_{3}^{(4)}a_{7/2}^{(5)}\\c_{7}^{(10)}&=(a_{7/2}^{(5)})^{2}\end{aligned}$&none&(464,464)\\
\hline
$[9,1^{2}]$&$[2^{2},1^{6}]$&$\{1,\tfrac{3}{2},3,\tfrac{7}{2},4,\tfrac{9}{2},5,\tfrac{11}{2},6\}$&
$\begin{aligned}c_{3}^{(4)}&=(a_{3/2}^{(2)})^{2}\\
c_{7/2}^{(5)}&=2a_{2}^{(3)}a_{3/2}^{(2)}\\
c_{4}^{(6)}&=(a_{2}^{(3)})^{2}\\
&+2a_{5/2}^{(4)}a_{3/2}^{(2)}\\
c_{9/2}^{(7)}&=2a_{2}^{(3)}a_{5/2}^{(4)}\\
&+2a_{3/2}^{(2)}a_{3}^{(5)}\\
c_{5}^{(8)}&=(a_{5/2}^{(4)})^{2}\\
&+2a_{2}^{(3)}a_{3}^{(5)}\\
c_{11/2}^{(9)}&=2a_{5/2}^{(4)}a_{3}^{(5)}\\
c_{6}^{(10)}&=(a_{3}^{(5)})^{2}
\end{aligned}$&$\UU(1)$&(380,381)\\
\hline
$[11]$&$[1^{10}]$&$\{1,\tfrac{3}{2},2,\tfrac{5}{2},3,\tfrac{7}{2},4,\tfrac{9}{2},5\}$&
\footnotesize
$\begin{aligned}
c'^{(4)}_{2}&\equiv c_{2}^{(4)}-\tfrac{1}{4}\left(c_{1}^{(2)}\right)^2\\
c'^{(5)}_{5/2}&\equiv c_{5/2}^{(5)}-\tfrac{1}{2}c_{3/2}^{(3)}c^{(2)}_{1}\\
c_{3}^{(6)}&=\tfrac{1}{2}c_{1}^{(2)}c'^{(4)}_{2}\\
&+\tfrac{1}{4}(c_{3/2}^{(3)})^{2}\\
c_{7/2}^{(7)}&=\tfrac{1}{2}c_{3/2}^{(3)}c'^{(4)}_{2}\\
&+\tfrac{1}{2}c_{1}^{(2)}c'^{(5)}_{5/2}\\
c_{4}^{(8)}&=\tfrac{1}{4}\left(c'^{(4)}_{2}\right)^{2}\\
&+\tfrac{1}{2}c_{3/2}^{(3)}c'^{(5)}_{5/2}\\
c_{9/2}^{(9)}&=\tfrac{1}{2}c'^{(4)}_{2}c'^{(5)}_{5/2}\\
c_{5}^{(10)}&=\tfrac{1}{4}\left(c'^{(5)}_{5/2}\right)^{2}
\end{aligned}$&none&(280,282)\\
\hline
\end{longtable}
\normalsize

\bibliographystyle{ytphys}
\bibliography{ref}

\end{document}